%% file: Nf3_EMFF.tex
\documentclass[12pt,prd,aps,floatfix]{revtex4}

\usepackage{graphicx}
\usepackage{amsmath}
\usepackage{amssymb}
\usepackage{longtable}

\newcommand{\be}{\begin{equation}}
\newcommand{\ee}{\end{equation}}
\newcommand{\bea}{\begin{eqnarray}}
\newcommand{\eea}{\end{eqnarray}}
\newcommand{\bi}{\begin{itemize}}
\newcommand{\ei}{\end{itemize}}
\newcommand{\ben}{\begin{enumerate}}
\newcommand{\een}{\end{enumerate}}
\newcommand{\bt}{\begin{tabbing}}
\newcommand{\et}{\end{tabbing}}
\newcommand{\nn}{\nonumber}

\newcommand{\sgn}{{\rm sgn}}
\newcommand{\calO}{{\mathcal O}}
\newcommand{\Hw}{H_{\rm W}}

\newcommand{\bfp}{{\bf p}}

\newcommand{\bfx}{{\bf x}}
\newcommand{\bfr}{{\bf r}}
\newcommand{\bfpp}{{{\bf p}^\prime}}
\newcommand{\bfxp}{{{\bf x}^\prime}}
\newcommand{\bfxpp}{{{\bf x}^{\prime\prime}}}
\newcommand{\pff}{F_V^{\pi^+}}
\newcommand{\pfflo}{F_{V,0}^{\pi^+}}
\newcommand{\pffnlo}{F_{V,2}^{\pi^+}}
\newcommand{\pffnnlo}{F_{V,4}^{\pi^+}}
\newcommand{\pffnlol}{F_{V,2,l}^{\pi^+}}
\newcommand{\pffnlob}{F_{V,2,b}^{\pi^+}}
\newcommand{\pffnnlol}{F_{V,4,l}^{\pi^+}}
\newcommand{\pffnnlor}{F_{V,4,r}^{\pi^+}}
\newcommand{\pffnnlob}{F_{V,4,b}^{\pi^+}}
\newcommand{\ff}{F_V^{P}}
\newcommand{\fflo}{F_{V,0}^P}
\newcommand{\ffnlo}{F_{V,2}^P}
\newcommand{\ffnnlo}{F_{V,4}^P}
\newcommand{\ffnnnlo}{F_{V,6}^P}
\newcommand{\ffnloL}{F_{V,2,L}^P}
\newcommand{\ffnloB}{F_{V,2,B}^P}
\newcommand{\ffnnloL}{F_{V,4,L}^P}
\newcommand{\ffnnloC}{F_{V,4,C}^P}
\newcommand{\ffnnloB}{F_{V,4,B}^P}
\newcommand{\pffnloL}{F_{V,2,L}^{\pi^+}}
\newcommand{\pffnloB}{F_{V,2,B}^{\pi^+}}
\newcommand{\pffnnloL}{F_{V,4,L}^{\pi^+}}
\newcommand{\pffnnloC}{F_{V,4,C}^{\pi^+}}

\newcommand{\pffnnnlo}{F_{V,6}^{\pi^+}}
\newcommand{\kpff}{F_{V}^{K^+}}
\newcommand{\kpfflo}{F_{V,0}^{K^+}}
\newcommand{\kpffnlo}{F_{V,2}^{K^+}}
\newcommand{\kpffnnlo}{F_{V,4}^{K^+}}
\newcommand{\kpffnloL}{F_{V,2,L}^{K^+}}
\newcommand{\kpffnloB}{F_{V,2,B}^{K^+}}
\newcommand{\kpffnnloL}{F_{V,4,L}^{K^+}}
\newcommand{\kpffnnloC}{F_{V,4,C}^{K^+}}

\newcommand{\kpffnnnlo}{F_{V,6}^{K^+}}
\newcommand{\knff}{F_V^{K^0}}
\newcommand{\knfflo}{F_{V,0}^{K^0}}
\newcommand{\knffnlo}{F_{V,2}^{K^0}}
\newcommand{\knffnnlo}{F_{V,4}^{K^0}}
\newcommand{\knffnloL}{F_{V,2,L}^{K^0}}
\newcommand{\knffnloB}{F_{V,2,B}^{K^0}}
\newcommand{\knffnnloL}{F_{V,4,L}^{K^0}}
\newcommand{\knffnnloC}{F_{V,4,C}^{K^0}}

\newcommand{\knffnnnlo}{F_{V,6}^{K^0}}
\newcommand{\pkpff}{F_{V}^{\{\pi^+,K^+\}}}
\newcommand{\pkpffnlo}{F_{V,2}^{\{\pi^+,K^+\}}}
\newcommand{\pkpffnloL}{F_{V,2,L}^{\{\pi^+,K^+\}}}

\newcommand{\pkpffnnlo}{F_{V,4}^{\{\pi^+,K^+\}}}
\newcommand{\pkpffnnloL}{F_{V,4,L}^{\{\pi^+,K^+\}}}
\newcommand{\pkpffnnloC}{F_{V,4,C}^{\{\pi^+,K^+\}}}
\newcommand{\pkpffnnloB}{F_{V,4,B}^{\{\pi^+,K^+\}}}
\newcommand{\crad}{{\langle r^2 \rangle_V^{P}}}
\newcommand{\cradpff}{{\langle r^2 \rangle_V^{\pi^+}}}
\newcommand{\cradpffnlo}{{\langle r^2 \rangle_{V,2}^{\pi^+}}}
\newcommand{\cradpffnlol}{{\langle r^2 \rangle_{V,2,l}^{\pi^+}}}
\newcommand{\cradpffnlob}{{\langle r^2 \rangle_{V,2,b}^{\pi^+}}}
\newcommand{\cradpffnnlo}{{\langle r^2 \rangle_{V,4}^{\pi^+}}}
\newcommand{\cradpffnnlol}{{\langle r^2 \rangle_{V,4,l}^{\pi^+}}}
\newcommand{\cradpffnnlor}{{\langle r^2 \rangle_{V,4,r}^{\pi^+}}}
\newcommand{\cradpffnnlob}{{\langle r^2 \rangle_{V,4,b}^{\pi^+}}}
\newcommand{\cradpffnnnlo}{{\langle r^2 \rangle_{V,6}^{\pi^+}}}
\newcommand{\cradnlo}{{\langle r^2 \rangle_{V,2}^P}}
\newcommand{\cradnloL}{{\langle r^2 \rangle_{V,2,L}^P}}
\newcommand{\cradnloB}{{\langle r^2 \rangle_{V,2,B}^P}}
\newcommand{\cradnnlo}{{\langle r^2 \rangle_{V,4}^P}}
\newcommand{\cradnnloL}{{\langle r^2 \rangle_{V,4,L}^P}}
\newcommand{\cradnnloC}{{\langle r^2 \rangle_{V,4,C}^P}}
\newcommand{\cradnnloB}{{\langle r^2 \rangle_{V,4,B}^P}}
\newcommand{\cradnnnlo}{{\langle r^2 \rangle_{V,6}^P}}
\newcommand{\cradpffnloL}{{\langle r^2 \rangle_{V,2,L}^{\pi^+}}}
\newcommand{\cradpffnloB}{{\langle r^2 \rangle_{V,2,B}^{\pi^+}}}

\newcommand{\cradpffnnloC}{{\langle r^2 \rangle_{V,4,C}^{\pi^+}}}

\newcommand{\cradkpff}{{\langle r^2 \rangle_V^{K^+}}}

\newcommand{\cradkpffnloL}{{\langle r^2 \rangle_{V,2,L}^{K+}}}
\newcommand{\cradkpffnloB}{{\langle r^2 \rangle_{V,2,B}^{K^+}}}

\newcommand{\cradkpffnnloC}{{\langle r^2 \rangle_{V,4,C}^{K^+}}}

\newcommand{\cradkpffnnnlo}{{\langle r^2 \rangle_{V,6}^{K^+}}}
\newcommand{\cradknff}{{\langle r^2 \rangle_V^{K^0}}}
\newcommand{\cradknffnlo}{{\langle r^2 \rangle_{V,2}^{K^0}}}
\newcommand{\cradknffnloL}{{\langle r^2 \rangle_{V,2,L}^{K^0}}}
\newcommand{\cradknffnloB}{{\langle r^2 \rangle_{V,2,B}^{K^0}}}

\newcommand{\cradknffnnloC}{{\langle r^2 \rangle_{V,4,C}^{K^0}}}

\newcommand{\cradknffnnnlo}{{\langle r^2 \rangle_{V,6}^{K^0}}}

\newcommand{\dt}{{\Delta x_4}}
\newcommand{\dtp}{{\Delta x_4^\prime}}

\newcommand{\Ap}{{\bar{A}(M_\pi^2)}}
\newcommand{\Ak}{{\bar{A}(M_K^2)}}
\newcommand{\Bttppt}{{\bar{B}_{22}(M_\pi^2,M_\pi^2,t)}}
\newcommand{\Bttkkt}{{\bar{B}_{22}(M_K^2,M_K^2,t)}}

\newcommand{\Cppt}{{c_{\pi^+,\pi t}^r}}
\newcommand{\Cpkt}{{c_{\pi^+,K t}^r}}
\newcommand{\Ctt}{{c_{t^2}^r}}
\newcommand{\Ckpt}{{c_{K^+,\pi t}^r}}
\newcommand{\Ckkt}{c_{K^+,K t}^r}
\newcommand{\Ckn}{{c_{K^0}^r}}

\newcommand{\Dkn}{{d_{K^0}}}

\begin{document}

\input{0.title}

\input{1.intro}

\input{2.simulation}

\input{3.ff}

\input{4.chiral_fit_su2}

\input{5.chiral_fit_su3}

\input{6.conclusion}

\appendix

\input{A.1-loop_integrals}

\input{B.2-loop_integrals}

\input{Z.reference}

\end{document}

%% file: 0.title.tex
\vspace*{-10mm}
\begin{flushright}
\normalsize
 KEK-CP-325        \\
 YITP-15-70        \\
\end{flushright}

\title{
Light meson electromagnetic form factors
from three-flavor lattice QCD with exact chiral symmetry
}

\author{
   S.~Aoki$^{a,b}$, 
   G.~Cossu$^{c}$, 
   X.~Feng$^{d}$, 
   S.~Hashimoto$^{c,e}$, 
   T.~Kaneko$^{c,e}$, 
   J.~Noaki$^{c}$ and 
   T.~Onogi$^{f}$
   (JLQCD Collaboration)
}

\affiliation{
   $^a$Yukawa Institute for Theoretical Physics,
   Kyoto University, 
   Kyoto 606-8502, Japan
   \\
   $^b$Center for Computational Sciences, University of Tsukuba, 
   Ibaraki 305-8577, Japan
   \\
   $^c$High Energy Accelerator Research Organization (KEK),
   Ibaraki 305-0801, Japan 
   \\
   $^d$Physics Department, Columbia University, 
   New York, NY 10027, USA
   \\
   $^e$School of High Energy Accelerator Science,
   SOKENDAI (The Graduate University for Advanced Studies),
   Ibaraki 305-0801, Japan
   \\
   $^f$Department of Physics, Osaka University, 
   Osaka 560-0043, Japan
}

\date{\today}

\begin{abstract}

We study the chiral behavior of the electromagnetic (EM) form factors of 
pion and kaon in three-flavor lattice QCD. 
In order to make a direct comparison of the lattice data
with chiral perturbation theory (ChPT), 
we employ the overlap quark action that has exact chiral symmetry.
Gauge ensembles are generated at a lattice spacing of 0.11~fm
with four pion masses ranging between $M_\pi \simeq 290$~MeV and 540~MeV
and with a strange quark mass $m_s$ close to its physical value.
We utilize the all-to-all quark propagator technique
to calculate the EM form factors with high precision.
Their dependence on $m_s$ and on the momentum transfer
is studied by using the reweighting technique and 
the twisted boundary conditions for the quark fields, respectively.
A detailed comparison with SU(2) and SU(3) ChPT reveals that
the next-to-next-to-leading order terms in the chiral expansion are important
to describe the chiral behavior of the form factors
in the pion mass range studied in this work.
We estimate the relevant low-energy constants and the charge radii,
and find reasonable agreement with phenomenological and experimental results.

\end{abstract}


\maketitle

%% file: 1.intro.tex

\section{Introduction}


Rapid increase of computational power
and improvements of simulation algorithms allow 
us to perform large-scale simulations of unquenched lattice QCD 
in the chiral regime,
where the non-perturbative dynamics is characterized by chiral symmetry. 
Chiral perturbation theory (ChPT)~\cite{ChPT:SU2:NLO,ChPT:SU3:NLO}
is an effective theory in this regime,
though its Lagrangian has unknown parameters,
called low-energy constants (LECs).
A detailed comparison between lattice QCD and ChPT 
may validate numerical lattice calculations
and analytical predictions of ChPT.
This also provides a first-principle determination of LECs,
and hence widens the applicability of ChPT to different physical observables.

In such a program,
chiral symmetry plays an essential role.
But, it is violated in most of the existing lattice calculations,
and the comparison had to be made after carefully taking the continuum limit.
Effects of the explicit violation 
by the use of conventional Wilson and staggered fermion formulations 
on the lattice
were studied at next-to-leading order 
(NLO) in ChPT~\cite{WChPT:SS,SChPT:LS,WChPT:RS,SChPT:AB,WChPT:A,SChPT:SV}: 
in general, 
it modifies the functional form of the ChPT expansion of physical observables, 
and introduces additional unknown LECs. 
It is therefore not clear how one can disentangle
the next-to-next-to-leading order (NNLO) corrections,
which are significant in kaon physics,
from the extra terms due to the explicit chiral violation.
Lattice QCD with exact chiral symmetry provides 
a clean framework 
for an unambiguous comparison between lattice QCD and ChPT.
The JLQCD and TWQCD collaborations have performed 
such simulations employing the overlap quark action~\cite{Overlap:NN,Overlap:N},
and studied the chiral behavior of various observables
in detail~\cite{JLQCD:overlap:summary}.


Pion and kaon electromagnetic (EM) form factors
are fundamental quantities in ChPT.
The charged
pion EM form factor $\pff$ is defined through the matrix element 
of the EM current $J_\mu$ sandwiched by the pion states
\bea
   \langle P(p^\prime) | J_\mu | P(p) \rangle 
   & = &
   \left( p + p^\prime \right)_\mu \ff(t),
   \hspace{5mm}
   t=(p-p^\prime)^2,
   \label{eqn:intro:ff}
   \\
   J_\mu 
   & = & 
  \frac{2}{3} \bar{u}\gamma_\mu u
 -\frac{1}{3} \bar{d}\gamma_\mu d
 -\frac{1}{3} \bar{s}\gamma_\mu s,
\eea
where $|P(p)\rangle$ specifies the light meson state
(charged pion $P\!=\!\pi^+$, to be explicit) of momentum $p$,
and $t=(p-p^\prime)^2$ is the momentum transfer.
This form factor is known up to NNLO both in 
SU(2) ChPT~\cite{ChPT:SU2:NLO,PFF:ChPT:Nf2:NNLO:GM,PFF:ChPT:SU2:NNLO:BCT},
where the dependence on the strange quark mass $m_s$ is implicitly encoded 
in LECs, and in SU(3) ChPT with strange mesons
as dynamical degrees of freedom~\cite{PFF:ChPT:SU3:NLO,PFF+KFF:ChPT:NNLO:Nf3}.
Detailed analyses of experimental data based on NNLO ChPT
have led to precise estimates of the charge radius~\cite{PFF:ChPT:SU2:NNLO:BCT,PFF+KFF:ChPT:NNLO:Nf3},
\bea
   \crad
   & = & 
   6 \left. \frac{\partial \ff(t)}{\partial t} \right|_{t=0},
   \label{intro:radius}
\eea
which can be used as a benchmark of lattice calculations.
Its dependence on the momentum transfer $t$ and 
mass of degenerate up and down quarks $m_l$ has been studied 
in unquenched lattice QCD~\cite{PFF:Nf2:DBW2+DW:RBC,PFF:Nf2:Plq+Clv:HKL,PFF:Nf3:impG+AT+DWF:LHP,PFF:Nf2:Plq+Clv:JLQCD,PFF:Nf2:Plq+Clv:QCDSF,PFF:Nf3:RG+DW:RBC+UKQCD,PFF:Nf2:Sym+tmW:ETMC,PFF:JLQCD:Nf2:RG+Ovr,PFF:Nf3:RG+Clv:PACS-CS,PFF:Nf2:impG+Clv:Mainz,PFF:Nf3:impG+HISQ:HPQCD}.
Recent detailed comparisons with SU(2) ChPT~\cite{PFF:Nf2:Sym+tmW:ETMC,PFF:JLQCD:Nf2:RG+Ovr,PFF:Nf3:RG+Clv:PACS-CS,PFF:Nf2:impG+Clv:Mainz,PFF:Nf3:impG+HISQ:HPQCD} show that 
lattice data at the pion mass $M_\pi\!\lesssim\!500$~MeV 
are described reasonably well by the NNLO chiral expansion,
and reproduce the experimental value of the pion charge radius.
The NNLO contribution turns out to be non-negligible
in accordance with the two-loop ChPT analysis~\cite{PFF:ChPT:SU2:NNLO:BCT}.
This test has not yet been extended to SU(3) ChPT,
in which the $m_s$ dependence of $\pff$ and $\cradpff$ 
is explicitly taken into account.


The EM form factors of the charged and neutral kaons are similarly defined 
through Eq.~(\ref{eqn:intro:ff}) with $P\!=\!K^+$ and $K^0$, respectively.
Since strange valence quarks are involved,
we need SU(3) ChPT 
to describe their chiral behavior~\footnote{
Another way is to treat kaons as heavy particles~\cite{HKChPT}.
One introduces effective interactions involving kaons,
which are restricted only by SU(2) chiral symmetry. 
The number of LECs increases and the EM form factors are not known at NNLO.
}.
These form factors are known up to NNLO~\cite{PFF+KFF:ChPT:NNLO:Nf3}.
The $m_s$ expansion is expected to have poorer convergence 
than that in terms of $m_l$ due to $m_s \!\gg\! m_l$.
A detailed examination of the convergence and 
first-principle determination of relevant LECs are helpful 
for a better understanding of kaon physics:
for instance, a phenomenologically important form factors of 
the $K\!\to\!\pi$ semileptonic decays share LECs 
with the EM form factors~\cite{KFF:weak:ChPT:SU3:NNLO:PS,KFF:weak:ChPT:SU3:NNLO:BT}.
There has been no lattice calculation nor detailed comparison with ChPT
to our knowledge.


In the present work,
we calculate the pion and kaon EM form factors 
in three-flavor lattice QCD.
We employ the overlap quark action~\cite{Overlap:NN,Overlap:N}
to maintain exact chiral symmetry 
for a direct comparison of our lattice data with ChPT up to NNLO.
The form factors are precisely calculated
using the all-to-all quark propagator~\cite{A2A:SESAM,A2A:TrinLat}.
We also utilize the reweighting technique~\cite{reweight:1,reweight:2} 
and the twisted boundary conditions~\cite{TBC}
to study their dependence on $m_s$ and $t$, respectively.
We compare their chiral behavior
with NNLO SU(2) and SU(3) ChPT in detail,
and present an estimate of the relevant LECs and charge radii.
Our preliminary analysis has been reported in Ref.~\cite{Lat10:JLQCD:Kaneko}.


This paper is organized as follows.
Section~\ref{sec:simulation} introduces 
our method to generate the gauge ensembles
and to calculate relevant light meson correlators. 
The EM form factors are extracted 
at the simulation points in Section~\ref{sec:ff}.
We then study the chiral behavior of the form factors 
based on NNLO SU(2) and SU(3) ChPT
in Sections~\ref{sec:chiral_fit:su2} and \ref{sec:chiral_fit:su3},
respectively.
We summarize our conclusions in Section~\ref{sec:conclusion}.

%% file: 2.simulation.tex

\section{Simulation method}
\label{sec:simulation}

\subsection{Configuration generation}


We simulate $N_f\!=\!2+1$ QCD, 
in which strange quark has a distinct mass 
from degenerate up and down quarks.
We employ the Iwasaki gauge action \cite{Iwasaki} 
and the overlap quark action \cite{Overlap:NN,Overlap:N}.
The Dirac operator of the latter is given by 
\bea
   D(m_q) 
   & = &
   \left( 1 - \frac{m_q}{2m_0} \right) D(0) + m_q,
   \label{eqn:sim:conf_gen:overlap}
   \\
   D(0) 
   & = & 
   m_0 \left( 1 + \gamma_5 \, \sgn \left[ \Hw (-m_0) \right] \right).
   \label{eqn:sim:conf_gen:overlap:m0}
\eea   
Here $m_q$ represents the quark mass, whereas 
$-m_0$ is the mass parameter of the Hermitian Wilson-Dirac operator $H_W$
appearing in the construction of the overlap fermion as a kernel.
We set $m_0\!=\!1.6$ 
so that 
the overlap-Dirac operator $D(m_q)$ 
has good locality~\cite{Prod_Run:JLQCD:Nf2:RG+Ovr}.
This action exactly preserves chiral symmetry at finite lattice spacing
\cite{lat_chial_sym}.
This enables us to directly compare the lattice results for the form factors
at a finite lattice spacing 
with ChPT in the continuum limit,
where the NNLO chiral expansion is available.

%
%

We introduce an auxiliary determinant~\cite{exW:Vranas,exW+extmW:JLQCD}
\bea
   \Delta_{\rm W} 
   & = & 
   \frac{\det[\Hw(-m_0)^2]}{\det[\Hw(-m_0)^2+\mu^2]}
   \hspace{5mm}
   (\mu=0.2)
   \label{eqn:sim:conf_gen:det}
\eea
%
%
into the Boltzmann weight in the generation of the gauge ensembles.
This suppresses exact- and near-zero modes of $\Hw(-m_0)$,
and hence remarkably reduces the computational cost 
without changing the continuum limit of the theory.
Another interesting property of $\Delta_{\rm W}$ 
is that the global topological charge $Q$ is unchanged 
during the update of the gauge fields 
with the Hybrid Monte Carlo (HMC) algorithm.
In this study, 
we simulate trivial topological sector, $Q=0$.
We note that local topological excitations are active,
and the topological susceptibility is consistent with the ChPT expectation~\cite{chi_t:JLQCD}.
The effect of the fixed global topology is a part of 
finite volume effect,
which is suppressed by the inverse of the space-time volume~\cite{fixed_Q:AFHO}.


We set the gauge coupling $\beta\!=6/g^2\!=\!2.30$,
where the lattice spacing 
determined from the $\Omega$ baryon mass is $a\!=\!0.112(1)$\,fm.
We perform simulations at four values of degenerate up and down quark mass $m_l$
that cover a range of $M_\pi\!\sim\!290$\,--\,540~MeV.
The gauge ensembles are generated at a strange quark mass
$m_s\!=\!0.080$, which is close to its physical value $m_{s,\rm phys}\!=\!0.081$.
The EM form factors at a different value $m_s\!=\!0.060$ 
are calculated by the reweighting method~\cite{reweight:1,reweight:2}.

We set a spatial lattice extent to $N_s \!=\! L/a \!=\! 24$ 
at $m_l\!\leq\!0.025$ and to 16 at $m_l\!\geq\!0.035$ 
in order to control finite volume effects 
by satisfying a condition $M_\pi L \gtrsim 4$.
The additional finite volume effect due to the fixed global topology 
turned out to be small in our previous study in $N_f\!=\!2$ QCD
on similar or even smaller lattice volumes.
The temporal lattice size is fixed to $N_t \!=\! T/a \!=\! 48$. 
At each combination of $m_l$ and $m_s$,
we generate 50 gauge configurations separated by 50 HMC trajectories.
The statistical error quoted in this article is estimated
by a single-elimination jackknife method.
Our simulation parameters are summarized in Table~\ref{tbl:method:param}.

\begin{ruledtabular}
\begin{table}[t]
\begin{center}
\caption{
   Simulation parameters. 
   Meson masses, $M_\pi$ and $M_K$ are in units of MeV.
}
\label{tbl:method:param}
\begin{tabular}{l|llll|llll}
   lattice               & $m_l$   & $m_s$   & $M_\pi$  & $M_K$  
                         & $\theta$ 
   \\ \hline
   $16^3 \!\times\! 48$  & 0.050   & 0.080   & 540(4)  & 617(4)
                         & 0.00, 0.40, 0.96, 1.60
   \\
   $16^3 \!\times\! 48$  & 0.035   & 0.080   & 453(4)  & 578(4)
                         & 0.00, 0.60, 1.28, 1.76
   \\
   $24^3 \!\times\! 48$  & 0.025   & 0.080   & 379(2)  & 548(3)
                         & 0.00, 1.68, 2.64
   \\
   $24^3 \!\times\! 48$  & 0.015   & 0.080   & 293(2)  & 518(3)
                         & 0.00, 1.68, 2.64
   \\ \hline
   $16^3 \!\times\! 48$  & 0.050   & 0.060   & 540(4)  & 567(4)
                         & 0.00, 0.40, 0.96, 1.60
   \\
   $16^3 \!\times\! 48$  & 0.035   & 0.060   & 451(7)  & 524(5)
                         & 0.00, 0.60, 1.28, 1.76
   \\
   $24^3 \!\times\! 48$  & 0.025   & 0.060   & 378(7)  & 492(7)
                         & 0.00, 1.68, 2.64
   \\
   $24^3 \!\times\! 48$  & 0.015  & 0.060    & 292(3)  & 459(4)
                         & 0.00, 1.68, 2.64
\end{tabular}
\end{center}
\vspace{0mm}
\end{table}
\end{ruledtabular}

\subsection{Calculation of meson correlators}

We employ the all-to-all quark propagator~\cite{A2A:SESAM,A2A:TrinLat} 
in order to improve statistical accuracy of the meson correlators.
Let us consider an expansion of the quark propagator $D(m_q)^{-1}$
in terms of the eigenmodes of the overlap operator $D(m_q)$,
where $m_q$ $(q=l,s)$ represents the valence quark mass.
Light meson observables including the EM form factors
are expected to large contributions from the low-lying modes.
We calculate this important part by 
\bea
   \left\{D(m_q)^{-1}\right\}_{\rm low}(x,y)
   & = &
   \sum_{k=1}^{N_e} \frac{1}{\lambda_k^{(q)}} u_k(x) u_k^{\dagger}(y),
   \label{eqn:sim:a2a_prop:low}
\eea
where 
$\lambda_k^{(q)}$ represents the $k$-th lowest eigenvalue of $D(m_q)$,
and $u_k$ is the normalized eigenvector associated with $\lambda_k^{(q)}$.
Note that 
the overlap action has advantages in solving the eigenvalue problem:
i) the eigenvector does not depend on $m_q$, 
which only changes the normalization and the additive shift of $D$
(see Eq.~(\ref{eqn:sim:conf_gen:overlap})),
and ii) the left and right eigenvectors are equal to each other,
since $D$ is normal.
We employ the implicitly restarted Lanczos algorithm
to calculate the low-modes, 
the number of which is $N_e\!=\!240$ (160) 
on the $24^3 \! \times \! 48$ ($16^3 \!\times\! 48$) lattice.

The remaining contribution from higher eigenmodes is evaluated stochastically 
by the noise method~\cite{noise} with the dilution technique \cite{A2A:TrinLat}.
We prepare a complex $Z_2$ noise vector for each configuration,
and split it into $N_d = 3 \times 4 \times N_t/2$ vectors
$\eta_d(x) (d\!=\!1,\ldots,N_d)$,
each of which has non-zero elements only for a single combination of 
color and spinor indices and at two consecutive time-slices.
The high-mode contribution can be estimated as 
\bea
   \left\{ D(m_q)^{-1}\right\}_{\rm high}(x,y) 
   & = & 
   \sum_{d=1}^{N_d} x_d^{(q)}(x)\,\eta_d^{\dagger}(y)
   \label{eqn:sim:a2a_prop:high}
\eea
by solving a linear equation for each diluted source
\bea
   D(m_q)\,x_d^{(q)}
   = 
   P_{\rm high} \, \eta_d
   \hspace{5mm}           
   (d=1,\ldots,N_d).
   \hspace{2mm}           
   \label{eqn:sim:a2a_prop:high:leq}
\eea
Here $P_{\rm high}\!=\!1-P_{\rm low}$,
and $P_{\rm low} \!=\! \sum_{k=1}^{N_e} u_k\, u_k^{\dagger}$ 
is the projector to the eigenspace spanned by the low-modes.

The typical size of the momentum transfer is 
$|t| \!\gtrsim\! (500~\mbox{MeV})^2$
on our lattice of size $L \sim 1.8$\,--\,2.7~fm,
if we insert the meson momenta by using the Fourier transformation
with the standard periodic boundary condition.
Our previous study in two-flavor QCD~\cite{PFF:JLQCD:Nf2:RG+Ovr}
suggested that 
the next-to-next-to-next-to-leading order (N$^3$LO) correction
to the pion form factor $\pff$ can be sizable in this region of $t$.
In order to suppress such higher order contributions,
which are not known in ChPT, 
we simulate near-zero momentum transfers $|t| \!\lesssim\! (300~\mbox{MeV})^2$ 
by employing the twisted boundary condition~\cite{TBC}
for the valence quarks
\bea
   q(\bfx+L \hat{k},x_4) 
   = 
   e^{i\theta} q(\bfx,x_4),
   \hspace{3mm}
   \bar{q}(\bfx+L \hat{k},x_4) 
   = 
   e^{-i\theta} \bar{q}(\bfx,x_4)
   \hspace{5mm}
   (k=1,2,3),
   \label{eqn:sim:msn_corr:tbc}
\eea
where $\hat{k}$ is a unit vector in the $k$-th direction.
We set a common twist angle $\theta$ in all three spatial directions
for simplicity.
This boundary condition induces a quark momentum
of $p_k = \theta/L \leq 2\pi/L$. 
We choose the angles listed in Table~\ref{tbl:method:param},
so that $|t| \lesssim (300~\mbox{MeV})^2$, 
where the N$^3$LO correction to $\pff$ is expected to be insignificant.

We calculate the all-to-all quark propagator for each choice of $\theta$. 
By combining 
Eqs.~(\ref{eqn:sim:a2a_prop:low}) and (\ref{eqn:sim:a2a_prop:high}), 
the all-to-all propagator can be expressed as 
\bea
   \left\{ D(m_q;\theta)^{-1} \right\}(x,y)
   & = &
   \sum_{k=1}^{N_v} v_{k,\theta}^{(q)}(x)\,w_{k, \theta}^{(q)\dagger}(y)
   \hspace{3mm}
   (q=l,s)
   \label{eqn:sim:a2a_prop:a2a_prop}
\eea
with the following two sets of vectors $v$ and $w$
\bea
   \left\{
      v_{1,\theta}^{(q)},...,v_{N_v,\theta}^{(q)}
   \right\}
   & = & 
   \left\{
      \frac{u_{1,\theta}}{\lambda_{1,\theta}^{(q)}},
      \ldots,
      \frac{u_{N_e,\theta}}{\lambda_{N_e,\theta}^{(q)}},
      x_{1,\theta}^{(q)}, \dots, x_{N_d,\theta}^{(q)}
   \right\},
   \label{eqn:sim:a2a_prop:a2a:vw_vectors:v}
   \\
   \left\{
      w_{1,\theta}^{(q)},...,w_{N_v,\theta}^{(q)}
   \right\}
   & = &
   \left\{
      u_{1,\theta}, \ldots, u_{N_e,\theta},
      \eta_{1,\theta}^{(q)}, \dots, \eta_{N_d,\theta}^{(q)}
   \right\},
   \label{eqn:sim:a2a_prop:a2a:vw_vectors:w}
\eea
where $N_v = N_e+N_d$.


Meson two-point functions 
with a temporal separation $\Delta x_4$ and a spatial momentum $\bfp$
can be expressed as 
\bea
    C^{\pi}_{\phi \phi^\prime}(\dt; \bfp)
    & =  &
    \frac{1}{N_t}
    \sum_{x_4=1}^{N_t}
    \sum_{\bfxp,\bfx} 
    \langle 
       {\calO}_{\pi,\phi^\prime}(\bfxp,x_4+\dt) {\calO}_{\pi,\phi}(\bfx,x_4)^{\dagger} 
    \rangle
    e^{-i\bfp (\bfxp-\bfx)}
    \nn \\
    & = &
    \frac{1}{N_t}
    \sum_{x_4=1}^{N_t}
    \sum_{k,k^\prime=1}^{N_v}
    {\calO}^{(l,l)}_{\gamma_5,\phi^\prime,kk^\prime,\theta\theta^\prime}(x_4+\dt) \, 
    {\calO}^{(l,l)}_{\gamma_5,\phi,k^\prime k,\theta^\prime \theta}(x_4).
    \label{eqn:sim:msn_corr:msn_corr_2pt:pi} 
    \\
    C^{K}_{\phi \phi^\prime}(\dt; \bfp)
    & =  &
    \frac{1}{N_t}
    \sum_{x_4=1}^{N_t}
    \sum_{\bfx,\bfxp} 
    \langle 
       {\calO}_{K,\phi^\prime}(\bfxp,x_4+\dt) {\calO}_{K,\phi}(\bfx,x_4)^{\dagger} 
    \rangle
    e^{-i\bfp (\bfxp-\bfx)}
    \nn \\
    & =  &
    \frac{1}{N_t}
    \sum_{x_4=1}^{N_t}
    \sum_{k,k^\prime=1}^{N_v}
    {\calO}^{(s,l)}_{\gamma_5,\phi^\prime,kk^\prime,\theta\theta^\prime}(x_4+\dt) \, 
    {\calO}^{(l,s)}_{\gamma_5,\phi,k^\prime k,\theta^\prime \theta}(x_4),
    \label{eqn:sim:msn_corr:msn_corr_2pt:K} 
\eea
where $p_i \!=\! (\theta^\prime - \theta)/L$ ($i=1,2,3$)
represents the meson momentum induced by the twisted boundary conditions.
Interpolating operators for $\pi^+$ and $K^+$ are given by 
\bea
   {\calO}_{\pi,\phi}(\bfx,t)
   & = &
   \sum_{\bfr} \phi(|\bfr|) \bar{d}(\bfx+\bfr,t) \gamma_5 u(\bfx,t),
   \\
   {\calO}_{K,\phi}(\bfx,t)
   & = &
   \sum_{\bfr} \phi(|\bfr|) \bar{s}(\bfx+\bfr,t) \gamma_5 u(\bfx,t),
   \hspace{5mm}
\eea
where $\phi(|\bfr|)$ is a smearing function.
Note that light quarks are degenerate and denoted by 
$l$ $(\!=\!u,d)$ in this paper.
The quantity
\bea
   {\calO}^{(q,q^\prime)}_{\Gamma,\phi,kk^\prime,\theta\theta^\prime}(x_4)
   & = & 
   \sum_{\bfx,\bfr}
   \phi(\bfr)\, 
   w_{k,\theta}^{(q)\dagger}(\bfx+\bfr,x_4) \, 
   \Gamma \,
   v_{k^\prime,\theta^\prime}^{(q^\prime)}(\bfx,x_4),
   \label{eqn:sim:corr:msn_op} 
\eea
can be considered as a smeared meson field
constructed from the $v$ and $w$ vectors
at a time-slice $x_4$.
In this study,
we employ the local and an exponential smearing functions,
namely 
$\phi_{l}(\bfr) \!=\! \delta_{\bfr,{\bf 0}}$
and $\phi_{s}(\bfr) \!=\! \exp[-0.4|\bfr|]$.

Three-point functions needed to calculate the EM form factors
can be constructed in a similar way.
For example, the kaon three-point function 
with the light-quark current $V^{(l)}_\mu = \bar{l}\gamma_\mu l$ is expressed as 
\bea
    & & 
    C^{K}_{{V^{(l)}_\mu},\phi\phi^\prime}
    (\dt, \dtp;\bfp,\bfp^\prime)
    \nn \\
    & = &
    \frac{1}{N_t}
    \sum_{x_4=1}^{N_t}
    \sum_{\bfx,\bfxp,\bfxpp} 
    \langle 
       {\calO}_{K,\phi^\prime}(\bfxpp,x_4+\dt+\dtp) 
       V_\mu^{(l)}(\bfxp,x_4+\dt)
       {\calO}_{K,\phi}(\bfx,x_4)^{\dagger} 
    \rangle
    \nn \\
    &&
    \hspace{90mm}
    \times 
    e^{-i\bfpp(\bfxpp-\bfxp)} e^{-i\bfp(\bfxp-\bfx)}
    \nn \\
    & = & 
    \frac{1}{N_t} \sum_{x_4=1}^{N_t}
    \sum_{k,k^\prime,k^{\prime\prime}=1}^{N_v}
    {\calO}^{(s,l)}_{\gamma_5,\phi^\prime,k^{\prime\prime} k^\prime,\theta^{\prime\prime} \theta^\prime}
    (x_4+\dt + \dtp) \, 
    {\calO}^{(l,l)}_{\gamma_\mu,\phi_l,k^\prime k,\theta^\prime \theta}
    (x_4+\dt)
    \nn \\[1mm]
    & &
    \hspace{90mm}
    \times
    {\calO}^{(l,s)}_{\gamma_5,\phi,kk^{\prime\prime},\theta \theta^{\prime\prime}}(x_4),
    \label{eqn:sim:msn_corr:msn_corr_3pt:K}  
\eea
where 
$\dt$ ($\dtp$) represents the temporal separation 
between the vector current and meson source (sink) operator.
The initial and final meson momenta are given by the twist angles as 
\bea
   p_i = \frac{\theta-\theta^{\prime\prime}}{L},
   \hspace{5mm}
   p^\prime_i = \frac{\theta^\prime-\theta^{\prime\prime}}{L}
   \hspace{5mm}
   (i=1,2,3).
\eea
Note that 
we need to apply different twist angles to the quark and anti-quark fields
in ${\cal O}_{P,\phi}$ and $V_\mu^{(q)}$
so that the mesons can carry non-zero momentum.


We only calculate connected diagrams 
because of the use of the twisted boundary condition.
The contribution of the disconnected diagram to $\pff$
vanishes due to charge conjugation symmetry~\cite{disconnected}.
As a numerical check, 
we calculate the disconnected contributions to $F_V^{\{\pi^+,K^+,K^0\}}$ 
with meson momenta $\bfp\!=\!(2\pi/L,0,0)$ and $\bfpp\!=\!(0,0,0)$
using the Fourier transformation and 
the periodic boundary condition also for the valence quarks.
The disconnected contributions turn out to be insignificant
with our statistical accuracy.

\begin{figure}[t]
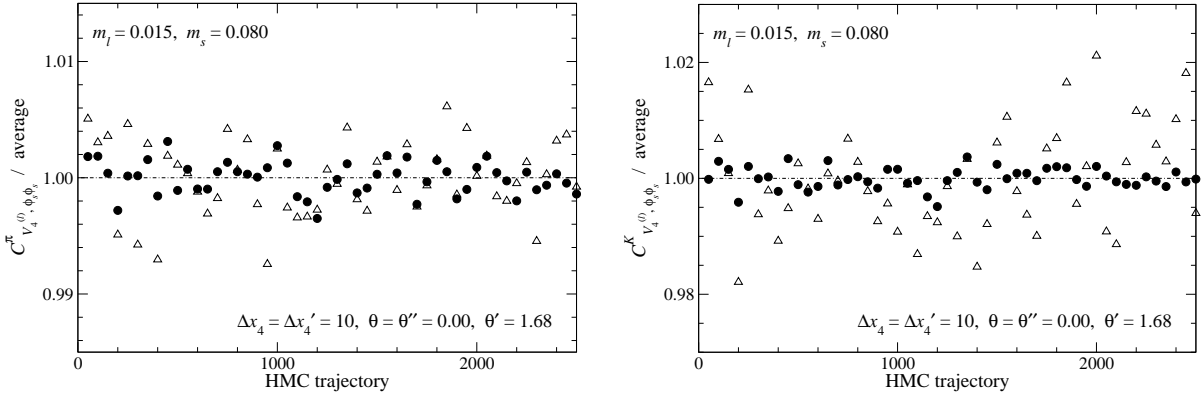

\begin{center}
   \includegraphics[angle=0,width=0.48\linewidth,clip]{jkd.p-p-v4_mud0_ms6_mval000_tbc010.dt10dtp10.eps}
   \hspace{3mm}
   \includegraphics[angle=0,width=0.48\linewidth,clip]{jkd.p-p-v4_mud0_ms6_mval600_tbc010.dt10dtp10.eps}

   \vspace{-3mm}
   \caption{
      Statistical fluctuation of three-point functions,
      $C^{\pi}_{V_4^{(l)},\phi_s \phi_s}(\dt,\dtp;\bfp,\bfpp)$ (left panel)
      and $C^{K}_{V_4^{(l)},\phi_s \phi_s}(\dt,\dtp;\bfp,\bfpp)$ (right panel),
      with $\dt\!=\!\dtp\!=\!10$, $\theta\!=\!\theta^{\prime\prime}\!=\!0.00$, 
      $\theta^\prime\!=\!1.68$
      at $(m_l,m_s)\!=\!(0.015,0.080)$.
      We plot the value at each jackknife sample 
      normalized by the statistical average.
      Triangles and circles are data before and after averaging over 
      the temporal location of the source operator $x_4$.
   }
   \label{fig:sim:msn_corr}
\end{center}
\end{figure}

By using the all-to-all propagator,
we can average the meson correlators over the location of the source operator,
{\it i.e.}
the summation over $\bfx$ and $x_4$
in Eqs.~(\ref{eqn:sim:msn_corr:msn_corr_2pt:pi}),
(\ref{eqn:sim:msn_corr:msn_corr_2pt:K})
and (\ref{eqn:sim:msn_corr:msn_corr_3pt:K}).
Figure~\ref{fig:sim:msn_corr} compares
the statistical fluctuation of the pion three-point function
with a certain choice of $\dt^{(\prime)}$ and $\bfp^{(\prime)}$.
We observe that 
an average over the temporal coordinate $x_4$ reduces the statistical error
of the pion (kaon) three-point functions by about a factor of two (four).

\subsection{Reweighting}

We use the gauge ensembles generated at the single value of $m_s\!=\!0.080$.
In order to study the $m_s$ dependence of the EM form factors,
the meson correlators are calculated 
at a different value $m_s^\prime\!=\!0.060$ 
by utilizing the reweighing technique~\cite{reweight:1,reweight:2}.
The kaon three-point function at $m_s^\prime$ 
is estimated on the gauge configurations at $m_s$ as 
\bea
  \langle C^{K}_{V_\mu^{(l)},\, \phi\phi^\prime} \rangle_{m_s^\prime}
  & = & 
  \langle C^{K}_{V_\mu^{(l)},\, \phi\phi^\prime} \, \tilde{w}(m_s^\prime,m_s) \rangle_{m_s},
   \label{eq:rew}
\eea
where $\langle \cdots \rangle_{m_s}$ represents the Monte Carlo average
at $m_s$, 
and $\tilde{w}$ is the reweighting factor for each configuration
\bea
   \tilde{w}(m_s^\prime,m_s)
   & = & 
   \frac{w(m_s^\prime,m_s)}{\langle w(m_s^\prime,m_s) \rangle_{m_s}},
   \hspace{5mm}
   w(m_s^\prime,m_s)
   = 
   \det\left[ \frac{D(m_s^\prime)}{D(m_s)} \right].
   \label{eqn:rw_fctr}
\eea
It is prohibitively time consuming 
to exactly calculate the quark determinant $\det[D(m_s^{(\prime)})]$.
Instead, we decompose $w$ into contributions from low- and high-modes
\bea
   w(m_s^\prime,m_s)
   & = & 
   w_{\rm low}(m_s^\prime,m_s) \,w_{\rm high}(m_s^\prime,m_s),
   \\
   w_{\rm low (high)}(m_s^\prime,m_s) 
   & = & 
   \det\left[ 
      P_{\rm low (high)} 
      \frac{D(m_s^\prime)}{D(m_s)} 
      P_{\rm low (high)}
   \right],
\eea
and the low mode contribution $w_{\rm low}$ is exactly calculated 
by using the low-lying eigenvalues.
We estimate the high-mode contribution $w_{\rm high}$ 
by a stochastic estimator for 
\begin{eqnarray}
 w_{\rm{high}}^2(m_s^\prime,m_s)
  =
  \frac{1}{N_r}
  \sum_{r=1}^{N_r} e^{-\frac{1}{2}(P_{\rm high}\xi_r)^\dagger 
                (\Omega-1) P_{\rm high} \xi_r},
  \label{eq:high}
\end{eqnarray}
with 
$\Omega 
 \equiv D(m_s)^\dagger \{D(m_s^\prime)^{-1}\}^\dagger D(m_s^\prime)^{-1}D(m_s)$.
We introduce $N_r$ normalized Gaussian random vectors
$\{\xi_1,...,\xi_{N_r}\}$.

\begin{figure}[tbp]
\begin{center}
   \includegraphics[width=0.48\textwidth,clip]{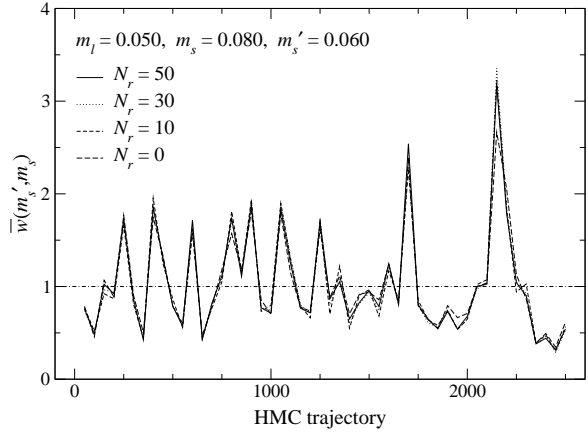}
   \vspace{-3mm}
   \caption{
      Monte Carlo history of reweighting factor $\tilde{w}(m_s^\prime,m_s)$
      at $m_l\!=\!0.050$
      with different numbers of the Gaussian random vector $N_r$.
   }
   \label{fig:sim:reweight:factor:nr-dep}
\end{center}
\end{figure}

\begin{figure}[tbp]
\begin{center}
   \centering
   \includegraphics[width=0.48\textwidth,clip]{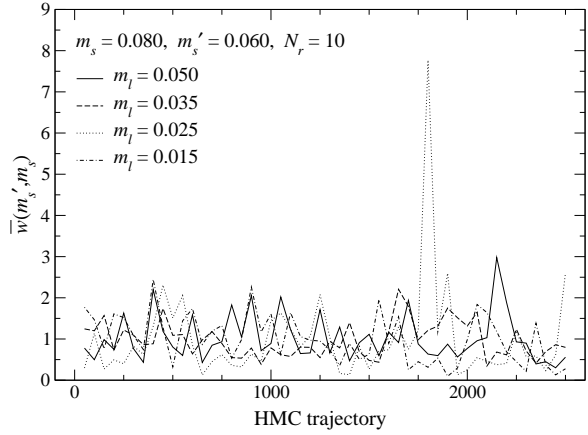}
   \vspace{-3mm}
   \caption{
      Monte Carlo history of reweighting factor $\tilde{w}(m_s^\prime,m_s)$
      calculated with $N_r\!=\!10$. 
      Different lines show data at different $m_l$.
   }
   \label{fig:sim:reweight:factor:mud-dep}
\end{center}
\end{figure}

At $m_l\!=\!0.050$,
we study how many Gaussian random vectors are needed 
to reliably estimate the high-mode contribution $w_{\rm{high}}$
for the reweighting from $m_s\!=\!0.080$ to $m_s^\prime\!=\!0.060$.
The normalized reweighting factor $\tilde{w}$ shows rather minor dependence on $N_r$, 
as shown in Fig.~\ref{fig:sim:reweight:factor:nr-dep}.
This suggests that 
$\tilde{w}$ is dominated by the low-mode contribution $w_{\rm low}$
for our choice of the number of low-modes $N_e$ 
and the lattice size $N_s^3 \times N_t$.
We do not need many random vectors
and set $N_r\!=\!10$ in this study.

Figure~\ref{fig:sim:reweight:factor:mud-dep} compares
$\tilde{w}$ at different values of $m_l$.
We observe that 
$\tilde{w}$ is typically in a range [0.5, 2.0]. 
There is no systematic trend 
in the magnitude of the statistical fluctuation of $\tilde{w}$,
as we decrease $m_l$.
We therefore consider that a large value $\tilde{w} \simeq 8$ 
observed at $m_l\!=\!0.025$ and at 1800-th HMC trajectory is accidental.

%% file: 3.ff.tex
\section{EM form factors and charge radii at simulation points}
\label{sec:ff}


\subsection{EM form factors}

Two- and three-point functions of the light mesons ($P\!=\!\pi, K$)
are dominated by the ground state contribution 
\bea
   C^{P}_{\phi\phi^\prime}(\dt;\bfp)
   & \xrightarrow[\dt \to \infty]{} &
   \frac{Z_{P,\phi^\prime}(\bfp)^*\, Z_{P, \phi}(\bfp)} 
        {2\, E_P(\bfp)}\,
   e^{-E_P(\bfp)\,\dt},
   \label{eqn:ff:ff:msn_corr_2pt}
   \\
   C^{P}_{J_\mu, \phi\phi^\prime}(\dt, \dtp;\bfp,\bfpp)
   & \xrightarrow[\dt,\dtp \to \infty]{} &
   \frac{Z_{P,\phi^\prime}(\bfpp)^*\, Z_{P, \phi}(\bfp)} 
   {4\, E_P(\bfpp) E_P(\bfp)}
   \frac{1}{Z_V}
   \langle P(p^\prime) | J_\mu | P(p) \rangle\,
   \nn \\
   &   &
   \hspace{25mm} 
   \times 
   e^{-E_P(\bfpp)\, \dtp} e^{-E_P(\bfp)\, \dt},
   \label{eqn:ff:ff:msn_corr_3pt}
\eea
in the limit of large temporal separations
between the meson source/sink operators and the EM current
$\dt,\dtp\!\to\!\infty$.
Here 
$Z_V$ is the renormalization factor for the vector current,
and 
$Z_{P,\phi}(\bfp)\!=\!\langle P(p) | {\calO}_{P,\phi} \rangle$
is the overlap of the meson interpolating field to the physical state.
We consider a ratio
\bea
   R_V^{PQ}(\dt,\dtp; \bfp, \bfp^\prime)
   & = &
   \frac{C^{P}_{J_4, \phi_s \phi_s}(\dt,\dtp; \bfp,\bfpp)\,
         C^{Q}_{\phi_s \phi_l}(\dt;{\bf 0})\, C^{Q}_{\phi_l \phi_s}(\dtp;{\bf 0})}
        {C^{Q}_{J_4, \phi_s \phi_s}(\dt,\dtp; {\bf 0},{\bf 0})\,
         C^{P}_{\phi_s \phi_l}(\dt;\bfp)\, C^{P}_{\phi_l \phi_s}(\dtp;\bfpp)},
   \label{eqn:ff:ff:ratio}
\eea       
with three choices of $(P,Q)\!=\!(\pi^+,\pi^+)$, $(K^+,K^+)$ and $(K^0,K^+)$.
Since 
$Z_{K^+,\phi}\!=\!Z_{K^0,\phi}$ with our simulation setup $m_u\!=\!m_d$, 
normalization factors $Z_{P,\phi_{\{l,s\}}}$ and $Z_V$
as well as the exponential damping factors $e^{-E_P(\bfp^{(\prime)})\dt^{(\prime)}}$
cancel in the ratio,
provided that they are dominated by the ground state contribution~\cite{dble_ratio}.
Therefore we can calculate the effective value of the EM form factors
through this ratio as 
\bea
   \ff(\dt,\dtp;t)
   & = &  
   \frac{\ff(\dt,\dtp;t)}{F_V^{Q}(\dt,\dtp;0)}
   = 
   \frac{2\,M_Q}{E_P(\bfp)+E_P(\bfpp)}
   R_V^{PQ}(\dt,\dtp; \bfp,\bfp^\prime),
   \label{eqn:ff:ff:dratio}
\eea
where 
we assume the vector current conservation $F_V^Q(0)\!=\!1$
($Q\!=\!\pi^+, K^+$), 
and use $M_P$ and $E_P$ determined 
by fitting two-point functions to Eq.~(\ref{eqn:ff:ff:msn_corr_2pt}).

\begin{figure}[t]
   \begin{center}
   \includegraphics[angle=0,width=0.48\linewidth,clip]%
                   {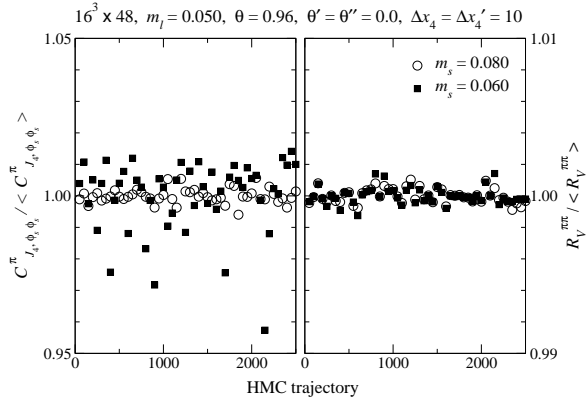}

   \vspace{-3mm}
   \caption{
      Pion three-point function $C^{\pi}_{J_4,\phi_s \phi_s}(\dt,\dtp;\bfp,\bfpp)$
      (left panel) and ratio $R_V^{\pi\pi}(\dt,\dtp;\bfp,\bfpp)$ (right panel)
      at each jackknife sample. 
      We plot data, which are normalized by their statistical average,
      at $m_l \!=\! 0.050$, $\theta \!=\! 0.96$, 
      $\theta^\prime \!=\! \theta^{\prime\prime} \!=\! 0.0$ 
      and $\dt \!=\! \dtp \!=\! 10$.
      Circles and squares are data 
      before ($m_s\!=\!0.080$) and after ($m_s\!=\!0.060$) reweighting. 
      The horizontal axis represents the HMC trajectory count 
      of the excluded configuration for the jackknife analysis.
      Note that the scale is much finer for the right panel than the left. 
   }
   \label{fig:ff:reweight:mud3}
\end{center}
\end{figure}

\begin{figure}[t]
\begin{center}
   \includegraphics[angle=0,width=0.48\linewidth,clip]%
                   {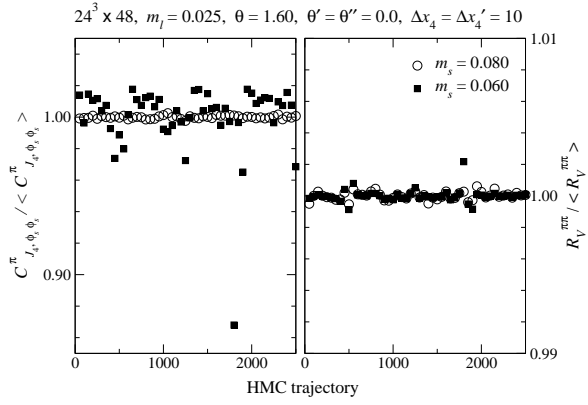}

   \vspace{-3mm}
   \caption{
      Same as Fig.~\protect\ref{fig:ff:reweight:mud3},  but for $m_l\!=\!0.025$.
      Note that the large value $\tilde{w} \!\sim\! 8$ 
      in Fig.~\protect\ref{fig:sim:reweight:factor:mud-dep}
      leads to a small (large) value of the three-point function
      (the ratio $R_V^{\pi\pi}$) at 1800th trajectory.
   }
   \label{fig:ff:reweight:mud1}
\end{center}
\end{figure}

Taking the ratio $R_V^{PQ}$ turns out to be effective 
also in reducing statistical fluctuation induced by reweighting.
The reweighting factor in our study 
is typically in a region $\tilde{w} \!\in\! [0.5,2.0]$,
and significantly enhances the statistical fluctuation
of the meson correlators.
In Fig.~\ref{fig:ff:reweight:mud3}, for instance, 
we observe about a factor of 5 increase 
in the statistical error of 
the pion three-point function $C^{\pi}_{J_4,\phi_s \phi_s}$ 
at $m_l\!=\!0.050$.
The enhanced fluctuation, however, largely cancels 
in the ratio $R_V^{PQ}$,
whose error increases only by $\approx$~15~\% by reweighting.
This is also the case at $m_l\!=\!0.025$,
where the reweighing factor in Fig.~\ref{fig:sim:reweight:factor:mud-dep}
takes occasionally a rather large value $\tilde{w}\!\simeq\!8$.
As suggested in Fig.~\ref{fig:ff:reweight:mud1},
the reweighting increases the error of $C^{\pi}_{J_4,\phi_s \phi_s}$ 
by about a factor of 24, 
which is however remarkably reduced to 1.6 in the ratio $R_V^{PQ}$. 

\begin{figure}[t]
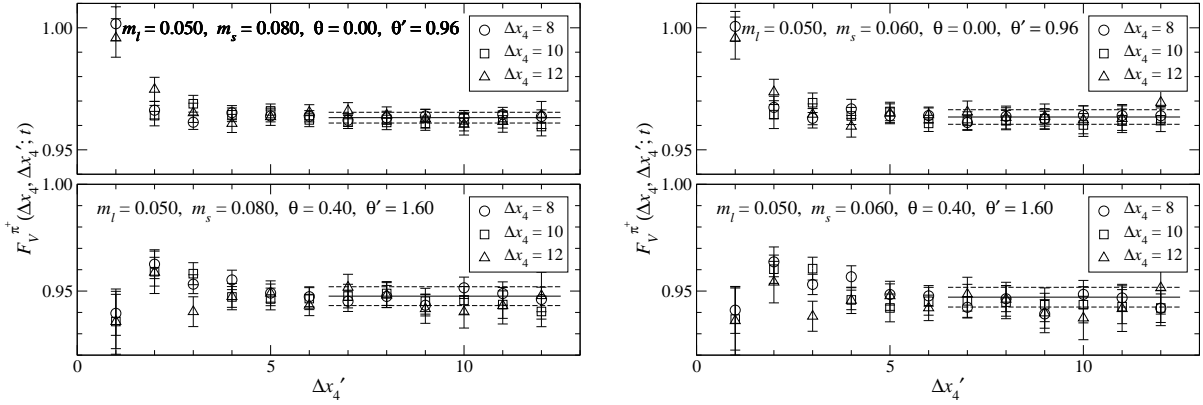

\begin{center}
\includegraphics[angle=0,width=0.48\linewidth,clip]%
                {pff_v4_drat_vs_dtsnk.mud3_ms6.eps}
\hspace{3mm}
\includegraphics[angle=0,width=0.48\linewidth,clip]%
                {pff_v4_drat_vs_dtsnk.mud3_ms4.eps}

\vspace{-3mm}
\caption{
   Effective value of pion EM form factor $\pff(\dt,\dtp,t)$ 
   at $m_l=0.050$.
   Left and right panels show data at $m_s\!=\!0.080$ and 0.060,
   whereas top and bottom panels are with 
   $(\theta,\theta^\prime,\theta^{\prime\prime}) \!=\! (0.00,0.96,0.00)$
   and $(0.40,1.60,0.00)$. 
}
\label{fig:ff:pff_eff:mud0050}
\end{center}
\end{figure}

\begin{figure}[t]
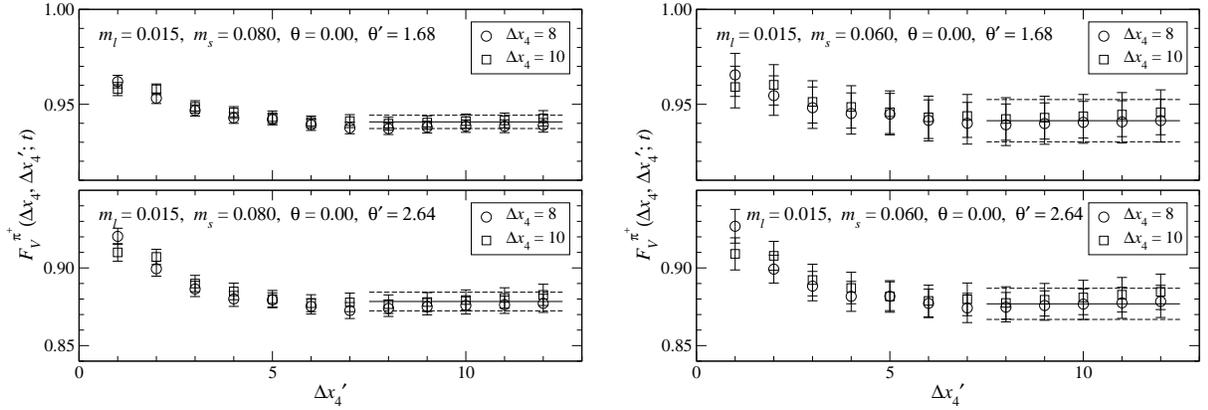

\begin{center}
\includegraphics[angle=0,width=0.48\linewidth,clip]%
                {pff_v4_drat_vs_dtsnk.mud0_ms6.eps}
\hspace{3mm}
\includegraphics[angle=0,width=0.48\linewidth,clip]%
                {pff_v4_drat_vs_dtsnk.mud0_ms4.eps}
\vspace{-3mm}
\caption{
   Effective value of pion EM form factor $\pff(\dt,\dtp,t)$ 
   at $m_l=0.015$.
   Left and right panels show data at $m_s\!=\!0.080$ and 0.060,
   whereas top and bottom panels are with 
   $(\theta,\theta^\prime,\theta^{\prime\prime}) \!=\! (0.00,1.68,0.00)$
   and $(0.00,2.64,0.00)$. 
}
\label{fig:ff:pff_eff:mud0015}
\end{center}
\end{figure}

\begin{figure}[t]
\begin{center}
\includegraphics[angle=0,width=0.48\linewidth,clip]%
                {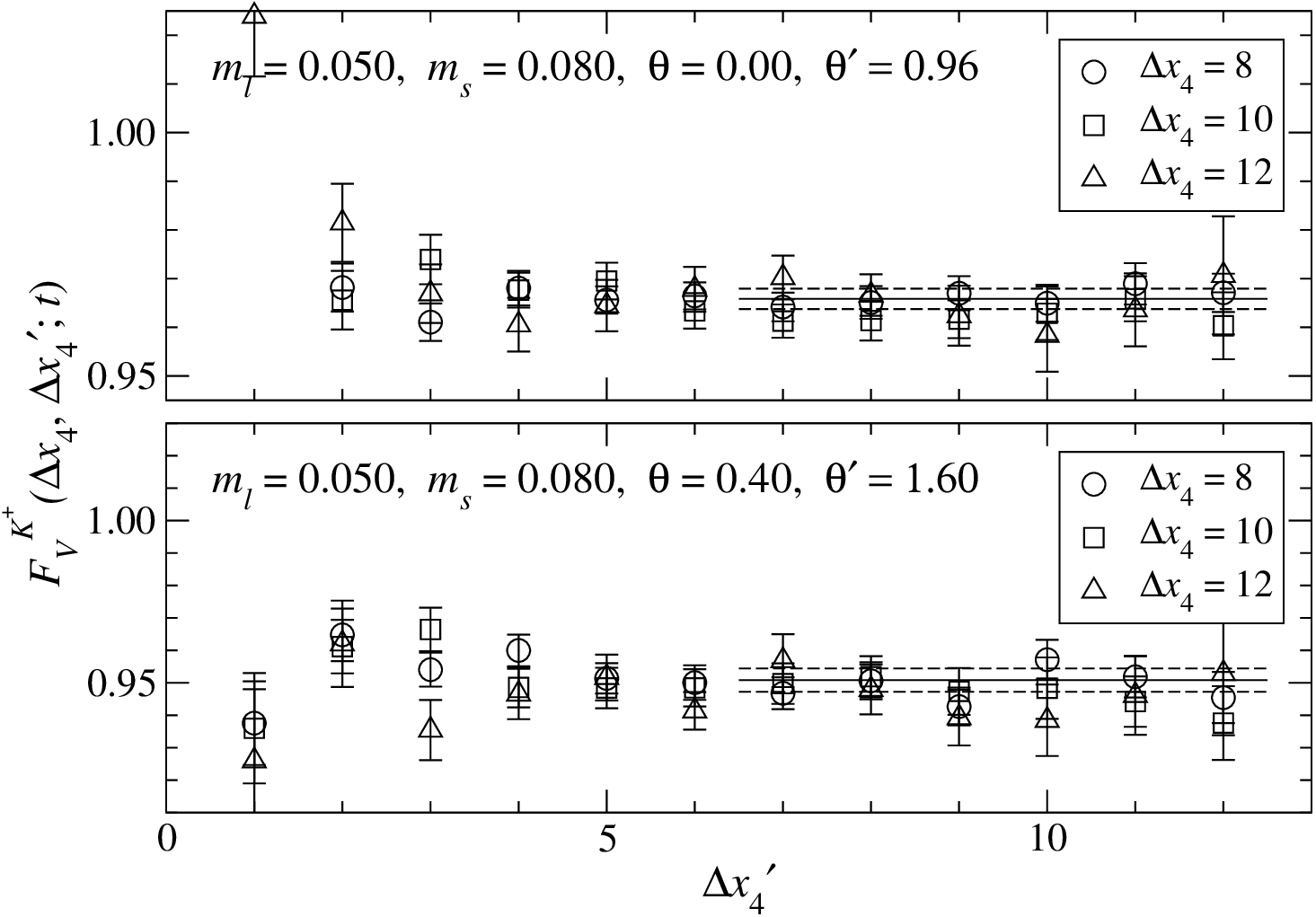}
\hspace{3mm}
\includegraphics[angle=0,width=0.48\linewidth,clip]%
                {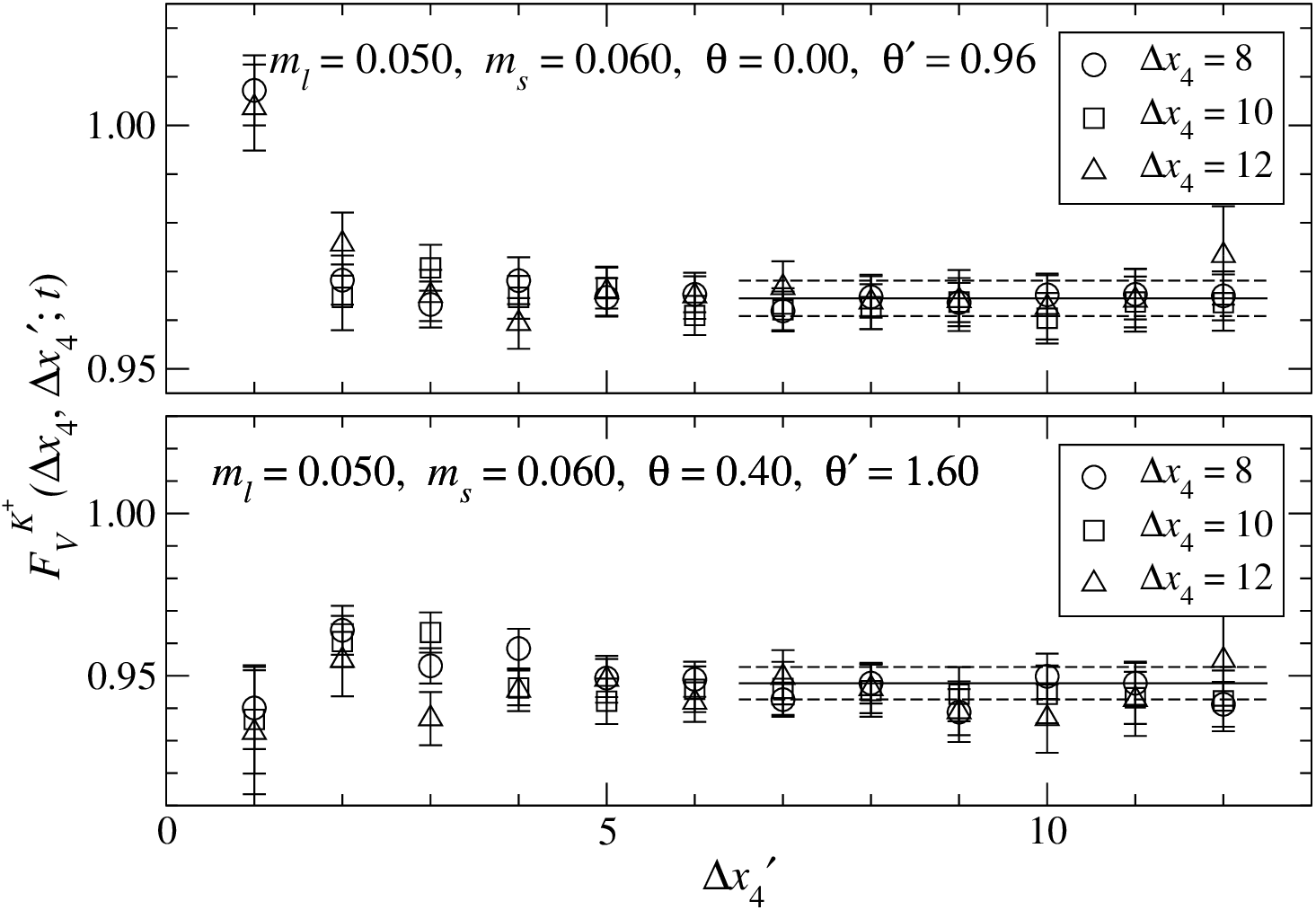}

\vspace{-3mm}
\caption{
   Effective value of charged kaon EM form factor $\kpff(\dt,\dtp,t)$
   at $m_l\!=\!0.050$.
}
\label{fig:ff:k+ff_eff:mud0050}
\end{center}
\end{figure}

\begin{figure}[t]
\begin{center}
\includegraphics[angle=0,width=0.48\linewidth,clip]%
                {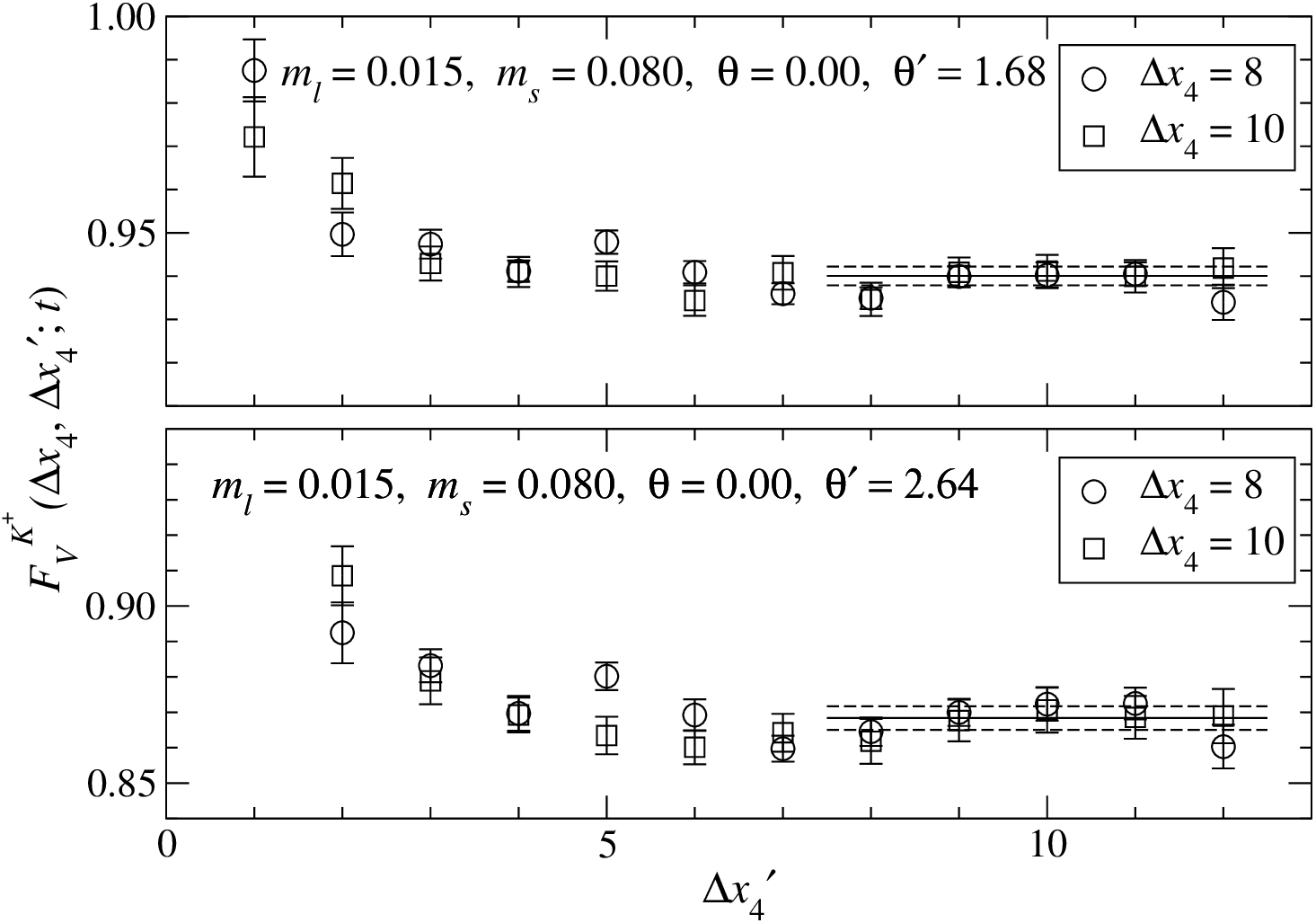}
\hspace{3mm}
\includegraphics[angle=0,width=0.48\linewidth,clip]%
                {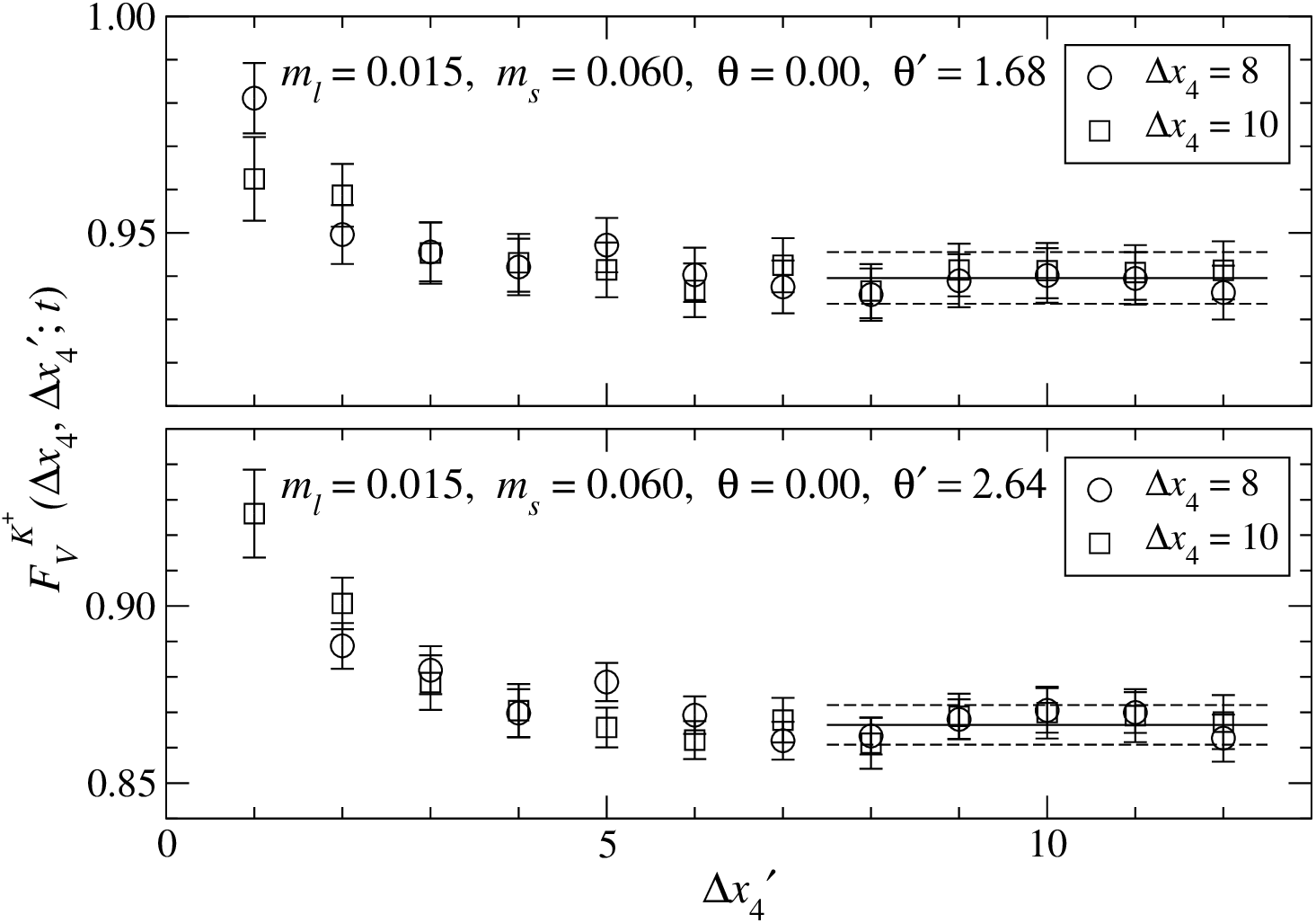}

\vspace{-3mm}
\caption{
   Effective value of charged kaon EM form factor $\kpff(\dt,\dtp,t)$
   at $m_l\!=\!0.015$.
}
\label{fig:ff:k+ff_eff:mud0015}
\end{center}
\end{figure}

\begin{figure}[t]
\begin{center}
\includegraphics[angle=0,width=0.48\linewidth,clip]%
                {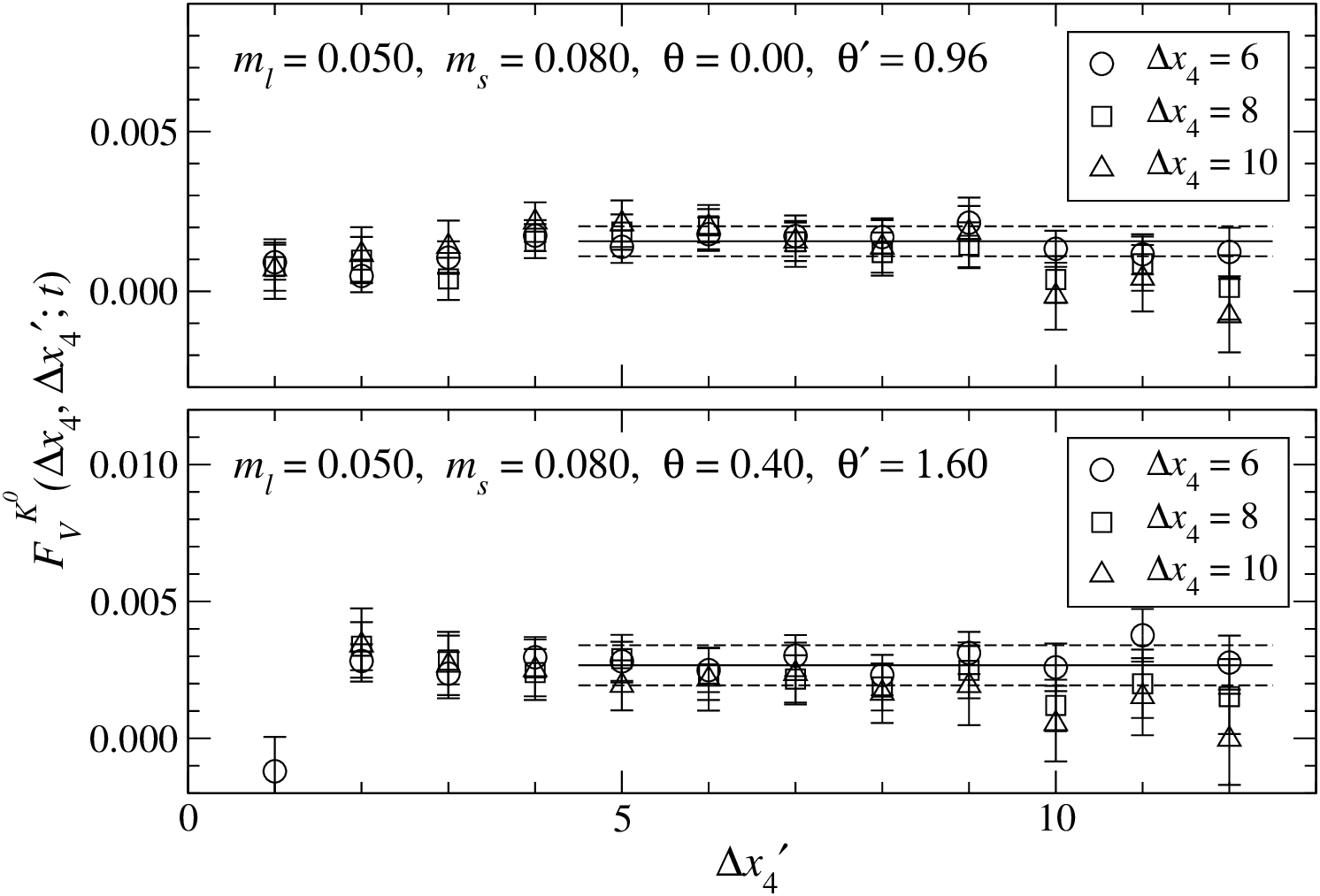}
\hspace{3mm}
\includegraphics[angle=0,width=0.48\linewidth,clip]%
                {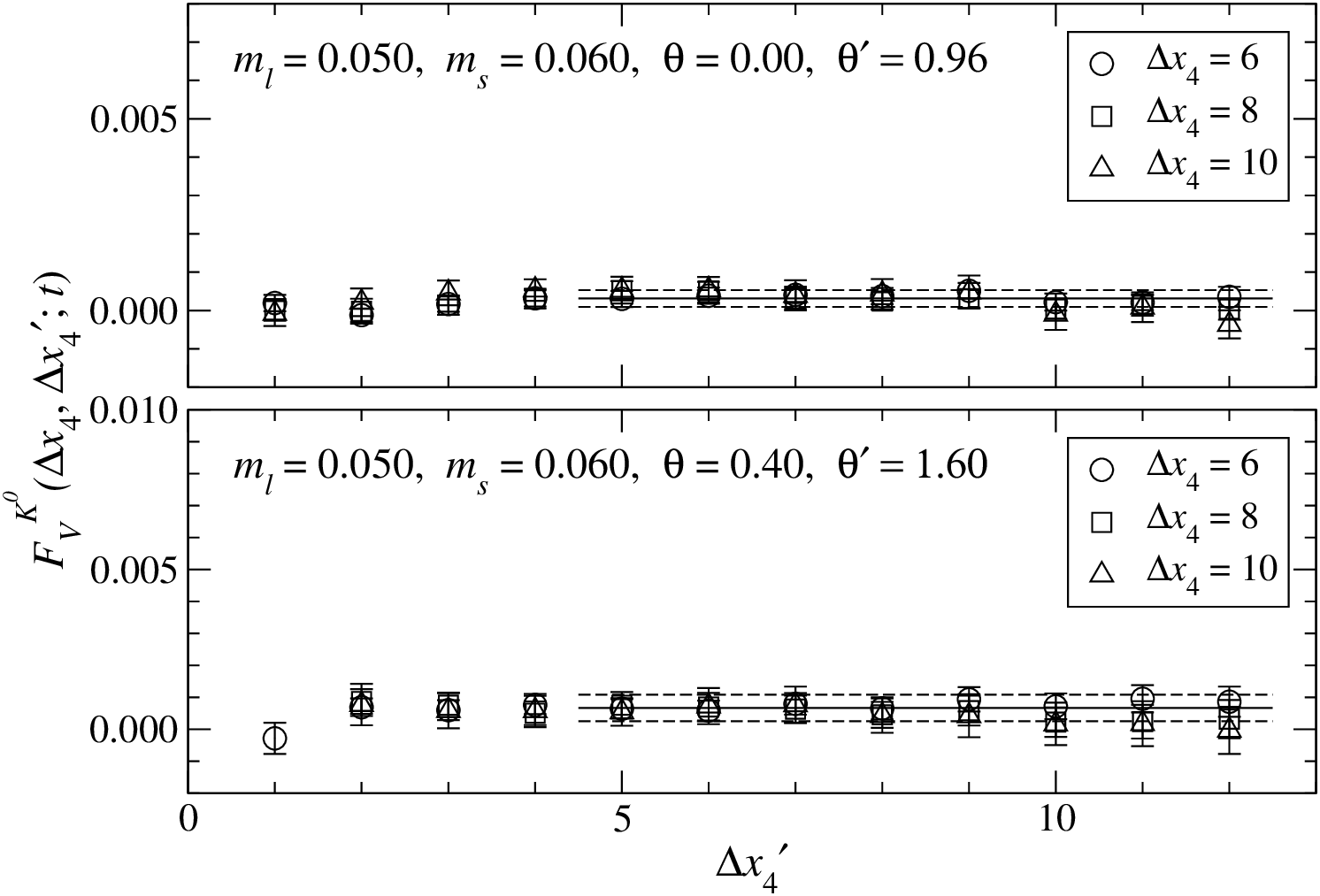}

\vspace{-3mm}
\caption{
   Effective value of neutral kaon EM form factor $\knff(\dt,\dtp,t)$
   at $m_l\!=\!0.050$.
}
\label{fig:ff:k0ff_eff:mud0050}
\end{center}
\end{figure}

\begin{figure}[t]
\begin{center}
\includegraphics[angle=0,width=0.48\linewidth,clip]%
                {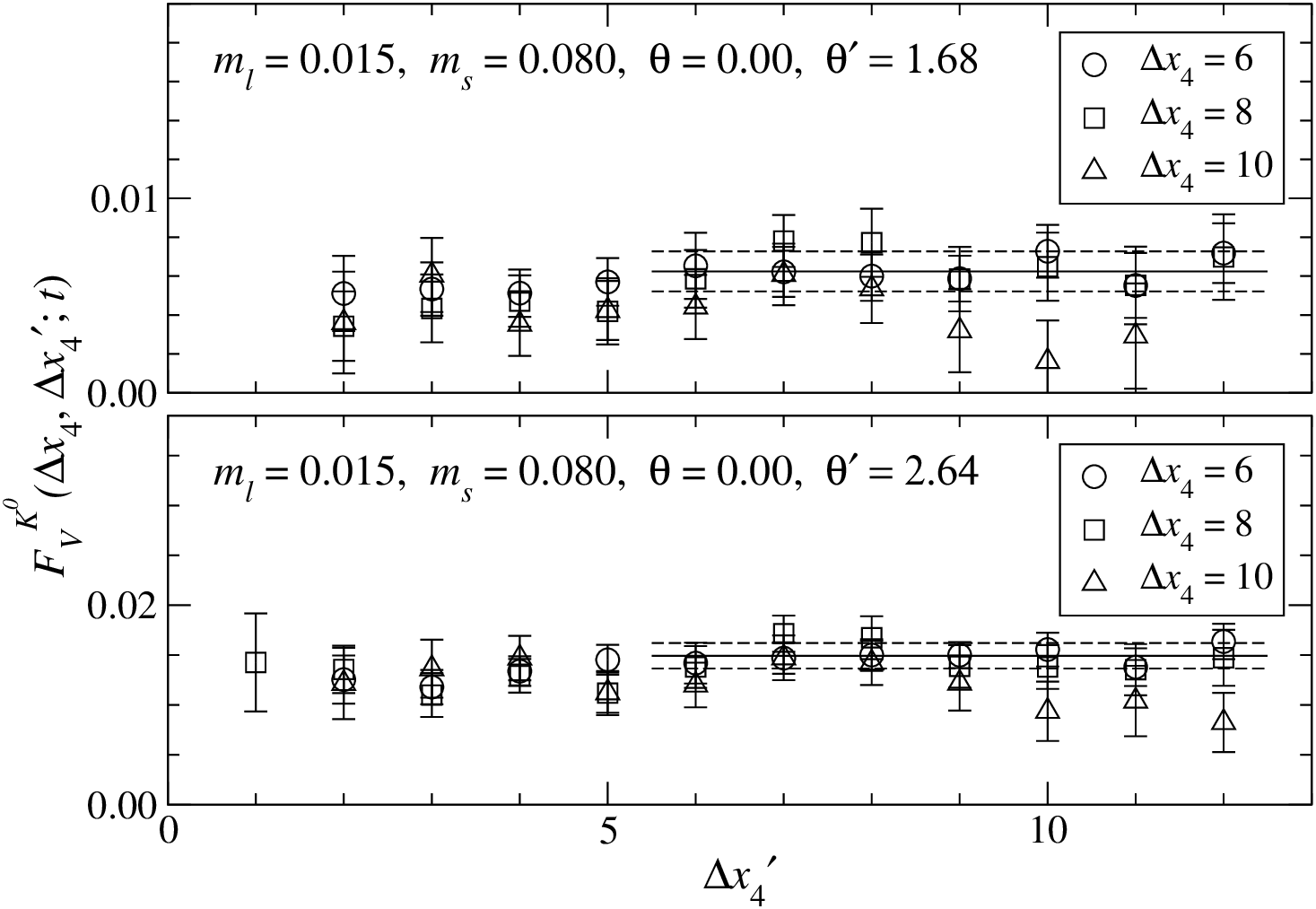}
\hspace{3mm}
\includegraphics[angle=0,width=0.48\linewidth,clip]%
                {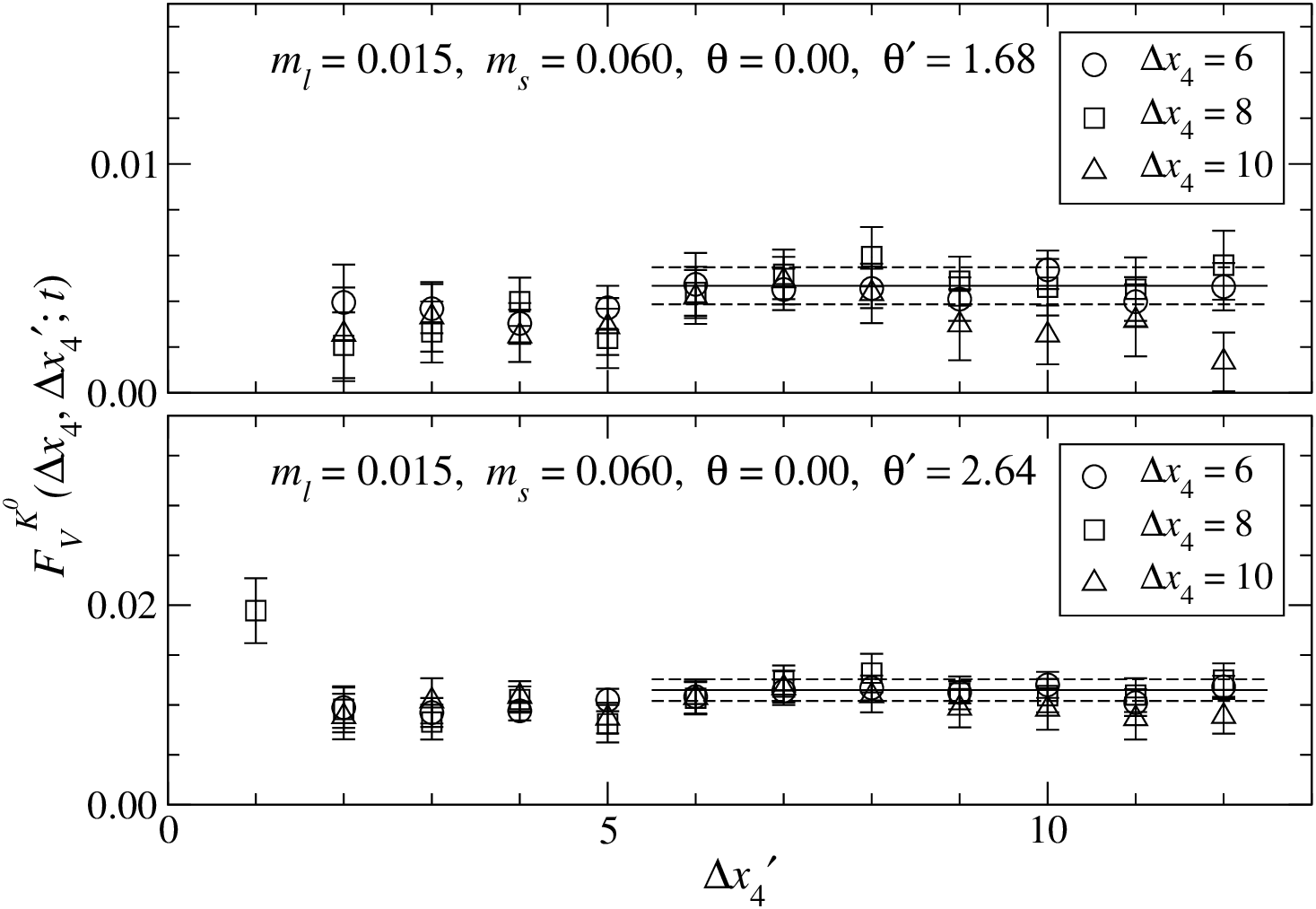}

\vspace{-3mm}
\caption{
   Effective value of neutral kaon EM form factor $\knff(\dt,\dtp,t)$
   at $m_l\!=\!0.015$.
}
\label{fig:ff:k0ff_eff:mud0015}
\end{center}
\end{figure}

We extract the EM form factor $\ff(t)$
by a constant fit to the effective value $\ff(\dt, \dtp; t)$.
Figures~\ref{fig:ff:pff_eff:mud0050}\,--\,\ref{fig:ff:k0ff_eff:mud0015}
show examples of this fit for 
$\pff$ (Figs.~\ref{fig:ff:pff_eff:mud0050}\,--\,\ref{fig:ff:pff_eff:mud0015}), 
$\kpff$ (Figs.~\ref{fig:ff:k+ff_eff:mud0050}\,--\,\ref{fig:ff:k+ff_eff:mud0015}),
and $\knff$ (Figs.~\ref{fig:ff:k0ff_eff:mud0050}\,--\,\ref{fig:ff:k0ff_eff:mud0015}).
We summarize numerical results 
in Tables~\ref{tbl:ff:ff:mud0050ms0080}\,--\,\ref{tbl:ff:ff:mud0015ms0060}.

The charged meson form factors are the sum of the contributions
with the light and strange quark currents
\bea
   \pff
   & \propto & 
   \frac{2}{3} \langle \pi \left| \bar{u}\gamma_\mu u \right| \pi \rangle
  +\frac{1}{3} \langle \pi \left| \bar{d}\gamma_\mu d \right| \pi \rangle
   = 
   \langle \pi \left| \bar{l}\gamma_\mu l \right| \pi \rangle,
\eea
\bea
   \kpff
   & \propto &
   \frac{2}{3}\langle K^+ \left| \bar{u}\gamma_\mu u \right| K^+ \rangle
  +\frac{1}{3}\langle K^+ \left| \bar{s}\gamma_\mu s \right| K^+ \rangle
   \nn \\
   & = &
   \frac{2}{3}\langle K^+ \left| \bar{l}\gamma_\mu l \right| K^+ \rangle
  +\frac{1}{3}\langle K^+ \left| \bar{s}\gamma_\mu s \right| K^+ \rangle.
\eea
Their normalizations are fixed as $\ff(0)\!=\!1$ ($P\!=\!\pi^+,K^+$)
from the vector current conservation.
Equation~(\ref{eqn:ff:ff:dratio}) implies that
what we study using $R_V^{PP}$
is a ratio $F_V^P(t)/F_V^P(0)$,
namely the finite $t$ correction to $F_V^P(t)$.
Since we explore near-zero momentum transfer $t\!\sim\!0$,
this correction is not large:
typically  $\ff(0)-\ff(t)\lesssim 0.1$ as seen in 
Tables~\ref{tbl:ff:ff:mud0050ms0080}\,--\,\ref{tbl:ff:ff:mud0015ms0060}.
Its statistical accuracy is typically 5~\% at $m_s\!=\!0.080$
and 10~\% at $m_s\!=\!0.060$. 
For these fitted values of $\ff$,  
we observe about a factor of two larger error after the reweighting
from $m_s\!=\!0.080$ to 0.060.

ChPT suggests that 
finite volume effects are exponentially suppressed as 
$\propto \! \exp[-M_\pi L]$~\cite{FVE:ChPT:TBC:GS},
which is roughly 2\,\% or less on the lattices with $M_\pi L \!\gtrsim\!4$. 
It is recently argued in Ref.~\cite{FVE:ChPT:TBC:BR} that
the twisted boundary condition breaks reflection symmetry 
and gives rise to an additional correction,
which is at the level of 0.1~\% 
for meson masses and decay constants at $M_\pi L \!\sim\!4$. 
These effects are well below the accuracy of 
the finite $t$ correction to $\ff$.
Yet another finite volume correction appears in our simulations
due to the fixed global topology.
We expect from our previous study on a similar volume~\cite{PFF:JLQCD:Nf2:RG+Ovr}
that this effect is also small compared to the statistical accuracy.

The neutral kaon form factor is the difference 
between the contributions of the light and strange quark currents
\bea
   \knff
   & \propto & 
  -\frac{1}{3} \langle \pi \left| \bar{l}\gamma_\mu l \right| \pi \rangle
  +\frac{1}{3} \langle \pi \left| \bar{s}\gamma_\mu s \right| \pi \rangle,
  \label{eqn:ff:k0ff:current_contribu}
\eea      
which vanishes at $t\!=\!0$. 
In the region of small $|t|$, 
$\knff(t)$ is close to zero as seen 
in Figs.~\ref{fig:ff:k0ff_eff:mud0050} and \ref{fig:ff:k0ff_eff:mud0015}.
The use of the all-to-all quark propagator
enables us to calculate this small form factor 
with an error of $\gtrsim \! 15$\,\%. 
The above mentioned finite volume corrections are negligible 
at this level of uncertainties.

\begin{ruledtabular}
\begin{table}[b]
\begin{center}
\caption{
   Fit results for EM form factors at $(m_l,m_s)\!=\!(0.050,0.080)$.
}
\label{tbl:ff:ff:mud0050ms0080}
\begin{tabular}{lll|lll}
   $\theta$ & $\theta^\prime$ & $\theta^{\prime\prime}$ &
   $\pff(t)$ & $\kpff(t)$  & $\knff(t)$ 
   \\ \hline
   0.00  & 0.40  & 0.00  & 0.9936(13)  & 0.9944(13)  & 0.00029(27)
   \\
   0.00  & 0.96  & 0.00  & 0.9632(22)  & 0.9659(21)  & 0.00157(47)
   \\
   0.00  & 1.60  & 0.00  & 0.9082(29)  & 0.9114(32)  & 0.00426(58)
   \\
   0.40  & 0.96  & 0.00  & 0.9875(33)  & 0.9900(29)  & 0.00044(64)
   \\
   0.40  & 1.60  & 0.00  & 0.9476(44)  & 0.9508(36)  & 0.00267(73)
   \\
   0.96  & 1.60  & 0.00  & 0.9837(66)  & 0.9870(54)  & 0.0009(10)
\end{tabular}
\end{center}
\vspace{0mm}
\end{table}
\end{ruledtabular}

\begin{ruledtabular}
\begin{table}[t]
\begin{center}
\caption{
   Fit results for EM form factors at $(m_l,m_s)\!=\!(0.050,0.060)$.
}
\label{tbl:ff:ff:mud0050ms0060}
\begin{tabular}{lll|lll}
   $\theta$ & $\theta^\prime$ & $\theta^{\prime\prime}$ &
   $\pff(t)$ & $\kpff(t)$  & $\knff(t)$ 
   \\ \hline
   0.00  & 0.40  & 0.00  & 0.9936(24)  & 0.9939(28)  & -0.00006(12)
   \\
   0.00  & 0.96  & 0.00  & 0.9634(30)  & 0.9645(36)  &  0.00031(22)
   \\
   0.00  & 1.60  & 0.00  & 0.9071(46)  & 0.9089(39)  &  0.00130(31)
   \\
   0.40  & 0.96  & 0.00  & 0.9878(42)  & 0.9888(52)  & -0.00016(34)
   \\
   0.40  & 1.60  & 0.00  & 0.9472(46)  & 0.9477(50)  &  0.00067(42)
   \\
   0.96  & 1.60  & 0.00  & 0.9823(61)  & 0.9830(78)  & -0.00004(61)
\end{tabular}
\end{center}
\vspace{3mm}
\end{table}
\end{ruledtabular}

\begin{ruledtabular}
\begin{table}[t]
\begin{center}
\caption{
   Fit results for EM form factors at $(m_l,m_s)\!=\!(0.035,0.080)$.
}
\label{tbl:ff:ff:mud0035ms0080}
\begin{tabular}{lll|lll}
   $\theta$ & $\theta^\prime$ & $\theta^{\prime\prime}$ &
   $\pff(t)$ & $\kpff(t)$  & $\knff(t)$ 
   \\ \hline
   0.00  & 0.60  & 0.00  & 0.9793(25)  & 0.9821(20)  &  0.00020(60)
   \\
   0.00  & 1.28  & 0.00  & 0.9244(54)  & 0.9302(41)  &  0.00288(80)
   \\
   0.00  & 1.76  & 0.00  & 0.8735(65)  & 0.8791(57)  &  0.00617(91)
   \\
   0.60  & 1.28  & 0.00  & 0.9666(76)  & 0.9712(61)  & -0.0017(22)
   \\
   0.60  & 1.76  & 0.00  & 0.9318(85)  & 0.9375(74)  &  0.0007(15)
   \\
   1.28  & 1.76  & 0.00  & 0.9627(19)  & 0.971(11)   & -0.0032(31)
\end{tabular}
\end{center}
\vspace{3mm}
\end{table}
\end{ruledtabular}

\begin{ruledtabular}
\begin{table}[t]
\begin{center}
\caption{
   Fit results for EM form factors at $(m_l,m_s)\!=\!(0.035,0.060)$.
}
\label{tbl:ff:ff:mud0035ms0060}
\begin{tabular}{lll|lll}
   $\theta$ & $\theta^\prime$ & $\theta^{\prime\prime}$ &
   $\pff(t)$ & $\kpff(t)$  & $\knff(t)$ 
   \\ \hline
   0.00  & 0.60  & 0.00  & 0.9805(34)  & 0.9794(42)  & -0.00032(44)
   \\
   0.00  & 1.28  & 0.00  & 0.9235(68)  & 0.9232(55)  &  0.00128(49)
   \\
   0.00  & 1.76  & 0.00  & 0.8717(87)  & 0.8711(70)  &  0.00266(71)
   \\
   0.60  & 1.28  & 0.00  & 0.9661(90)  & 0.9695(81)  & -0.0016(18)
   \\
   0.60  & 1.76  & 0.00  & 0.929(11)   & 0.9287(92)  & -0.0002(15)
   \\
   1.28  & 1.76  & 0.00  & 0.957(21)   & 0.965(12)   & -0.0022(16)
\end{tabular}
\end{center}
\vspace{5mm}
\end{table}
\end{ruledtabular}

\begin{ruledtabular}
\begin{table}[t]
\begin{center}
\caption{
   Fit results for EM form factors at $(m_l,m_s)\!=\!(0.025,0.080)$.
}
\label{tbl:ff:ff:mud0025ms0080}
\begin{tabular}{lll|lll}
   $\theta$ & $\theta^\prime$ & $\theta^{\prime\prime}$ &
   $\pff(t)$ & $\kpff(t)$  & $\knff(t)$ 
   \\ \hline
   0.00  & 1.68  & 0.00  & 0.9432(20)  & 0.9435(14)  & 0.00574(50)
   \\
   0.00  & 2.64  & 0.00  & 0.8777(34)  & 0.8748(23)  & 0.01219(94)
   \\
   1.68  & 2.64  & 0.00  & 0.9934(77)  & 0.9799(37)  & 0.00197(82)
\end{tabular}
\end{center}
\vspace{0mm}
\end{table}
\end{ruledtabular}

\begin{ruledtabular}
\begin{table}[t]
\begin{center}
\caption{
   Fit results for EM form factors at $(m_l,m_s)\!=\!(0.025,0.060)$.
}
\label{tbl:ff:ff:mud0025ms0060}
\begin{tabular}{lll|lll}
   $\theta$ & $\theta^\prime$ & $\theta^{\prime\prime}$ &
   $\pff(t)$ & $\kpff(t)$  & $\knff(t)$ 
   \\ \hline
   0.00  & 1.68  & 0.00  & 0.9398(95)  & 0.9400(75)  & 0.00426(42)
   \\
   0.00  & 2.64  & 0.00  & 0.874(13)   & 0.8715(85)  & 0.00828(53)
   \\
   1.68  & 2.64  & 0.00  & 0.992(20)   & 0.983(15)   & 0.00178(66)
\end{tabular}
\end{center}
\vspace{0mm}
\end{table}
\end{ruledtabular}

\begin{ruledtabular}
\begin{table}[t]
\begin{center}
\caption{
   Fit results for EM form factors at $(m_l,m_s)\!=\!(0.015,0.080)$.
}
\label{tbl:ff:ff:mud0015ms0080}
\begin{tabular}{lll|lll}
   $\theta$ & $\theta^\prime$ & $\theta^{\prime\prime}$ &
   $\pff(t)$ & $\kpff(t)$  & $\knff(t)$ 
   \\ \hline
   0.00  & 1.68  & 0.00  & 0.9407(35)  & 0.9400(22)  & 0.0062(10)
   \\
   0.00  & 2.64  & 0.00  & 0.8784(60)  & 0.8684(33)  & 0.0149(13)
   \\
   1.68  & 2.64  & 0.00  & 0.995(12)   & 0.9790(62)  & 0.0020(24)
\end{tabular}
\end{center}
\vspace{0mm}
\end{table}
\end{ruledtabular}

\begin{ruledtabular}
\begin{table}[t]
\begin{center}
\caption{
   Fit results for EM form factors at $(m_l,m_s)\!=\!(0.015,0.060)$.
}
\label{tbl:ff:ff:mud0015ms0060}
\begin{tabular}{lll|lll}
   $\theta$ & $\theta^\prime$ & $\theta^{\prime\prime}$ &
   $\pff(t)$ & $\kpff(t)$  & $\knff(t)$ 
   \\ \hline
   0.00  & 1.68  & 0.00  & 0.941(11)  & 0.9396(60)  & 0.00467(81)
   \\
   0.00  & 2.64  & 0.00  & 0.877(10)  & 0.8664(56)  & 0.0115(11)
   \\
   1.68  & 2.64  & 0.00  & 0.997(22)  & 0.985(11)   & 0.0001(20)
\end{tabular}
\end{center}
\vspace{0mm}
\end{table}
\end{ruledtabular}

\clearpage

\subsection{charge radii}
\label{subsec:r2}


In this article, 
we determine the charge radii $\crad$ 
of the light mesons ($P\!=\!\pi^+,K^+,K^0$)
at the physical quark masses 
from ChPT-based parametrizations of $\ff$.
In this subsection, 
we assume a $t$ dependence of $\ff$ 
based on phenomenological models,
and estimate the radii at simulated quark masses.

\begin{figure}[t]
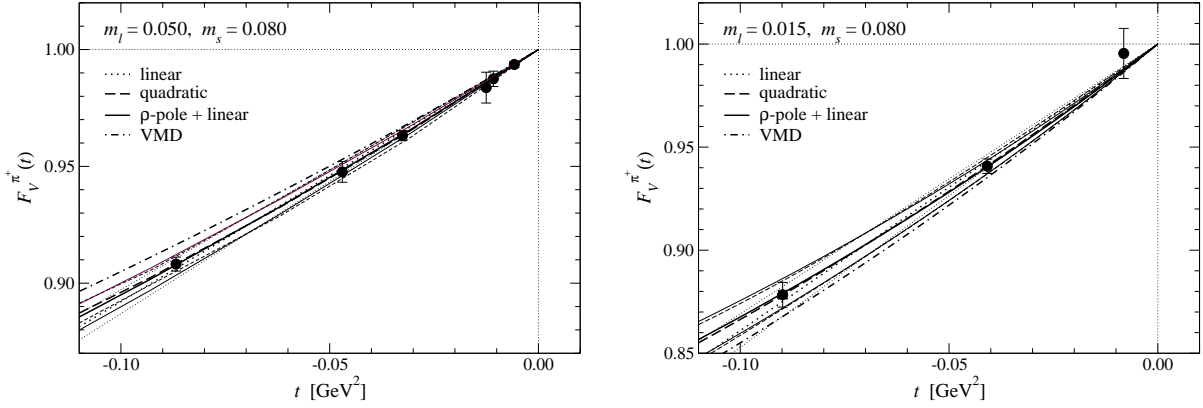

\begin{center}
\includegraphics[angle=0,width=0.48\linewidth,clip]%
                {pff_em_vs_q2_mud3_ms6.phys.eps}
\hspace{3mm}
\includegraphics[angle=0,width=0.48\linewidth,clip]%
                {pff_em_vs_q2_mud0_ms6.phys.eps}

\vspace{-3mm}
\caption{
   Pion EM form factor $\pff(t)$ as a function of momentum transfer $t$.
   The left and right panels show data at 
   $(m_l,m_s)\!=\!(0.050,0.080)$ and (0.015,0.080), respectively.
   Thick dotted and dashed lines show linear and quadratic fits, 
   whereas the fit based on VMD is plotted by the thick solid line.
   Their errors are plotted by thin lines. 
   The thick dot-dashed line shows the $t$ dependence expected from VMD.
}
\label{fig:ff:q2_interp:r2_pff_vs_q2}
\end{center}
\end{figure}

\begin{figure}[t]
\begin{center}
\includegraphics[angle=0,width=0.48\linewidth,clip]%
                {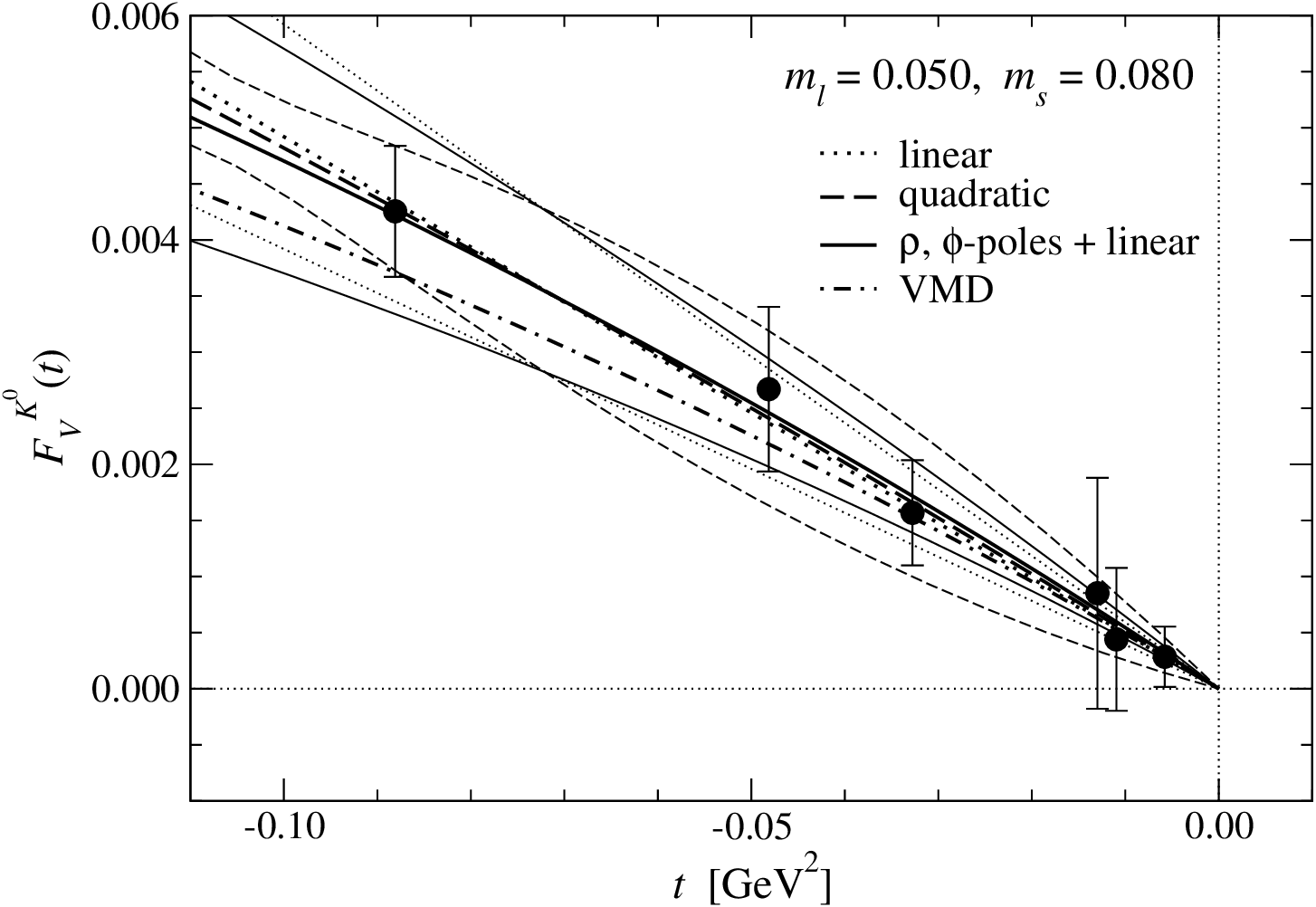}
\hspace{3mm}
\includegraphics[angle=0,width=0.48\linewidth,clip]%
                {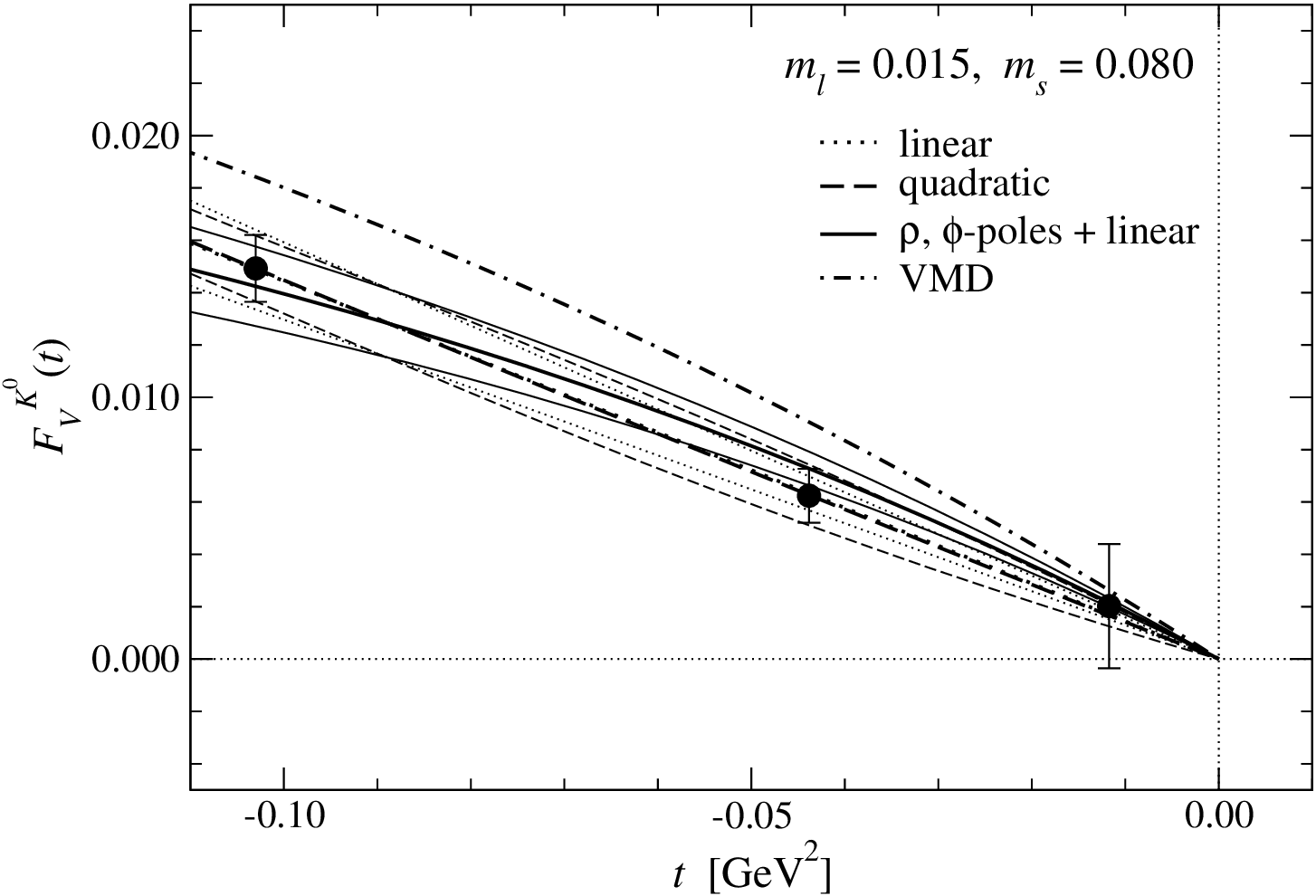}

\vspace{-3mm}
\caption{
   Neutral kaon EM form factor $\knff(t)$ 
   as a function of momentum transfer $t$.
}
\label{fig:ff:q2_interp:r2_k0ff_vs_q2}
\end{center}
\end{figure}

Figures~\ref{fig:ff:q2_interp:r2_pff_vs_q2}\,--\,\ref{fig:ff:q2_interp:r2_k0ff_vs_q2}
show the results for $\ff(t)$ as a function of the momentum transfer $t$.
We observe that their $t$ dependence 
is reasonably well described by the vector meson dominance (VMD) hypothesis 
(in the plots shown by dot-dashed curves)
\bea
   \pff(t)
   & = &
   \frac{1}{1-t/M_\rho^2},
   \\
   \kpff(t)
   & = &
   \frac{2}{3} \frac{1}{1-t/M_\rho^2} 
  +\frac{1}{3} \frac{1}{1-t/M_\phi^2},
   \\
   \knff(t)
   & = &
  -\frac{1}{3} \frac{1}{1-t/M_\rho^2} 
  +\frac{1}{3} \frac{1}{1-t/M_\phi^2},
   \label{eqn:ff:q2_interp:vmd:knff}
\eea
where $M_\rho$ and $M_\phi$ represent
the light and strange vector meson masses 
calculated at the simulated quark masses.
The small deviation may be attributed
to the effects of higher poles and cuts,
and is approximated by a polynomial correction in the following analysis.
Because quadratic and higher order corrections turn out to be insignificant
in the region of small $t$,
we employ the following fitting forms 
\bea
   \pff(t)
   & = &
   \frac{1}{1-t/M_\rho^2} + a_\pi t,
   \label{eqn:ff:q2_interp:pff_vs_q2}
   \\
   \kpff(t)
   & = &
   \frac{2}{3} \frac{1}{1-t/M_\rho^2} 
  +\frac{1}{3} \frac{1}{1-t/M_\phi^2}
  + a_{K^+} t,
   \label{eqn:ff:q2_interp:k+ff_vs_q2}
   \\
   \knff(t)
   & = &
  -\frac{1}{3} \frac{1}{1-t/M_\rho^2} 
  +\frac{1}{3} \frac{1}{1-t/M_\phi^2}
  + a_{K^0} t
   \label{eqn:ff:q2_interp:k0ff_vs_q2}
\eea
to estimate the charge radii defined in Eq.~(\ref{intro:radius}).
We also carry out linear and quadratic fits
\bea
   \ff(t)
   & = &
   b_0^P + b_1^P t \ (+ b_2^P t^2)
\eea
with $b_0^{\pi^+} = b_0^{K^+} = 1$ and $b_0^{K^0} = 0$.
The systematic uncertainty due to the choice of the parametrization form~(\ref{eqn:ff:q2_interp:pff_vs_q2})\,--\,(\ref{eqn:ff:q2_interp:k0ff_vs_q2})
is estimated by the difference in $\crad$ from these polynomial fits.

\begin{figure}[t]
\begin{center}
\includegraphics[angle=0,width=0.48\linewidth,clip]%
                {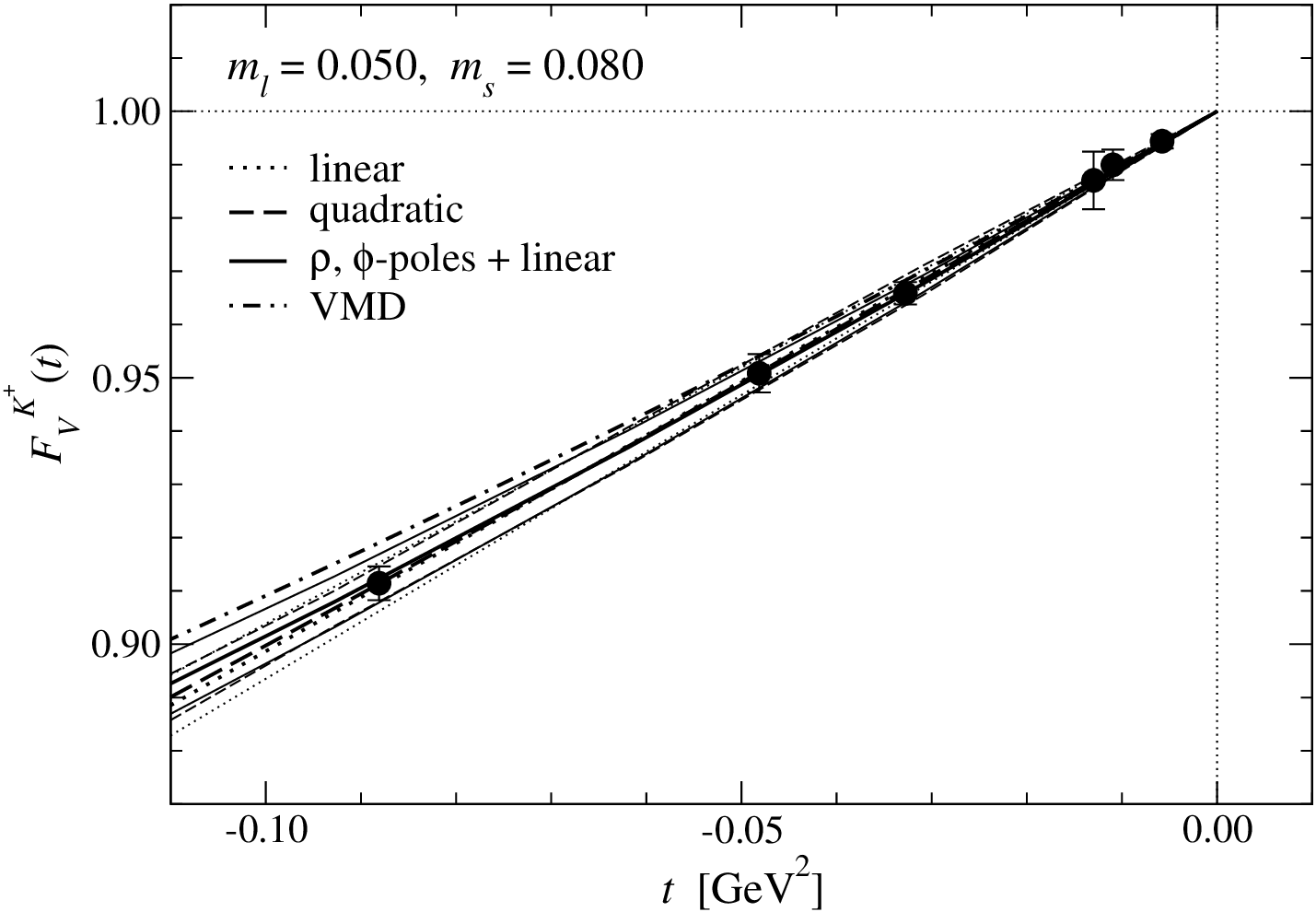}
\hspace{3mm}
\includegraphics[angle=0,width=0.48\linewidth,clip]%
                {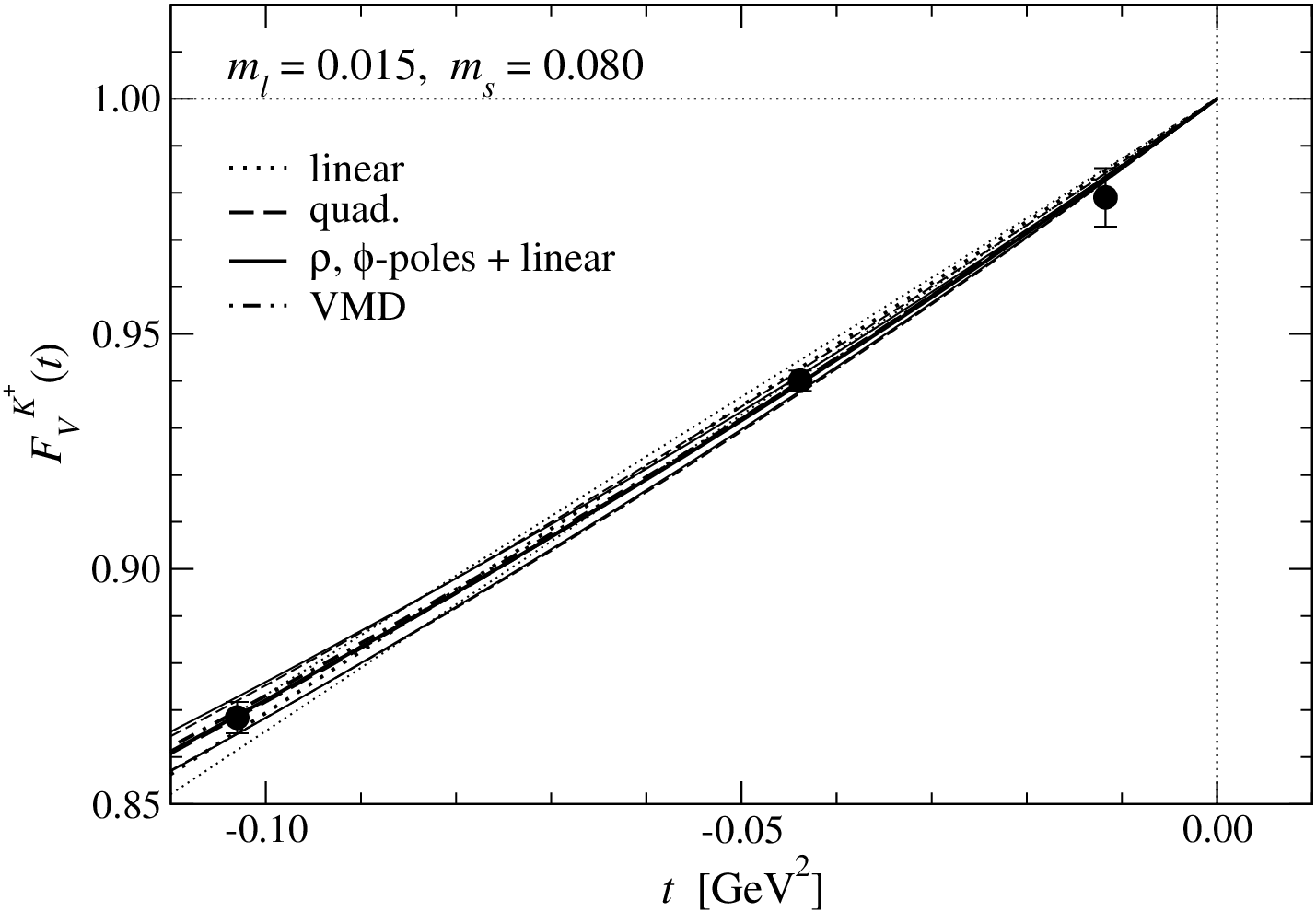}

\vspace{-3mm}
\caption{
   Charged kaon EM form factor $\kpff(t)$ 
   as a function of momentum transfer $t$.
}
\label{fig:ff:q2_interp:r2_k+ff_vs_q2}
\end{center}
\end{figure}

In Figs~\ref{fig:ff:q2_interp:r2_pff_vs_q2}\,--\,\ref{fig:ff:q2_interp:r2_k0ff_vs_q2},
we also plot fit curves with these parametrizations.
Numerical results for $\crad$ are summarized 
in Table~\ref{tbl:ff:q2_interp:radii}.
The radii have larger systematic error 
on the larger lattice, namely at $m_l\!\leq\!0.025$,
simply because 
we simulate only three values of $t$ in order to reduce the computational cost.
At each simulation point,
our data favor a smaller radius for the heavier charged meson $K^+$ 
than for the lighter one $\pi^+$,
though the difference is not large.
The radius of the neutral meson $K^0$ 
is much smaller than those for the charged mesons.
(Notice the scale of the vertical axis in Fig.~\ref{fig:ff:q2_interp:r2_k0ff_vs_q2}.)
These are qualitatively in accordance with ChPT and experiments.
We give quantitative comparisons in the next sections.

\begin{ruledtabular}
\begin{table}[t]
\begin{center}
\caption{
   Charge radii $\crad$ at simulated quark masses.
}  
\label{tbl:ff:q2_interp:radii}
\begin{tabular}{ll|llllll}
   $m_l$  & $m_s$  & $\cradpff$~[fm$^2$]  
                   & $\cradkpff$~[fm$^2$]    
                   & $\cradknff$~[fm$^2$]  
   \\ \hline
   0.050  & 0.080  &  $0.268(12)\left(^{+0}_{-17}\right)$
                   &  $0.251(12)\left(^{+0}_{-15}\right)$
                   & $-0.0129(23)\left(^{+15}_{-0}\right)$
   \\
   0.050  & 0.060  &  $0.270(16)\left(^{+2}_{-16}\right)$
                   &  $0.263(15)\left(^{+0}_{-17}\right)$
                   & $-0.0036(12)\left(^{+6}_{-0}\right)$
   \\ \hline
   0.035  & 0.080  &  $0.339(23)\left(^{+0}_{-23}\right)$
                   &  $0.305(18)\left(^{+0}_{-20}\right)$
                   & $-0.0157(28)\left(^{+31}_{-0}\right)$
   \\
   0.035  & 0.060  &  $0.344(31)\left(^{+0}_{-22}\right)$
                   &  $0.333(23)\left(^{+0}_{-22}\right)$
                   & $-0.0072(20)\left(^{+20}_{-0}\right)$
   \\ \hline
   0.025  & 0.080  &  $0.334(10)\left(^{+0}_{-32}\right)$
                   &  $0.317(6)\left(^{+0}_{-29}\right)$
                   & $-0.0345(23)\left(^{+58}_{-0}\right)$
   \\
   0.025  & 0.060  &  $0.346(43)\left(^{+0}_{-34}\right)$
                   &  $0.332(28)\left(^{+0}_{-32}\right)$
                   & $-0.0256(15)\left(^{+56}_{-0}\right)$
   \\ \hline
   0.015  & 0.080  &  $0.366(19)\left(^{+0}_{-42}\right)$
                   &  $0.343(9)\left(^{+0}_{-39}\right)$
                   & $-0.045(3)\left(^{+11}_{-0}\right)$
   \\
   0.015  & 0.060  &  $0.368(36)\left(^{+0}_{-48}\right)$
                   &  $0.354(17)\left(^{+0}_{-44}\right)$
                   & $-0.0349(28)\left(^{+0}_{-89}\right)$
\end{tabular}
\end{center}
\vspace{5mm}
\end{table}
\end{ruledtabular}

%% file: 4.chiral_fit_su2.tex
\section{Chiral extrapolation based on SU(2) ChPT}
\label{sec:chiral_fit:su2}


In this section, we fit our data of the pion EM form factor $\pff(t)$ 
to the NNLO formula in SU(2) ChPT as a function of $M_\pi$ and $t$.
We observe in Ref.~\cite{Spectrum:Nf2:RG+Ovr:JLQCD} that
the chiral expansion of the pion mass and decay constant 
shows better convergence
by using the expansion parameter 
$\xi_\pi \!=\! M_\pi^2 / (4\pi F_\pi)^2$ 
rather than $x\!=\! 2 B m_l / (4 \pi F )^2$,
where $B$ and $F$ are LECs in the LO chiral Lagrangian:
$F$ is the decay constant in the SU(2) chiral limit,
and $B$ appears in the LO relation $M_\pi\!=\!2Bm_l$.
We employ this ``$\xi$-expansion'' throughout this paper
to describe the quark mass dependence of the form factors. 
A typical functional form of the chiral logarithms at $n$-loops 
is $\xi_\pi^n \ln^m\left[ M_\pi^2 / \mu^2 \right]$  
$(m \leq n)$.
We set the renormalization scale $\mu\!=\!M_\rho$.

\begin{figure}[t]
\begin{center}
   \vspace{10mm}
   \includegraphics[angle=0,width=0.15\linewidth,clip]%
                   {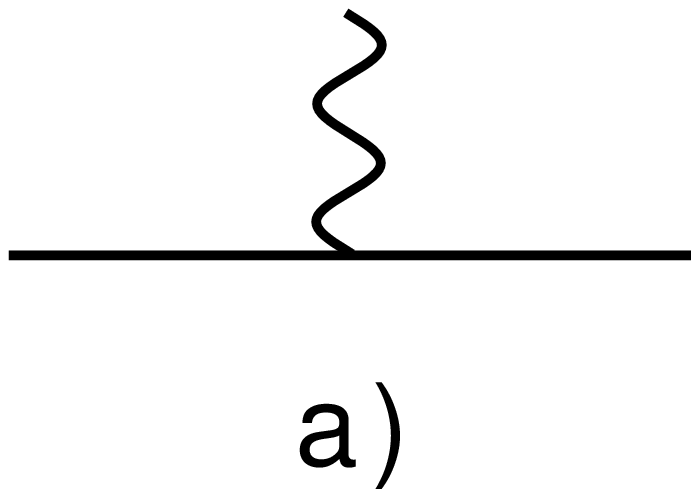}
   \hspace{5mm}
   \includegraphics[angle=0,width=0.15\linewidth,clip]%
                   {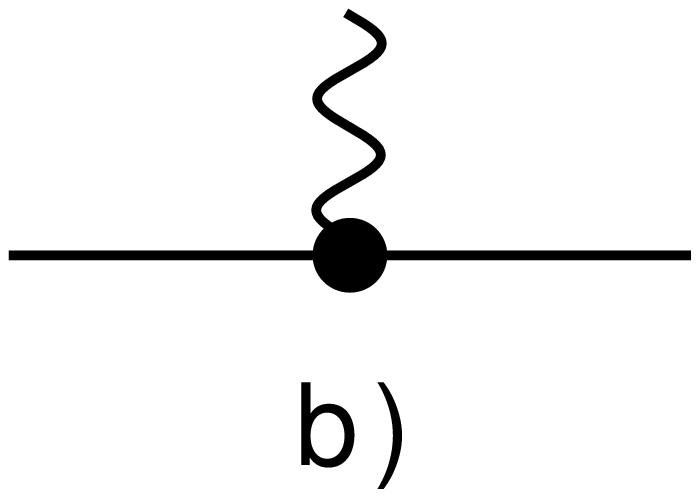}
   \hspace{5mm}
   \includegraphics[angle=0,width=0.15\linewidth,clip]%
                   {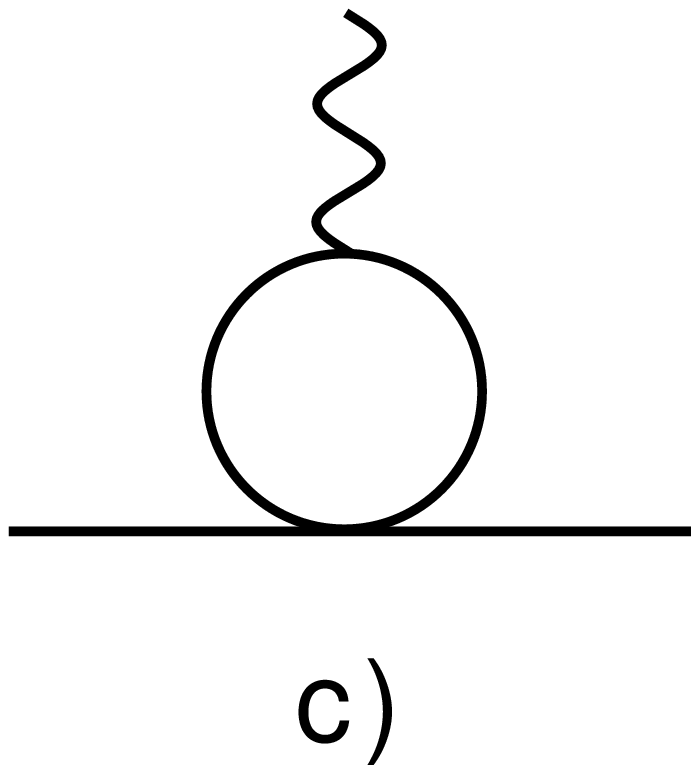}

   \vspace{10mm}
   \includegraphics[angle=0,width=0.15\linewidth,clip]%
                   {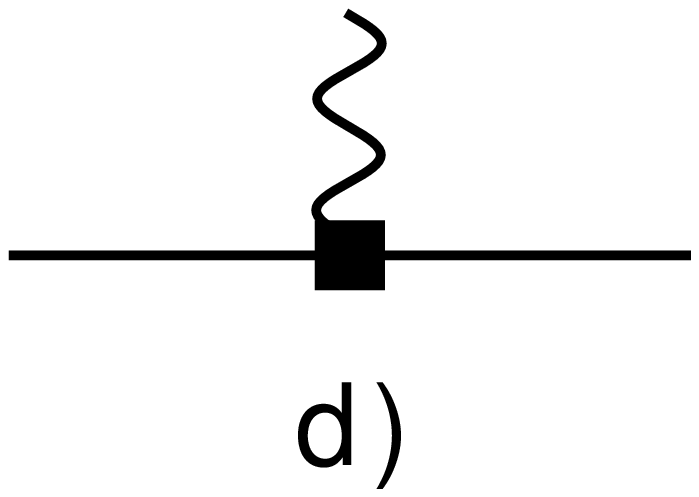}
   \hspace{5mm}
   \includegraphics[angle=0,width=0.15\linewidth,clip]%
                   {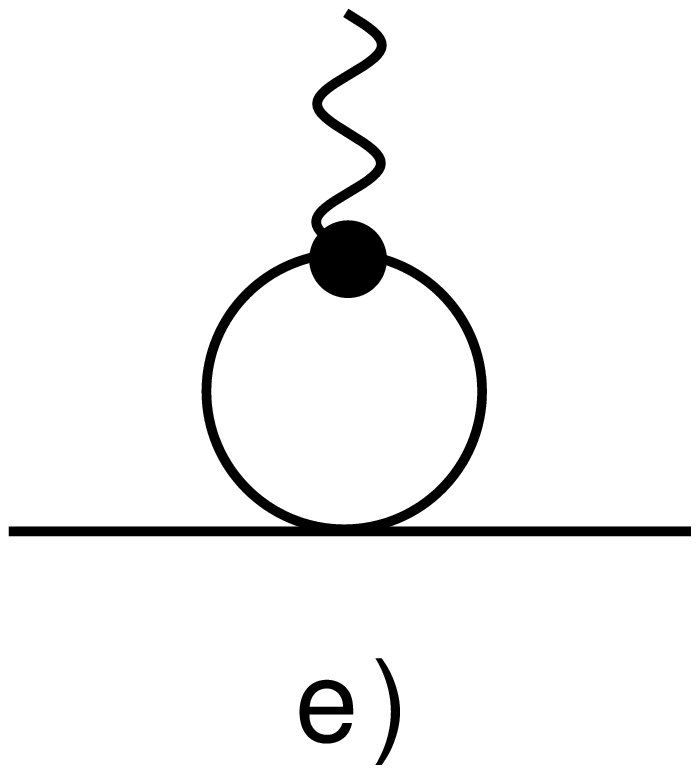}
   \hspace{5mm}
   \includegraphics[angle=0,width=0.15\linewidth,clip]%
                   {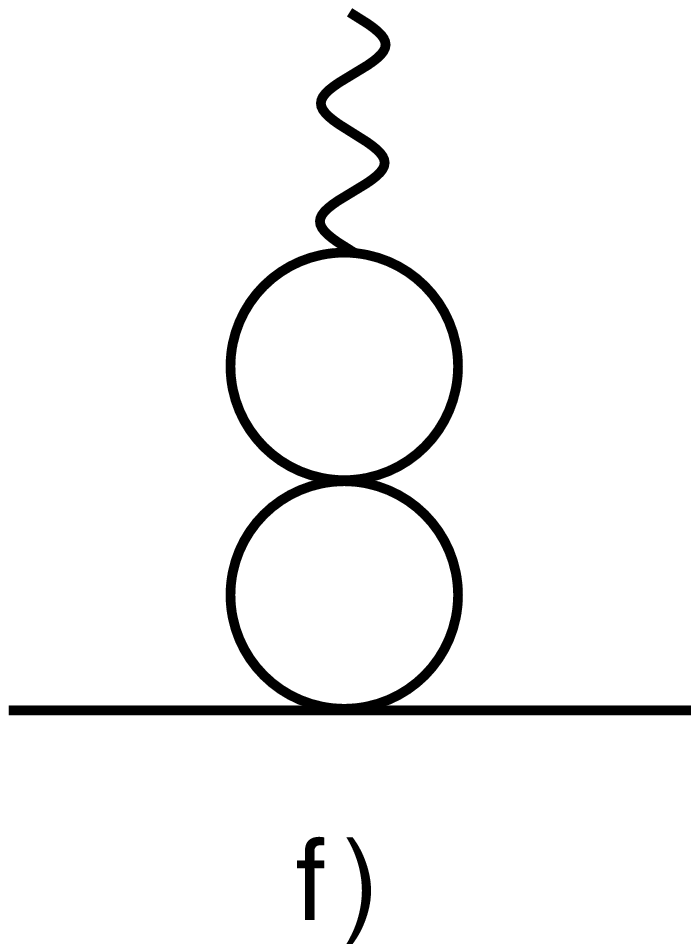}

   \vspace{0mm}
   \caption{
      Example of LO (a), NLO (b-c) and NNLO (d-f) diagrams.
      Straight and wavy lines represent the Nambu-Goldstone (NG) boson
      and photon, respectively. 
      The solid circle (square) represents a vertex 
      from $O(p^4)$ ($O(p^6)$) chiral Lagrangian $\mathcal{L}_4$ ($\mathcal{L}_6$).
   }
   \label{fig:chiral_fit:su2:diagram} 
\end{center}
\vspace{0mm}
\end{figure}

We denote the NNLO chiral expansion as 
\bea
   \pff(t)
   & = & 
   \pfflo + \pffnlo(t) + \pffnnlo(t).
   \label{eqn:chiral_fit:su2:pff}
\eea
The LO contribution $\pfflo$ arises from the diagram 
shown in Fig.\ref{fig:chiral_fit:su2:diagram}\,-\,a, 
and $\pfflo \!=\! \pff(0) \!=\! 1$ from the vector current conservation.
Examples of the NLO (NNLO) diagrams leading to $\pffnlo$ ($\pffnnlo$)
are shown in Figs.\ref{fig:chiral_fit:su2:diagram}\,-\,b and c (d, e and f).
These are expressed as~\cite{PFF:ChPT:SU2:NNLO:BCT}
\bea
   \pffnlo(t)
   & = & 
   \left\{
      \left( - N l_6^r - \frac{1}{18} \right) s 
      - \frac{N}{6} s L 
      + \frac{N}{6} (s-4) \bar{J}(s)
   \right\} 
   \xi_\pi,
   \label{eqn:chiral_fit:su2:pff:nlo}
   \\
   \pffnnlo(t)
   & = & 
   N^2 \left\{ P_{V,4}(s) + U_{V,4}(s) \right\} \xi_\pi^2,
   \label{eqn:chiral_fit:su2:pff:nnlo:P+U}
   \\
   P_{V,4}(s)
   & = & 
   \left\{
      - \frac{1}{2} k_{1,2} - \frac{1}{12} k_4 + \frac{1}{2} k_6 
   \right.
   \nn \\
   & & 
   \hspace{10mm}
   \left.
      - l_4^r \left( 2 l_6^r + \frac{1}{9N} \right)
      + \frac{23}{36 N} L + \frac{5}{576N} + \frac{37}{864N^2} + r_{V,1}^r 
   \right\} s 
   \nn \\
   & & 
   + 
   \left\{
      \frac{1}{12} k_{1,2} + \frac{1}{24} k_6 
    + \frac{1}{9N} \left( 
                      l_{1,2}^r + \frac{1}{2} l_6^r - \frac{1}{12} L 
                     -\frac{1}{384} - \frac{47}{192N} 
                   \right)
    + r_{V,2}^r    
   \right\} 
   s^2,
   \hspace{5mm}
\eea
\bea   
   U_{V,4}(s)
   & = & 
   \left\{ 
      - \frac{1}{3} l_{1,2}^r (s^2-4s) + \frac{1}{3} l_4^r (s-4)
      - \frac{1}{6} l_6^r (s^2-4s)
   \right.
   \nn \\
   & & 
   \hspace{10mm}
   \left.
      - \frac{1}{36} (s^2 + 8s - 48) L 
      +\frac{1}{N}\left( 
                     \frac{7}{108}s^2 - \frac{97}{108}s + \frac{3}{4} 
                  \right)
   \right\}
   \bar{J}(s)
   \nn \\
   & &
 + \frac{1}{9} K_1(s) + \frac{1}{9} \left( \frac{1}{8}s^2 - s + 4 \right) K_2(s)
 + \frac{1}{6} \left( s - \frac{1}{3} \right) K_3(s) - \frac{5}{3} K_4(s),
\eea
where 
\bea
   &&
   N = (4\pi)^2, 
   \hspace{5mm}
   s = \frac{t}{M_\pi^2},
   \hspace{5mm}
   L = \frac{1}{N} \ln \left[ \frac{M_\pi^2}{\mu^2} \right],
   \hspace{5mm}
   k_i = \left(4 l_i^r - \gamma_i  L \right) L, 
\eea   
with $\gamma_1\!=\!1/3$, $\gamma_2\!=\!2/3$, 
$\gamma_{1,2}\!=\gamma_1-\gamma_2/2\!=\!0$, 
$\gamma_4\!=\!2$, and $\gamma_6\!=\!-1/3$.
Here 
$l_i^r$ denotes the LECs in the NLO chiral Lagrangian $\mathcal{L}_4$. 
In the following, we refer to $l_i^r$'s and $\mathcal{L}_4$ 
as $O(p^4)$ couplings and $O(p^4)$ chiral Lagrangian, respectively.
Note that $M_{\{\pi,K\}}^2$ and $t$ are $O(p^2)$ quantities 
in the chiral order counting.
We define $l_{1,2}^r \!=\! l_1^r - l_2^r/2$,
because $l_1^r$ and $l_2^r$ appear in $\pff$ 
only through this linear combination.
The loop integral functions are defined as 
\bea
   \bar{J}(s) 
   & = & 
   h(s)z(s) + \frac{2}{N},
   \\
   K_1(s) 
   & = & 
   z(s) h(s)^2,
   \\
   K_2(s)
   & = & 
   z(s)^2 h(s)^2 - \frac{4}{N^2},
   \\
   K_3(s)
   & = &
   N \frac{z(s)h(s)^3}{s} + \frac{1}{16} \frac{h(s)}{s} - \frac{1}{32N},
   \\
   K_4(s)
   & = & 
   \frac{1}{sz(s)}
   \left\{ 
      \frac{1}{N} \bar{J}(s) + \frac{1}{2} K_1(s)  + \frac{1}{3} K_3(s) 
    + \frac{(\pi^2-6)s}{12N^2} 
   \right\},
\eea
using 
\bea
   z(s) = 1 - \frac{4}{s},
   \hspace{5mm}
   h(s) = \frac{1}{N\sqrt{z(s)}} \ln \left[ 
                                        \frac{\sqrt{z(s)}-1}{\sqrt{z(s)+1}} 
                                     \right].
\eea 
Therefore, 
$P_{V,4}(s)$ in Eq.~(\ref{eqn:chiral_fit:su2:pff:nnlo:P+U})
represents the NNLO contribution polynomial in $s \propto t$,
whereas $U_{V,4}(s)$ is the remaining one 
involving non-analytic loop functions in terms of $s$.


The chiral expansion (\ref{eqn:chiral_fit:su2:pff}) involves 
five unknown parameters:
three $O(p^4)$ couplings $l_6^r$, $l_{1,2}^r$, $l_4^r$,
and two couplings $r_{V,1}^r$ and $r_{V,2}^r$
from the $O(p^6)$ (NNLO) Lagrangian $\mathcal{L}_6$.
In order to obtain a stable chiral fit, 
we treat only $l_6^r$, $r_{V,1}^r$ and $r_{V,2}^r$ as fitting parameters,
because 
i) $l_6^r$ is the only free parameter 
appearing in the possibly large NLO correction, 
and ii) $r_{V,1}^r$ and $r_{V,2}^r$ from $\mathcal{L}_6$ are poorly known
and should be determined on the lattice.


\begin{ruledtabular}
\begin{table}[t]
\begin{center}
\caption{
   Input values for $O(p^4)$ couplings in SU(2) ChPT.
}
\label{tbl:chiral_fit:su2:input}
\begin{tabular}{ll}
   $\bar{l}_{1,2}$  & $\bar{l}_4$
   \\ \hline
   -2.55(60)       & 4.3(0.3)
\end{tabular}
\end{center}
\vspace{0mm}
\end{table}
\end{ruledtabular}

The $O(p^4)$ couplings, $l_{1,2}^r$ and $l_4^r$, appear only at NNLO.
We fix them to a phenomenological estimate
summarized in Table~\ref{tbl:chiral_fit:su2:input},
where we quote a scale-invariant combination 
\bea
   \bar{l}_i
   & = &
   \frac{2N}{\gamma_i} l_i^r - NL.
\eea
The input value for $l_{1,2}^r$ is obtained from 
a phenomenological analysis of the $\pi\pi$ scattering~\cite{ChPT:LECs:SU2}.
The value of $l_4^r$ suggested in Ref.~\cite{ChPT:LECs:SU2+SU3}
covers a phenomenological estimate
as well as 
lattice averages for $2 \!\leq\! N_f \!\leq\! 4$ 
obtained by the Flavor Lattice Averaging Group~\cite{FLAG2}.
The uncertainty due to this choice of the inputs 
is estimated by repeating our analysis
with $l_{1,2}^r$ and $l_4^r$
shifted by their uncertainty quoted in Table~\ref{tbl:chiral_fit:su2:input}.


\begin{figure}[t]
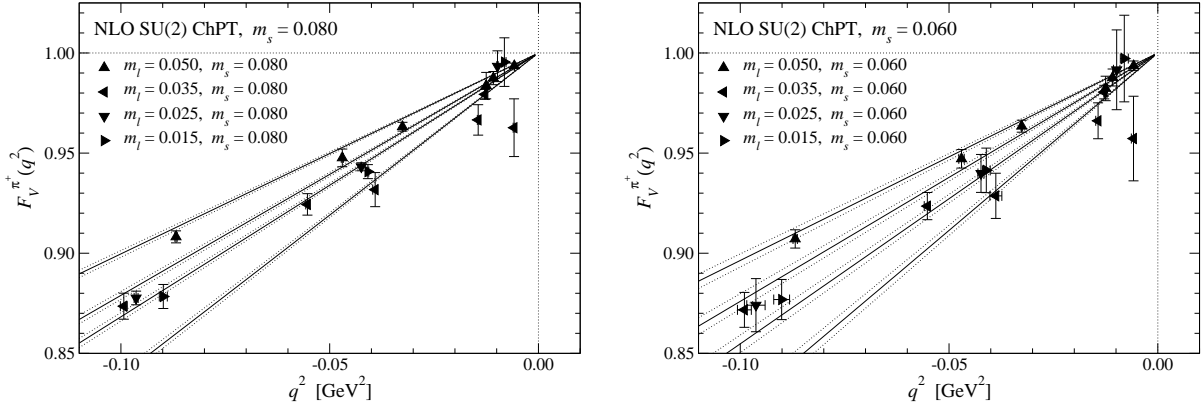

\begin{center}
\includegraphics[angle=0,width=0.48\linewidth,clip]%
                {pff_em_vs_q2_ms0080.phys.nlo.eps}
\hspace{3mm}
\includegraphics[angle=0,width=0.48\linewidth,clip]%
                {pff_em_vs_q2_ms0060.phys.nlo.eps}

\vspace{-3mm}
\caption{
   Chiral extrapolation of $\pff$ using NLO SU(2) ChPT formula
   at $m_s\!=\!0.080$ (left panel) and 0.060 (right panel).
   The data at four different $m_l$ are plotted as a function of $t$.
   Solid and dotted lines show the NLO fit curve and its statistical error.
   The lines correspond to $m_l\!=\!0.050, 0.035, 0.025$, and 0.015
   from top to bottom, respectively.
}
\label{fig:chiral_fit:su2:pff_vs_q2:nlo}
\end{center}
\end{figure}

\begin{figure}[t]
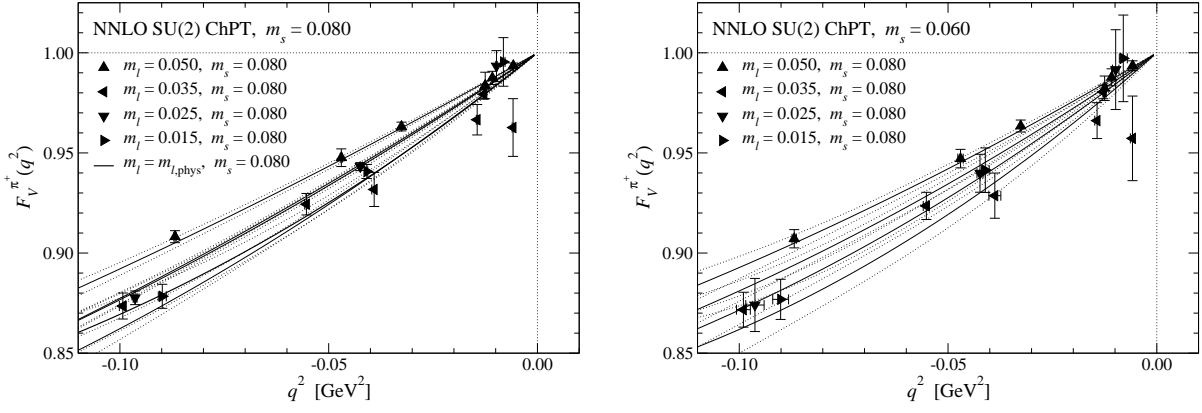

\begin{center}
\includegraphics[angle=0,width=0.48\linewidth,clip]%
                {pff_em_vs_q2_ms0080.phys.nnlo.l124fixed.eps}
\hspace{3mm}
\includegraphics[angle=0,width=0.48\linewidth,clip]%
                {pff_em_vs_q2_ms0060.phys.nnlo.l124fixed.eps}

\vspace{-3mm}
\caption{
   Chiral extrapolation of $\pff$ using NNLO SU(2) ChPT formula.
   Thin solid lines show the NNLO fit curve at simulated $m_l$.
   In the left panel, we also plot the fit curve at the physical 
   light quark mass $m_{l,\rm phys}$ by the thick solid line. 
   Note that $m_s\!=\!0.080$ is close to $m_{s,\rm phys}$.
}
\label{fig:chiral_fit:su2:pff_vs_q2:nnlo}
\end{center}
\end{figure}

Figure~\ref{fig:chiral_fit:su2:pff_vs_q2:nlo}
shows the chiral extrapolation using the NLO expression at each $m_s$.
The lattice data at the largest and smallest $m_l$ tend to deviate 
from the fit curve and 
lead to large values of $\chi^2/{\rm d.o.f} \!\sim\! 1.9$\,--\,2.9.
Note that
$l_6^r$ is the only free parameter appearing at NLO
and may be too few to describe both the $m_l$ and $t$ dependences.
The NNLO fit shown in Fig.~\ref{fig:chiral_fit:su2:pff_vs_q2:nnlo} 
describes our data better 
and $\chi^2/{\rm d.o.f}$ is significantly reduced to 0.9\,--\,1.2.

\begin{figure}[t]
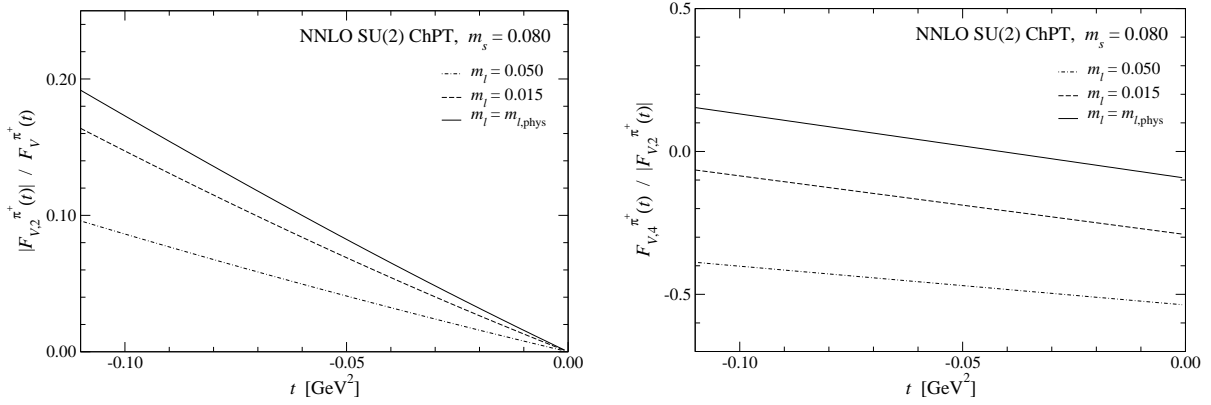

\begin{center}
\includegraphics[angle=0,width=0.48\linewidth,clip]%
                {pff_em_vs_q2_ms0080.phys.l124fixed.nlo_over_full.eps}
\hspace{3mm}
\includegraphics[angle=0,width=0.48\linewidth,clip]%
                {pff_em_vs_q2_ms0080.phys.l124fixed.nnlo_over_nlo.eps}
\vspace{-3mm}
\caption{
   Convergence of chiral expansion at $m_s\!=\!0.080$.
   Left panel: ratio of the NLO contribution to the total 
               $|F_{V,2}^{\pi^+}|/\pff$.  
               The dot-dashed, dashed and solid lines show
               data at $m_l\!=\!0.050$, 0.015 and $m_{l,\rm phys}$, respectively.
   Right panel: ratio of the NLO and NNLO contributions
                $F_{V,4}^{\pi^+}/|F_{V,2}^{\pi^+}|$.
}
\label{fig:chiral_fit:su2:convergence:pff}
\end{center}
\end{figure}


The convergence of this NNLO expansion seems reasonable
around the physical strange quark mass $m_s\!\sim\!m_{s,\rm phys}$
as plotted in Fig.~\ref{fig:chiral_fit:su2:convergence:pff}.
We observe that the NLO contribution $\pffnlo$ is at most 20\,\% 
of the total value $\pff$ in our simulated region of $t$ and $m_l$.
The slightly worse convergence at lighter $m_l$ is 
because $\pffnlo$ is proportional to $F_\pi^{-2}$ in the $\xi$-expansion.
The magnitude of the NNLO contribution relative to NLO is about 0.5
at our largest $m_l$.
It however decreases to $\lesssim 0.1$\,--\,0.2 
around our lightest $m_l$ and down to $m_{l,\rm phys}$.


\begin{figure}[t]
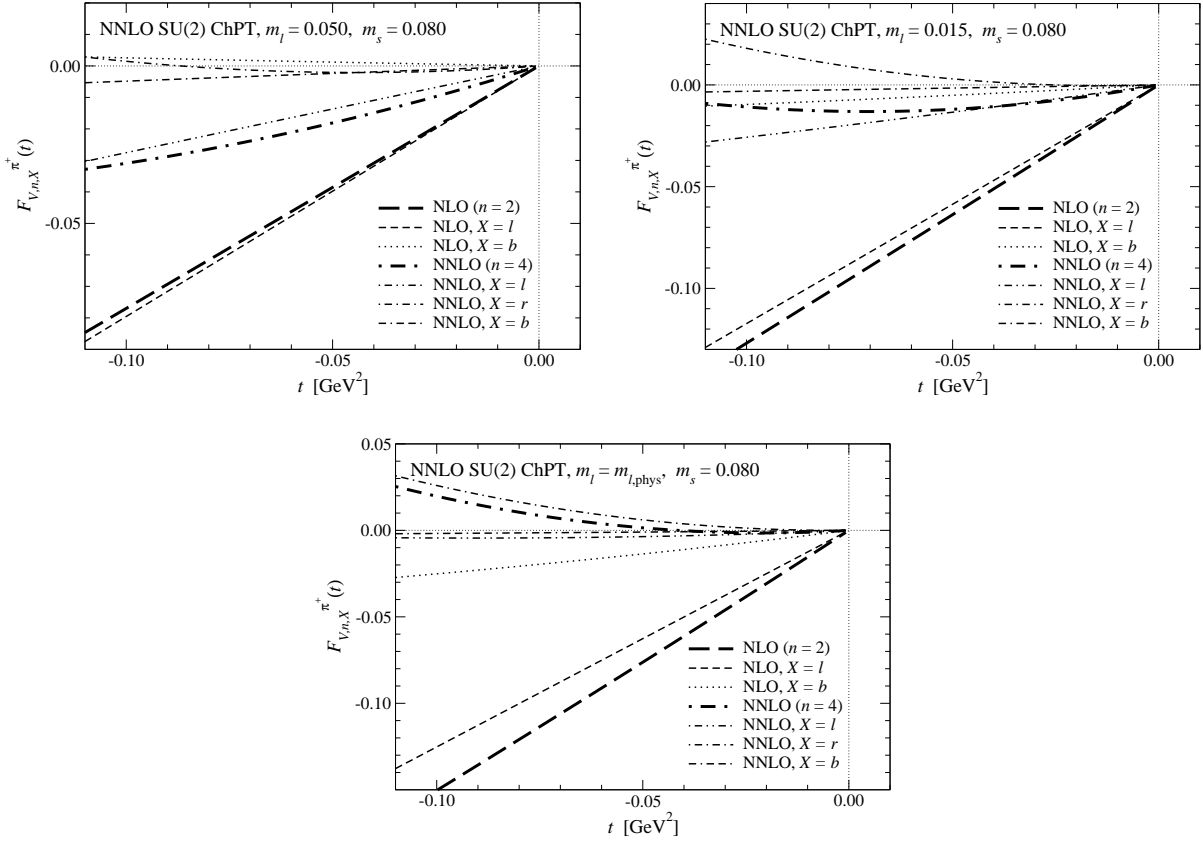

\begin{center}
\includegraphics[angle=0,width=0.48\linewidth,clip]%
                {pff_em_vs_q2.mud0050_ms0080.phys.l124fixed.cntrb.eps}
\hspace{3mm}
\includegraphics[angle=0,width=0.48\linewidth,clip]%
                {pff_em_vs_q2.mud0015_ms0080.phys.l124fixed.cntrb.eps}

\vspace{5mm}
\includegraphics[angle=0,width=0.48\linewidth,clip]%
                {pff_em_vs_q2.mudphys_ms0080.phys.l124fixed.cntrb.eps}

\vspace{-3mm}
\caption{
   LEC-(in)dependent contributions to $\pff$
   in our chiral fit at $m_s\!=\!0.080$ based on NNLO SU(2) ChPT.
   Top left and right panels show data at our simulation points
   $m_l\!=\!0.050$ and 0.015.
   The bottom panel is at the physical light quark mass $m_{l,\rm phys}$.
}
\label{fig:chiral_fit:su2:contribu:pff}
\end{center}
\end{figure}

For a more detailed look,
we decompose the NLO and NNLO contributions
into LEC-dependent and independent parts
and rewrite the chiral expansion (\ref{eqn:chiral_fit:su2:pff}) as 
\bea
   \pff(t)
   & = & 
   \pfflo 
   + \pffnlol(t) + \pffnlob(t) 
   + \pffnnlol(t) + \pffnnlor(t) + \pffnnlob(t).
   \label{eqn:chiral_fit:su2:pff:contribu}
\eea
Here $\pffnlol$ ($\pffnlob$) represents 
the $l_i^r$-dependent (independent) NLO term, 
which arises from the diagrams shown 
in Fig.~\ref{fig:chiral_fit:su2:diagram}\,-\,b (c). 
The $r_{V,i}^r$- and $l_i^r$- dependent NNLO terms, $\pffnnlor$ and $\pffnnlol$, 
mainly come from the tree diagrams involving an $\mathcal{L}_6$ vertex
and the one-loop diagrams with an $\mathcal{L}_4$ vertex, respectively.
An example of these diagrams is shown in 
Figs.~\ref{fig:chiral_fit:su2:diagram}\,-\,d and e.
The LEC-independent NNLO term $\pffnnlob$ is from two-loop diagrams 
such as Fig.~\ref{fig:chiral_fit:su2:diagram}\,-\,f.
Figure~\ref{fig:chiral_fit:su2:contribu:pff} compares these terms
at $m_l\!=\!0.050$, 0.015, and $m_{l,\rm phys}$.
We observe that
the NLO contribution $\pffnlo$ is largely dominated
by the $l_i^r$-dependent analytic term $\pffnlol$.
%
The NNLO contribution $\pffnnlo$ is dominated by the $l_i^r$-dependent
term $\pffnnlol$ at our largest $m_l$,
whereas 
$r_{V,i}^r$-dependent term $\pffnnlor$ tends to dominate $\pffnnlo$
at smaller $m_l$.
Therefore 
the uncertainty 
due to the use of the phenomenological input for $l_{1,2}^r$ and $l_4^r$
may not be large for our results at physical $m_l$, 
such as the charge radius $\cradpff$
(see Eq.~(\ref{eqn:chiral_fit:su2:r2:pff})).
Compared to these LEC-dependent contributions,
$\pffnlob$ and $\pffnnlob$ coming from 
genuine loop diagrams (namely without $\mathcal{L}_{\{4,6\}}$ vertices)
are rather small.


\begin{ruledtabular}
\begin{table}[t]
\begin{center}
\caption{
   Numerical results of chiral fit based on NNLO SU(2) ChPT
   at $m_s\!=\!0.080$ and $0.060$. 
   For the LECs,
   we quote the values at the renormalization scale $\mu\!=\!M_\rho$.
   The first error is statistical, and the second is 
   systematic one due to the choice of the input $l_{1,2}^r$ and $l_4^r$.
   We also quote results extrapolated to $m_{s,\rm phys}$.
}
\label{tbl:chiral_fit:su2:nnlo}
\begin{tabular}{l|llll}
   $m_s$  &  $l_6^r\!\times\!10^3$  
          &  $r_{V,1}^r\!\times\!10^5$ & $r_{V,2}^r\!\times\!10^5$ 
          &  $\cradpff$~[fm$^2$]
   \\ \hline
   0.080  &  -10.65(94)(15)  & 5.9(5.9)(3.5)
          &   19.9(9.3)(0.1) & 0.395(26)(3)
   \\
   0.060  &  -10.9(2.4)(0.2) & 7(14)(4)
          &   31(19)(0)      & 0.403(67)$\left(^{+6}_{-3}\right)$
   \\ \hline
   $m_{s,\rm phys}$ & -10.64(94)(15) & 5.9(5.9)(3.5) 
                  & 19.4(9.4)(0.1)    & 0.395(26)(3)
\end{tabular}
\end{center}
\vspace{0mm}
\end{table}
\end{ruledtabular}

\begin{figure}[t]
\begin{center}
\includegraphics[angle=0,width=0.48\linewidth,clip]%
                {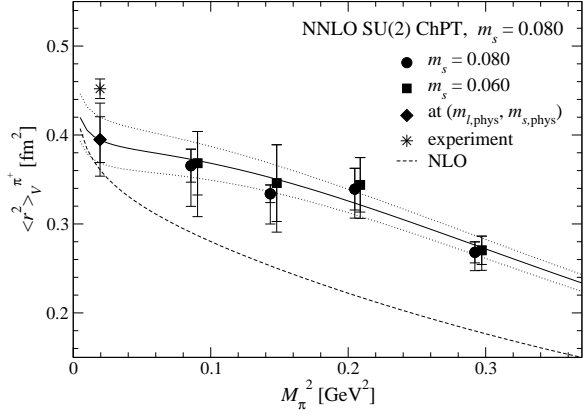}

\vspace{-3mm}
\caption{
   Pion charge radius $\cradpff$ as a function of $M_\pi^2$. 
   The solid line represents $\cradpff$ at $m_s\!=\!0.080$ 
   reproduced from the NNLO SU(2) ChPT fit. 
   The dashed line shows the NLO contribution.
   We plot the values in Table~\ref{tbl:ff:q2_interp:radii}
   by solid circles ($m_s\!=\!0.080$) and squares (0.060).
   The diamond and star are the value extrapolated to the physical point 
   and the experimental value~\cite{PDG:2014}, respectively.
}
\label{fig:chiral_fit:su2:r2_vs_Mpi2}
\end{center}
\end{figure}

Numerical results of the NNLO fits 
at the simulated strange quark masses
are summarized in Table~\ref{tbl:chiral_fit:su2:nnlo}.
We estimate the charge radius $\cradpff$ by using 
these results in the NNLO ChPT expression~\cite{PFF:ChPT:SU2:NNLO:BCT}
\bea
   M_\pi^2\,
   \cradpff
   & = & 
   N \left( -6l_6^r - L - \frac{1}{N} \right) \xi_\pi
  +N^2 \left\{
          -3 k_{1,2} - \frac{1}{2} k_4 + 3k_6 - 12 l_4^r l_6^r
   \right.
   \nn \\
   &  &
       \left.       
          +\frac{1}{N} \left( 
                          -2l_4^r + \frac{31}{6}L 
                          +\frac{13}{192} +\frac{181}{48N} \right)
          +6r_{V,1}^r 
       \right\} \xi_\pi^2. 
\eea
As plotted in Fig.~\ref{fig:chiral_fit:su2:r2_vs_Mpi2}, 
the NNLO fit reproduces the values in Table~\ref{tbl:ff:q2_interp:radii},
which are evaluated at simulation points assuming $t$-dependence 
of Eqs.~(\ref{eqn:ff:q2_interp:pff_vs_q2})\,--\,(\ref{eqn:ff:q2_interp:k0ff_vs_q2}), reasonably well. 
This figure also shows that 
the NNLO contribution is significant in our simulation region 
$M_\pi \!\gtrsim\! 300$~MeV 
($M_\pi^2 \!\gtrsim\! 0.09~\mbox{GeV}^2$ in the horizontal axis of the figure).
This is consistent with our previous finding 
in two-flavor QCD~\cite{PFF:JLQCD:Nf2:RG+Ovr}.

\begin{figure}[t]
\begin{center}
\includegraphics[angle=0,width=0.48\linewidth,clip]%
                {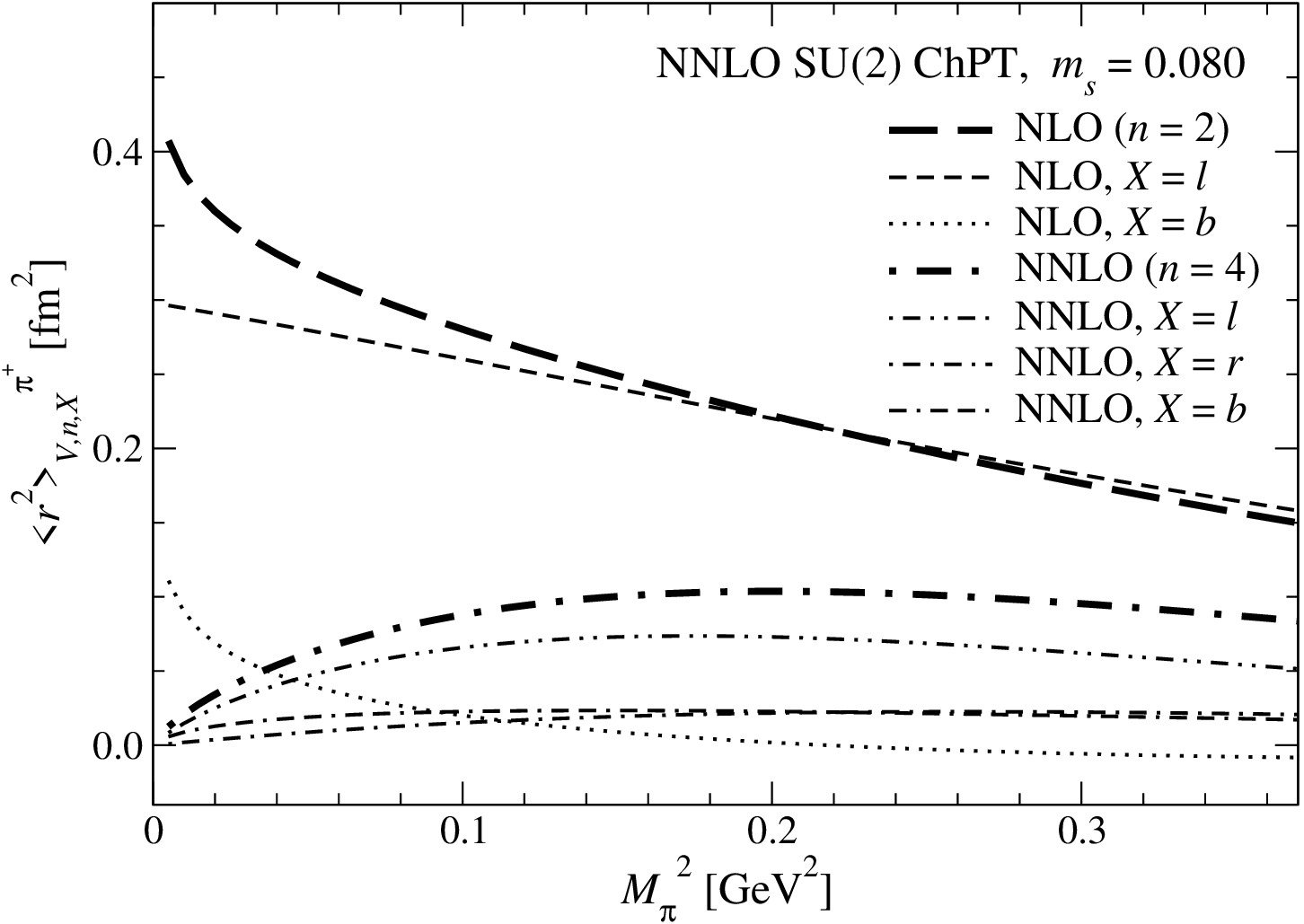}

\vspace{-3mm}
\caption{
   LEC-(in)dependent contributions at NLO and NNLO to $\cradpff$.
}
\label{fig:chiral_fit:su2:r2_vs_Mpi2:contribu}
\end{center}
\end{figure}


Similar to the decomposition of $\pff$ 
in Eq.~(\ref{eqn:chiral_fit:su2:pff:contribu}), 
we express the chiral expansion of $\cradpff$ as
\bea
   \cradpff
   & = & 
   \cradpffnlo + \cradpffnnlo, 
   \\
   \cradpffnlo 
   & = & 
   \cradpffnlol + \cradpffnlob,
   \hspace{5mm}
   \cradpffnnlo 
   =
   \cradpffnnlol + \cradpffnnlor + \cradpffnnlob.
   \label{eqn:chiral_fit:su2:r2:pff:contribu}
\eea
Namely, $\cradpffnlol$, $\cradpffnnlol$  and $\cradpffnnlor$
depend on $l_i^r$ and $r_{V,i}^r$, 
whereas $\cradpffnlob$ and $\cradpffnnlob$ are independent of the LECs.
These contributions are plotted as a function of $M_\pi^2$ 
in Fig.~\ref{fig:chiral_fit:su2:r2_vs_Mpi2:contribu}.
The NLO contribution is largely dominated 
by the analytic term $\cradpffnlol$, 
as $\pffnlol$ dominates $\pffnlo$.
The charge radius has been considered as a good quantity 
to observe the one-loop chiral logarithm 
$\frac{1}{NF_{(\pi)}^2} \ln[ M_\pi^2/\mu^2]$,
which is not suppressed by a multiplicative factor $M_\pi^2$ 
and hence diverges toward the chiral limit.
In our notation, 
this is included in the NLO loop correction $\cradpffnlob$ 
but becomes significant only at $M_\pi \!\lesssim\! 300$~MeV,
namely below our simulation points.
In addition, the enhancement of $\cradpffnlob$ 
is partly compensated by the decrease of the NNLO contribution,
particularly of $\cradpffnnlol$.
Therefore, we may be able to clearly observe the logarithmic singularity 
only near the chiral limit.
Our work in the so-called $\epsilon$-regime~\cite{PFF:JLQCD:Nf3:RG+Ovr:e-regime}
is an interesting step in this direction.


The NNLO contribution $\cradpffnnlo$ turns out to be 
a 30\,--\,50\% correction at the simulated values of $M_\pi^2$
and becomes small, $\lesssim\!10$~\%, only near the physical point.
The two-loop term $\cradpffnnlob$ is rather small. 
The analytic term $\cradpffnnlor$ vanishes towards the chiral limit,
whereas the similar term $\pffnnlor$ is not a small correction to $\pff$.
This is because $O(t^2)$ term of $\pff$ with $r_{V,2}^r$ 
does not contribute to $\cradpff$,
and $\cradpffnnlor\!=\!6 N r_{V,1}^r \xi_\pi / F_\pi^2$
is suppressed by $M_\pi^2$ in the chiral limit.
Hence the $l_i^r$-dependent term $\cradpffnnlol$ gives 
the largest contribution at NNLO.
Note that this term has non-trivial $M_\pi^2$ dependence: 
roughly constant down to $M_\pi\!\simeq\!400$~MeV and non-linearly
decreases towards the chiral limit. 
It is therefore important to correctly take account of the NNLO contributions
for a reliable chiral extrapolation of $\cradpff$.

\begin{figure}[t]
\begin{center}
\includegraphics[angle=0,width=0.48\linewidth,clip]%
                {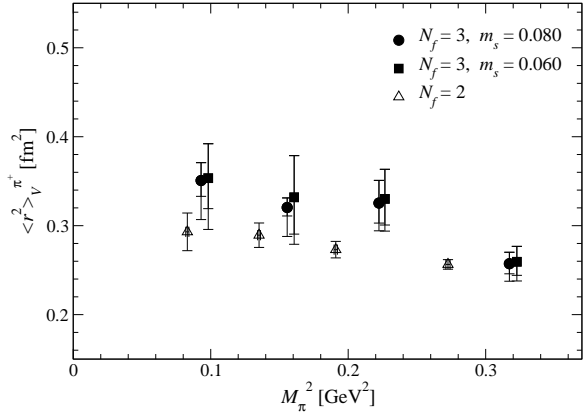}

\vspace{-3mm}
\caption{
   Comparison of $\cradpff$ between three-flavor QCD
   (solid circles and squares) 
   and two-flavor QCD (open triangles)~\cite{PFF:JLQCD:Nf2:RG+Ovr}.
   The latter was obtained on $16^3\times32$ at $a=0.118(2)$~fm
   with four times higher statistics, but $|t|\!\gtrsim\!(\mbox{500 MeV})^2$
   without the twisted boundary conditions.
   For a fair comparison, we use the lattice spacing determined 
   from $r_0\!=\!0.49$~fm~\cite{r0} 
   to convert all data to physical units.
}
\label{fig:chiral_fit:su2:r2_vs_nf}
\end{center}
\end{figure}


In SU(2) ChPT,
the $m_s$ dependence of physical quantities is encoded in that of LECs. 
We need to extrapolate our results to the physical strange quark mass 
$m_{s,\rm phys}$ in order to obtain information about the real world.
As far as the pion observables $\pff$ and $\cradpff$ are concerned, 
the $m_s$ dependence turns out to be mild
as suggested by the good stability of $\cradpff$ 
between $m_s\!=\!0.080$ and 0.060 as shown in Fig.~\ref{fig:chiral_fit:su2:r2_vs_Mpi2}.
This is confirmed also in Fig.~\ref{fig:chiral_fit:su2:r2_vs_nf},
which shows that the difference in $\cradpff$ between three- and 
two-flavor QCD is not large.

\begin{figure}[t]
\begin{center}
\includegraphics[angle=0,width=0.48\linewidth,clip]%
                {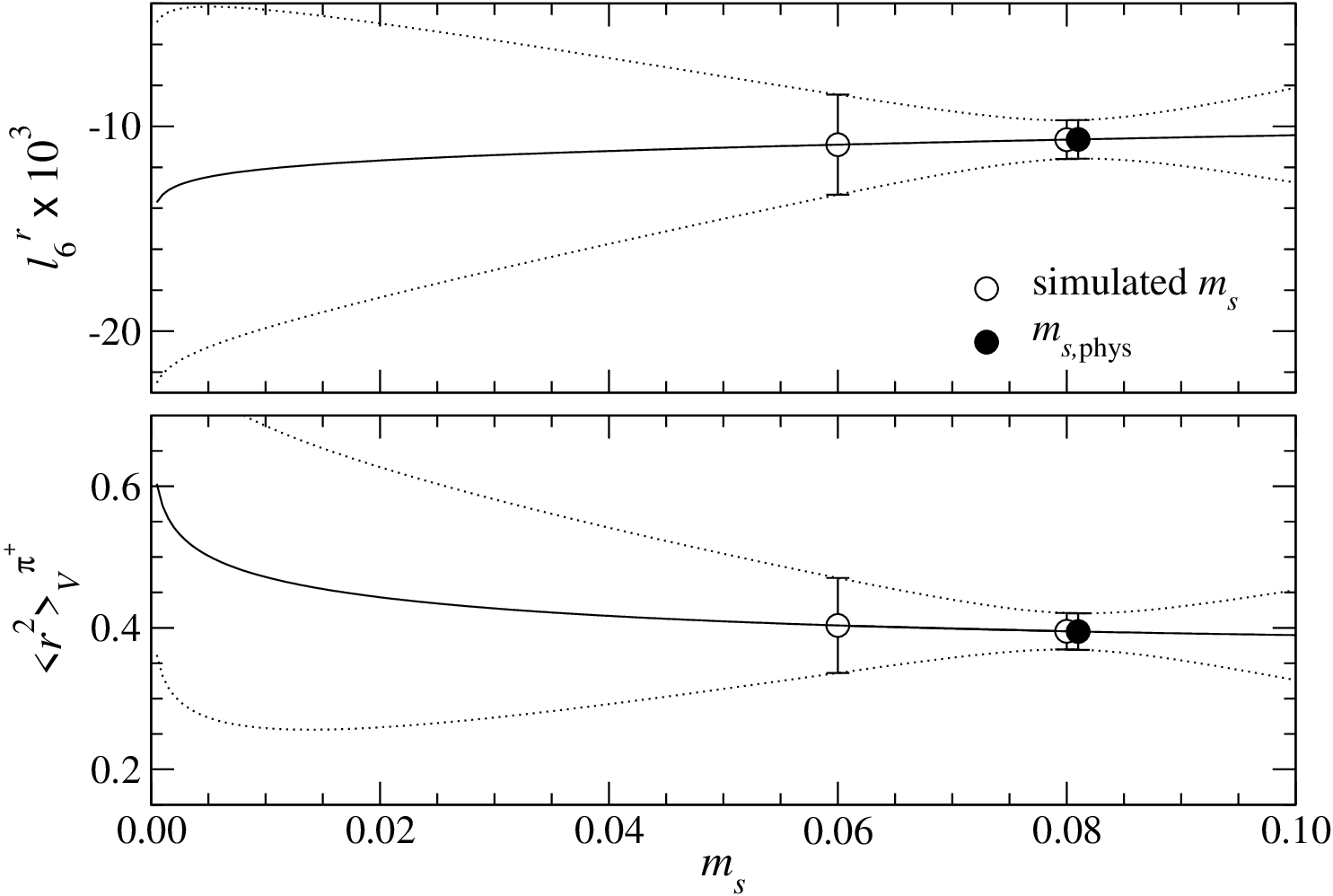}

\vspace{-3mm}
\caption{
   Extrapolation of $l_6^r$ (top panel) and $\cradpff$ (bottom panel)
   to $m_{s,\rm phys}$.
}
\label{fig:chiral_fit:su2:ms-fit}
\end{center}
\end{figure}

For the extrapolation of $l_6^r$ and $\cradpff$, 
we parametrize their $m_s$ dependence 
by a linear function including the NLO chiral logarithm~\cite{ChPT:SU3:NLO} 
\bea
   l_6^r 
   & = &
   a_{l,0} + \frac{1}{12N}\ln\left[ M_K^2 \right] + a_{l,1}\, m_s,
   \\
   \cradpff 
   & = &
   a_{r^2,0} - \frac{1}{2NF_\pi^2}\ln\left[ M_K^2 \right] + a_{r^2,1}\, m_s.
\eea  
Figure~\ref{fig:chiral_fit:su2:ms-fit} shows that
the logarithmic term $\ln[M_K^2]$ becomes significant only near the $m_s\!=\!0$ limit,
and that the simulated value $m_s\!=\!0.080$ is close to $m_{s,\rm phys}$. 
Moreover, the $m_s$ dependence is rather mild as discussed above. 
The extrapolation therefore does not deteriorate the statistical accuracy,
and is stable against the choice of the parametrization form:
for instance, removing the logarithmic term and/or
including an $O(m_s^2)$ correction. 
These observations lead us to employ a simple linear form 
\bea
   r_{V,i}^r 
   & = &
   a_{r_i,0} + a_{r_i,1} m_s
\eea  
for $r_{V,i}^r$, which has the large statistical error.

The extrapolated values are summarized in Table~\ref{tbl:chiral_fit:su2:nnlo}. 
We obtain
\bea
   \cradpff
   & = & 
   0.395(26)(3)(32)~\mbox{fm}^2,
   \label{eqn:chiral_fit:su2:r2:pff}
\eea
where the first error is statistical,
and the second one is the systematic error due to 
the choice of the input values of $l_i^r$.
The third is the discretization error at our finite lattice spacing,
which is conservatively estimated by an order counting 
$O((a\Lambda_{\rm QCD})^2)\!\sim\!8$\,\% with $\Lambda_{\rm QCD}\!=\!500$~MeV.
Our result of $\cradpff$ 
is consistent with the experimental value 
$\cradpff\!=\!0.452(11)~\mbox{fm}^2$~\cite{PDG:2014}
within estimated uncertainties.
Note that the systematic error due to the choice of the inputs 
$l_{1,2}^r$ and $l_4^r$ is rather small for this quantity,
because only the NNLO $l_i^r$-dependent terms, $\pffnnlol$ and $\cradpffnnlol$,
contain these inputs and decrease towards the physical point.

For the $O(p^4)$ coupling, we obtain
\bea
   \bar{l}_6
   & = & 
   13.49(89)(14)(81)
   \hspace{5mm}
   (l_6^r = -10.64(94)(15)(86) \!\times\! 10^{-3}).
\eea
This is consistent with
our estimate $\bar{l}_6\!=\!11.9(1.2)$ in two-flavor QCD~\cite{PFF:JLQCD:Nf2:RG+Ovr} 
as well as 
with phenomenological estimates
$16.0(0.9)$~\cite{PFF:ChPT:SU2:NNLO:BCT} 
from the experimental data of $\pff$, 
and 15.2(4) obtained together with the $\pi\!\to\!e\nu\gamma$ decay 
and the $V\!-\!A$ spectral function~\cite{ChPT:l6:GPP,ChPT:LECs:SU2+SU3}.
Our results for the $O(p^6)$ couplings at $\mu\!=\!M_\rho$ are 
\bea
   r_{V,1}^r 
   & = & 
   5.9(5.9)(3.5)(0.5) \times 10^{-5},
   \\
   r_{V,2}^r
   & = &
   19.4(9.4)(0.1)(1.6) \times 10^{-5}.
\eea

%% file: 5.chiral_fit_su3.tex
\section{Chiral extrapolation based on SU(3) ChPT}
\label{sec:chiral_fit:su3}



In this section, 
we extend our analysis to SU(3) ChPT, 
which is applicable also to the kaon EM form factors $\kpff$ and $\knff$.
According to Ref.~\cite{PFF+KFF:ChPT:NNLO:Nf3} and 
similar to Eq.~(\ref{eqn:chiral_fit:su2:pff:contribu}),
we write the chiral expansion of the EM form factors of the light mesons 
($P\!=\!\pi^+,K^+,K^0$) as 
\bea
   \ff(t)
   & = &
   \fflo + \ffnlo(t) + \ffnnlo(t) + \ffnnnlo(t),
   \label{eqn:chiral_fit:su3:ff}
   \\
   \ffnlo(t)
   & = & 
   \ffnloL(t) + \ffnloB(t),
   \hspace{5mm}
   \ffnnlo(t)
   = 
   \ffnnloL(t) + \ffnnloC(t) + \ffnnloB(t).
   \label{eqn:chiral_fit:su3:ff:contribu}
\eea
Here $\fflo$, $\ffnloB$ and $\ffnnloB$ are LEC-independent 
LO, NLO and NNLO contributions,
whereas $\ffnloL$, $\ffnnloL$ and $\ffnnloC$ depend on the LECs.
Because $m_s\!\gg\!m_l$, the chiral expansion in SU(3) ChPT 
may have poorer convergence than in SU(2) ChPT. 
Hence we include a possible higher order correction $\ffnnnlo$, 
the explicit form of which is not known in ChPT. 
The vector current conservation states that
the LO contribution for the charged mesons is
\bea
   \pfflo = \kpfflo = 1.
\eea
The NLO analytic term
\bea
   \pffnloL(t) = \kpffnloL(t) = \frac{2}{F_\pi^2} L_9^r t
   \label{eqn:chiral_fit:su3:pff:nlo_l}
\eea
arises from the diagram Fig.~\ref{fig:chiral_fit:su2:diagram}\,-\,b
with a vertex from $\mathcal{L}_4$,
which involves the $O(p^4)$ coupling $L_9^r$. 
In contrast, these contributions vanish,
\bea
   \knfflo = \knffnloL(t) = 0,
\eea
for the neutral kaon EM form factor, 
which is the difference of the light and strange quark currents
as written in Eq.~(\ref{eqn:ff:k0ff:current_contribu}).


The term $\ffnloB$ represents 
the LEC-independent NLO contribution coming from 
one-loop diagrams, such as Fig.~\ref{fig:chiral_fit:su2:diagram}\,-\,c,
and is given by
\bea
   F_\pi^2\, \pffnloB(t)
   & = &
   \Ap + \frac{1}{2} \Ak - 2 \Bttppt - \Bttkkt,
   \label{eqn:chiral_fit:su3:pff:nlo_b}
   \\
   F_\pi^2\, \kpffnloB(t)
   & = &
   \frac{1}{2} \Ap + \Ak - \Bttppt - 2 \Bttkkt,
   \\
   F_\pi^2\, \knffnloB(t)
   & = &
   - \frac{1}{2} \Ap + \frac{1}{2} \Ak
                     + \Bttppt - \Bttkkt,
   \label{eqn:chiral_fit:su3:k0ff:nlo_b}
\eea
where $\bar{A}$ ($\bar{B}_{22}$)
represents $t$-independent (dependent) one-loop integral function.
Their definition and expression
are summarized in Appendix~\ref{sec:appndx:1-loop_int}.


The LEC-independent NNLO term $\ffnnloB$ involves two-loop integrals, 
and hence its expression is rather complicated. 
See Appendix~\ref{sec:appndx:2-loop_int} for more details.
We note, however, that this term in the $\xi$-expansion
does not contain any free parameter,
and is not an obstacle to obtaining a stable chiral extrapolation.


The $L_i^r$-dependent NNLO term $\ffnnloL$ mainly comes from 
one-loop diagrams with one vertex from $\mathcal{L}_4$,
such as Fig.~\ref{fig:chiral_fit:su2:diagram}\,-\,e  . 
This term can be expressed with $L_i^r$ and the one-loop integral functions as 
\bea
   F_\pi^4\, 
   \pffnnloL(t)
   & = & 
   8 M_\pi^2\, (2 L_4^r + L_5^r ) \Ap
  +4 M_\pi^2 L_5^r \Ak
  +t L_9^r \, \left\{6 \Ap + 3 \Ak\right\}
   \nn \\
   & & 
  +\left\{ 
      -16 (2 L_4^r + L_5^r) M_\pi^2 + 4 (4L_1^r - 2L_2^r + 2L_3^r - L_9^r ) t
   \right\}
    \Bttppt 
   \nn \\
   & &
  +( -8 L_5^r M_\pi^2 + 4 L_3^r t - 2 L_9^r t )
    \Bttkkt, 
  \\
   F_\pi^4\, 
   \kpffnnloL(t)
   & = &
   (16 L_4^r M_K^2 + 8 L_5^r M_\pi^2 ) \Ak + 4 L_5^r M_\pi^2 \Ap
  +L_9^r t \, \left\{ 5 \Ap + 4 \Ak \right\}
   \nn \\
   & &
  +\left\{
      -32 L_4^r M_K^2 - 16 L_5^r M_\pi^2 + 4(4L_1^r - 2L_2^r + 2L_3^r - L_9^r)t
   \right\}
   \Bttkkt
   \nn \\
   & &
  +(-8 L_5^r M_\pi^2 + 4 L_3^r t - 2 L_9^r t) \Bttppt
  +16 L_5^r L_9^r (M_\pi^2 - M_K^2) t,
  \\
   F_\pi^4\, \knffnnloL(t)
   & = &
   (4 L_5^r M_\pi^2 + L_9^r t) \left\{ - \Ap + \Ak \right\}
   \nn \\
   & &
  +\left\{  8 L_5^r M_\pi^2 - 2(2L_3^r - L_9^r)t \right\} 
   \left\{ \Bttppt - \Bttkkt \right\}.
   \label{eqn:chiral_fit:su3:k0ff:nnlo_l}
\eea


\begin{ruledtabular}
\begin{table}[t]
\begin{center}
\caption{
   Input values for $O(p^4)$ couplings in SU(3) ChPT 
   taken from Ref.~\cite{ChPT:LECs:SU2+SU3}. 
   The central value and first error are 
   from the authors' preferred fit ``BE14'', 
   whereas we assign the difference from the other fit (see text)
   as the second error.
}
\label{tbl:chiral_fit:su3:input}
\begin{tabular}{lllll}
   $L_1^r\!\times\!10^3$   & $L_2^r\!\times\!10^3$   & 
   $L_3^r\!\times\!10^3$   & $L_4^r\!\times\!10^3$   & 
   $L_5^r\!\times\!10^3$   
   \\ \hline
   0.53(6)(+11)  & 0.81(4)(-22)  & -3.07(20)(+27)  & 0.3(0)(+0.46)  &  
   1.01(6)(-51)
\end{tabular}
\end{center}
\vspace{0mm}
\end{table}
\end{ruledtabular}

Together with Eq.~(\ref{eqn:chiral_fit:su3:pff:nlo_l}),
we have the single $O(p^4)$ coupling $L_9^r$ at NLO,
and additional five $L_{\{1\,-\,5\}}$ at NNLO.
Similar to our analysis in SU(2) ChPT,
we treat $L_9^r$ as a fitting parameter,
and fix others to a phenomenological estimate. 
In Ref.~\cite{ChPT:LECs:SU2+SU3},
the authors present two types of the NNLO ChPT fit 
of experimental data, such as the meson masses and decay constants.
A fit called ``BE14'' fixes $L_4^r$ to a fiducial value $0.3\!\times\!10^{-3}$,
since this is difficult to determine due to the strong (anti-)correlation 
with $F_0$. 
(We note that the renormalization scale is set to $\mu\!=\!M_\rho$
also in this section.)
The other fit without the constraint on $L_4^r$ obtains 
$L_4^r\!=\!0.76(18)\!\times\!10^{-3}$, 
which is slightly higher than that for BE14. 
In our analysis, 
we employ the authors' preferred fit BE14
and consider the difference between BE14 and the free-fit 
as an additional uncertainty of $L_i^r$. 
These input values are summarized in Table~\ref{tbl:chiral_fit:su3:input}.

\begin{figure}[t]
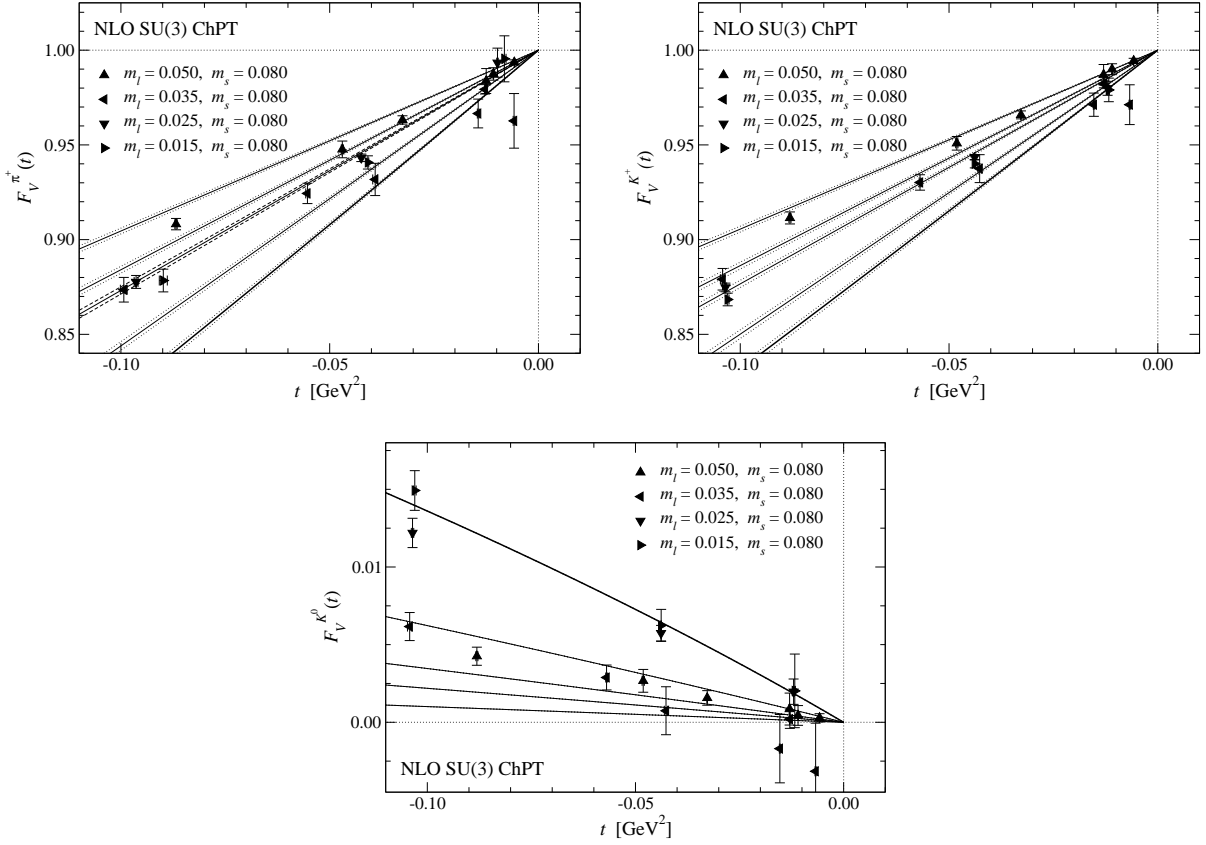

\begin{center}
   \includegraphics[angle=0,width=0.48\linewidth,clip]%
                   {pff_em_vs_q2_ms0080.phys.su3.w_Fpi.nlo.eps}
   \hspace{3mm}
   \includegraphics[angle=0,width=0.48\linewidth,clip]%
                   {k+ff_em_vs_q2_ms0080.phys.su3.w_Fpi.nlo.eps}

   \vspace{5mm}
   \includegraphics[angle=0,width=0.48\linewidth,clip]%
                   {k0ff_em_vs_q2_ms0080.phys.su3.w_Fpi.nlo.eps}
   \vspace{0mm}
   \caption{
      Chiral extrapolations of $\pff$ (top left panel), 
      $\kpff$ (top right panel) and $\knff$ (bottom panel) 
      based on NLO SU(3) ChPT.
      Triangles and thin lines show 
      our data and their fit curves at $m_s\!=\!0.080$. 
      We also plot the fit curve at the physical point 
      $(m_{l,\rm phys},m_{s,\rm phys})$ by the thick lines.
      Note that there is no tunable parameter for $\knff$ at NLO.
   }
   \label{fig:chiral_fit:su3:pff_vs_q2:nlo}
\end{center}
\vspace{0mm}
\end{figure}


The most important issue to obtain a stable chiral extrapolation 
is how to deal with $O(p^6)$ couplings $C_i^r$~\cite{ChPT:SU3:NNLO}
in the NNLO analytic term $\ffnnloC$,
since these couplings are in general poorly known in phenomenology.
The three NNLO analytic terms have six independent parameter dependences 
\bea
   F_\pi^4\,
   \pffnnloC(t)
   & = & 
   -4 \Cppt\, M_\pi^2\, t  - 8 \Cpkt\, M_K^2\, t  - 4\Ctt\, t^2,
   \label{eqn:chiral_fit:su3:pff:nnlo-c}
   \\
   F_\pi^4\,
   \kpffnnloC(t)
   & = &
   -4 \Ckpt\, M_\pi^2\, t  - 4 \Ckkt\, M_K^2\, t  - 4\Ctt\, t^2,
   \label{eqn:chiral_fit:su3:kpff:nnlo-c}
   \\
   F_\pi^4\,
   \knffnnloC(t)
   & = &
  -\frac{8}{3}\, \Ckn\, (M_K^2 - M_\pi^2)\, t,
   \label{eqn:chiral_fit:su3:knff:nnlo-c}
\eea 
and seven $C_i^r$'s enter into these six coefficients through 
the $\mathcal{L}_6$ vertex in Fig.~\ref{fig:chiral_fit:su2:diagram}\,-\,d
\bea
   \Cppt 
   & = & 
   4 C_{12}^r + 4 C_{13}^r + 2 C_{63}^r + C_{64}^r + C_{65}^r + 2 C_{90}^r, 
   \\
   \Cpkt
   & = & 
   4 C_{13}^r + C_{64}^r,
   \\
   \Ctt
   & = & 
   C_{88}^r - C_{90}^r, 
\eea
\bea
   \Ckpt 
   & = &
   4 C_{13}^r + \frac{2}{3} C_{63}^r + C_{64}^r - \frac{1}{3} C_{65}^r, 
   \\
   \Ckkt
   & = &     
   4 C_{12}^r + 8 C_{13}^r + \frac{4}{3} C_{63}^r + 2 C_{64}^r 
                 + \frac{4}{3} C_{65}^r + 2 C_{90}^r,
   \\
   \Ckn
   & = &
   2C_{63}^r - C_{65}^r.
\eea 
Hence our chiral fit can not determine all these $O(p^6)$ couplings separately,
but the six coefficients.
We note that these are not totally independent:
\bea
  \Ckpt 
  & = & 
  \Cpkt + \frac{1}{3} \Ckn,
  \label{eqn:chiral_fit:su3:ckpt}
  \\
  \Ckkt 
  & = &
  \Cppt + \Cpkt - \frac{1}{3} \Ckn.
  \label{eqn:chiral_fit:su3:ckkt}
\eea
We carry out simultaneous fit to $\pff$, $\kpff$ and $\knff$,
in which four coefficients $\Cppt$, $\Cpkt$, $\Ctt$ and $\Ckn$ 
are treated as fitting parameters.


Our chiral fit based on NLO SU(3) ChPT is plotted 
in Fig.~\ref{fig:chiral_fit:su3:pff_vs_q2:nlo}.
Similar to the analysis in SU(2) ChPT, 
the NLO formula is poorly fitted to our data 
resulting in a rather large value of $\chi^2/{\rm d.o.f} \sim 8.3$.
Note that SU(3) chiral symmetry constrains
the dependence of the form factors on $m_l$, $m_s$ and $t$, 
and allows only single tunable parameter at NLO;
namely $L_9^r$ to describe $t$ dependence of $\pff$ and $\kpff$.
Consequently, the NLO formula fails to reproduce 
the $m_l$ dependence particularly of $\knff$. 


\begin{figure}[t]
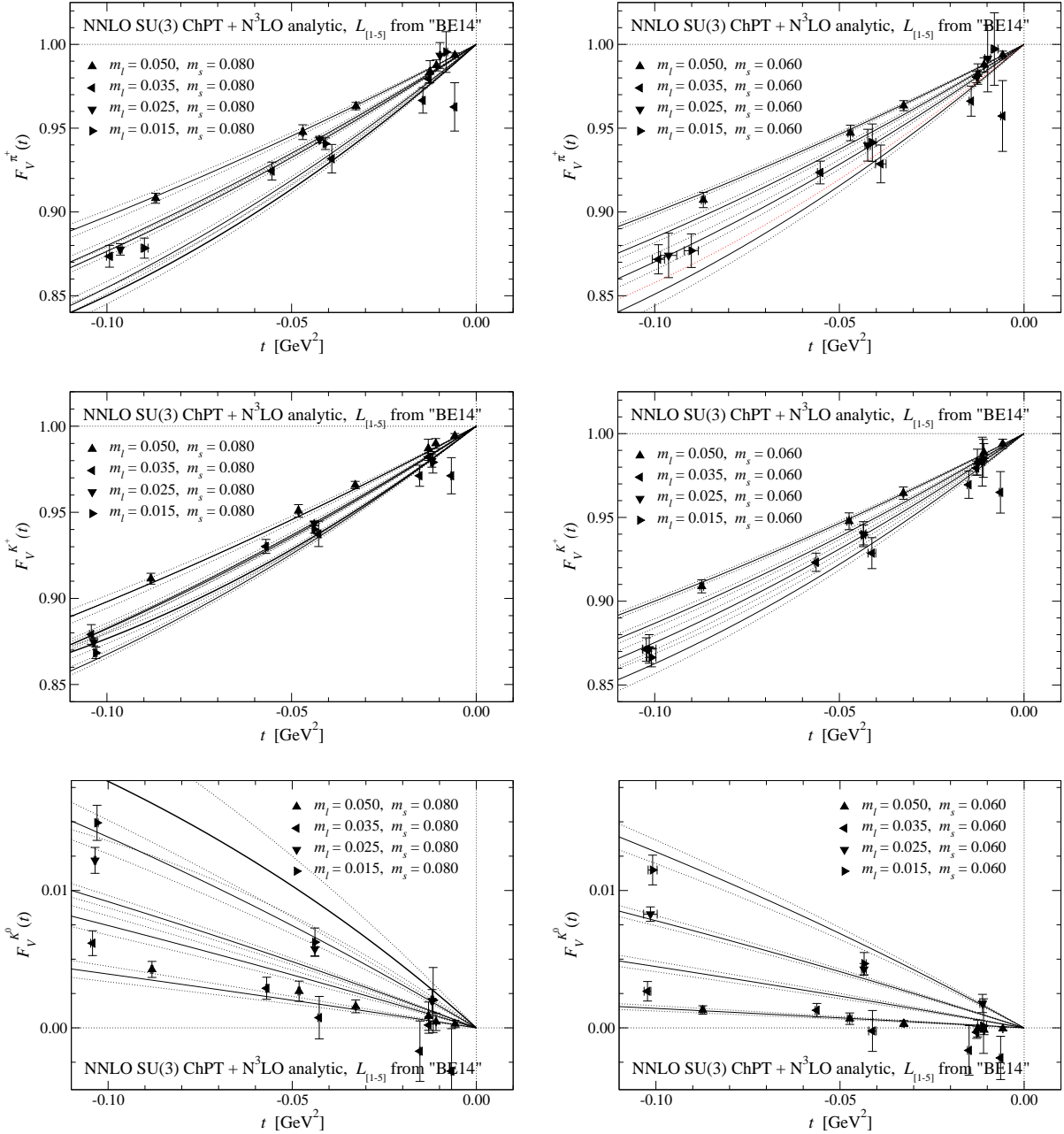

\begin{center}
   \includegraphics[angle=0,width=0.48\linewidth,clip]%
                   {pff_em_vs_q2_ms0080.phys.su3.w_Fpi.nnlo+n3lo-analy.be14.eps}
   \hspace{3mm}
   \includegraphics[angle=0,width=0.48\linewidth,clip]%
                   {pff_em_vs_q2_ms0060.phys.su3.w_Fpi.nnlo+n3lo-analy.be14.eps}

   \vspace{5mm}
   \includegraphics[angle=0,width=0.48\linewidth,clip]%
                   {k+ff_em_vs_q2_ms0080.phys.su3.w_Fpi.nnlo+n3lo-analy.be14.eps}
   \hspace{3mm}
   \includegraphics[angle=0,width=0.48\linewidth,clip]%
                   {k+ff_em_vs_q2_ms0060.phys.su3.w_Fpi.nnlo+n3lo-analy.be14.eps}

   \vspace{5mm}
   \includegraphics[angle=0,width=0.48\linewidth,clip]%
                   {k0ff_em_vs_q2_ms0080.phys.su3.w_Fpi.nnlo+n3lo-analy.be14.eps}
   \hspace{3mm}
   \includegraphics[angle=0,width=0.48\linewidth,clip]%
                   {k0ff_em_vs_q2_ms0060.phys.su3.w_Fpi.nnlo+n3lo-analy.be14.eps}

   \vspace{0mm}
   \caption{
      Chiral extrapolations of $\pff$ (top panels), 
      $\kpff$ (middle panels) and $\knff$ (bottom panels) 
      based on NNLO SU(3) ChPT.
      The left and right panels show our data at $m_s\!=\!0.080$ and 0.060.
      In the left panel for $m_s\!=\!0.080\!\sim\!m_{s,\rm phys}$,
      we also plot the fit curve at the physical point 
      $(m_{l,\rm phys},m_{s,\rm phys})$ by the thick lines.   
   }
   \label{fig:chiral_fit:su3:pff_vs_q2:nnlo}
\end{center}
\vspace{0mm}
\end{figure}

The value of $\chi^2/{\rm d.o.f}$ is largely decreased to 2.3
by taking account of the NNLO contribution.
We observe that
simulation data of $\knff$ tend to deviate from the NNLO fit curve
and give rise to a large part of $\chi^2$.
Since $\knff$ has only single free parameter $\Ckn$ even at NNLO,
we also test a fitting form with an N$^3$LO analytic correction
\bea
   \pffnnnlo  = \kpffnnnlo = 0,
   \hspace{5mm}
   \knffnnnlo = \frac{\Dkn}{F_\pi^6}\, M_\pi^2\, (M_K^2-M_\pi^2)\, t.
   \label{eqn:chiral_fit:su3:n3lo}
\eea
Note that the factor $(M_K^2-M_\pi^2)\,t$ in $\knffnnnlo$ is needed 
to satisfy $\knff(0)\!=\!0$ (vector current conservation) 
and $\knff(t)\!=\!0$ in the SU(3) symmetric limit
(see, Eq.~(\ref{eqn:ff:k0ff:current_contribu})).
This fit is plotted in Fig.~\ref{fig:chiral_fit:su3:pff_vs_q2:nnlo}
and leads to a slightly smaller $\chi^2/{\rm d.o.f.}\!=\!1.8$.
Including more terms at N$^3$LO and even higher orders
reduces $\chi^2$ only slightly,
and the fitting parameters in these corrections are poorly determined. 
We therefore employ the NNLO ChPT fit 
including the N$^3$LO correction~(\ref{eqn:chiral_fit:su3:n3lo}) 
in the following discussion.


Numerical results of the fit 
are summarized in Table~\ref{tbl:chiral_fit:su3:nnlo}.
We estimate the systematic error 
due to the choice of the input $L_{\{1,\cdots,5\}}$ 
by shifting each of $L_{\{1,\cdots,5\}}$ by its uncertainty
quoted in Table~\ref{tbl:chiral_fit:su3:input}.
In our analysis,
the choice of $L_3$ and $L_5$ tends to lead to the largest deviation
in the fitting results. 
This systematic uncertainty from $L_{\{1,\cdots,5\}}$ 
is generally well below the statistical error,
because the $L_i^r$-dependent term $\pkpffnnloL$ is
not a dominant contribution at NNLO (see below).

\begin{ruledtabular}
\begin{table}[t]
\begin{center}
\caption{
   Numerical results of chiral fit based on NNLO SU(3) ChPT.
   LECs are the values at the renormalization scale $\mu\!=\!M_\rho$.
   The first error is statistical, and the second is 
   systematic one due to the choice of the input $L_{\{1,\cdots,5\}}^r$.
   We also quote $\Ckpt$ and $\Ckkt$ calculated using 
   Eqs.~(\ref{eqn:chiral_fit:su3:ckpt})\,--\,(\ref{eqn:chiral_fit:su3:ckkt}).
}
\label{tbl:chiral_fit:su3:nnlo}
\begin{tabular}{lllllll}
   $L_9^r\!\times\!10^3$  & 
   $\Cppt\!\times\!10^5$  & $\Cpkt\!\times\!10^5$  & $\Ctt\!\times\!10^5$  & 
   \\ \hline
   $4.6(1.1)\left(^{+0.1}_{-0.5}\right)$   &
   $-1.95(84)\left(^{+38}_{-21}\right)$     &
   $-1.4(1.2)\left(^{+0.1}_{-0.7}\right)$  &
   $-6.4(1.1)(0.1)$                      &
   \\ \hline   
   $\Ckn\!\times\!10^5$   & $\Dkn\!\times\!10^7$  &
   $\Ckpt\!\times\!10^5$  & $\Ckkt\!\times\!10^5$
   \\ \hline
   $0.15(62)\left(^{+12}_{-7}\right)$      &
   $-37(12)(2)$                          &
   $-1.3(1.2)\left(^{+0.1}_{-0.7}\right)$   &
   $-3.4(1.9)\left(^{+0.1}_{-0.3}\right)$    
   \\ \hline
\end{tabular}
\end{center}
\vspace{0mm}
\end{table}
\end{ruledtabular}


\begin{figure}[t]
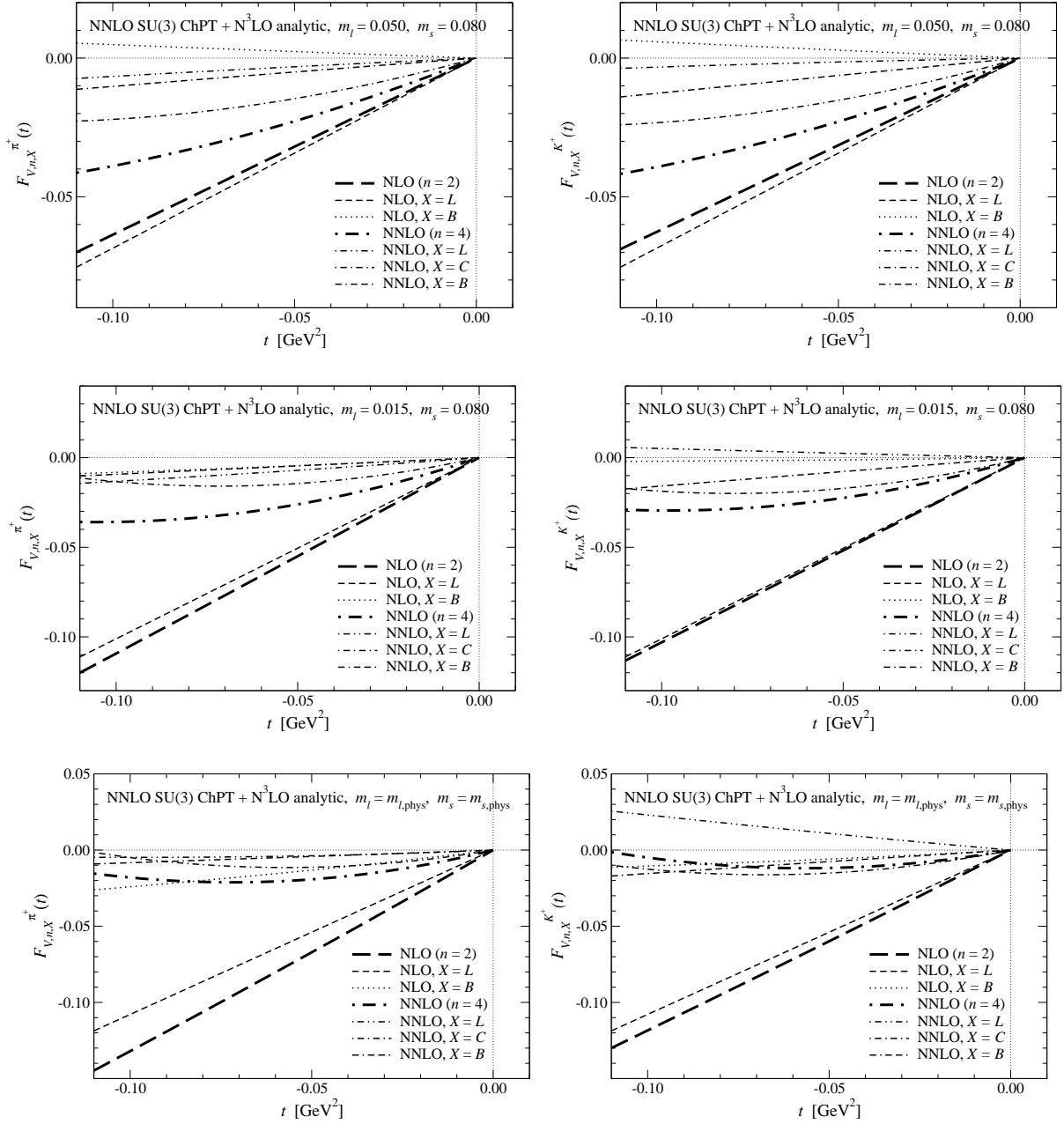

\begin{center}
   \includegraphics[angle=0,width=0.48\linewidth,clip]%
                   {pff_em_vs_q2_mud0050_ms0080.phys.su3.w_Fpi.nnlo+n3lo-analy.be14.cntrb.eps}
   \hspace{3mm}
   \includegraphics[angle=0,width=0.48\linewidth,clip]%
                   {k+ff_em_vs_q2_mud0050_ms0080.phys.su3.w_Fpi.nnlo+n3lo-analy.be14.cntrb.eps}
   \vspace{5mm}

   \includegraphics[angle=0,width=0.48\linewidth,clip]%
                   {pff_em_vs_q2_mud0015_ms0080.phys.su3.w_Fpi.nnlo+n3lo-analy.be14.cntrb.eps}
   \hspace{3mm}
   \includegraphics[angle=0,width=0.48\linewidth,clip]%
                   {k+ff_em_vs_q2_mud0015_ms0080.phys.su3.w_Fpi.nnlo+n3lo-analy.be14.cntrb.eps}

   \vspace{5mm}
   \includegraphics[angle=0,width=0.48\linewidth,clip]%
                   {pff_em_vs_q2_mudphys_msphys.phys.su3.w_Fpi.nnlo+n3lo-analy.be14.cntrb.eps}
   \vspace{3mm}
   \includegraphics[angle=0,width=0.48\linewidth,clip]%
                   {k+ff_em_vs_q2_mudphys_msphys.phys.su3.w_Fpi.nnlo+n3lo-analy.be14.cntrb.eps}

   \vspace{-3mm}
   \caption{
      LEC-(in)dependent NLO and NNLO contributions
      in our chiral fit based on NNLO SU(3) ChPT.  
      The left and right panels show data for $\pff$ and $\kpff$,
      whereas 
      top, middle and bottom panels are for 
      $(m_l,m_s)\!=\!(0.050,0.080)$, (0.015,0.080)
      and the physical point $(m_{l,\rm phys},m_{s,\rm phys})$,
      respectively.
   }
   \label{fig:chiral_fit:su3:contribu:pff+kpff}
\end{center}
\end{figure}

\begin{figure}[t]
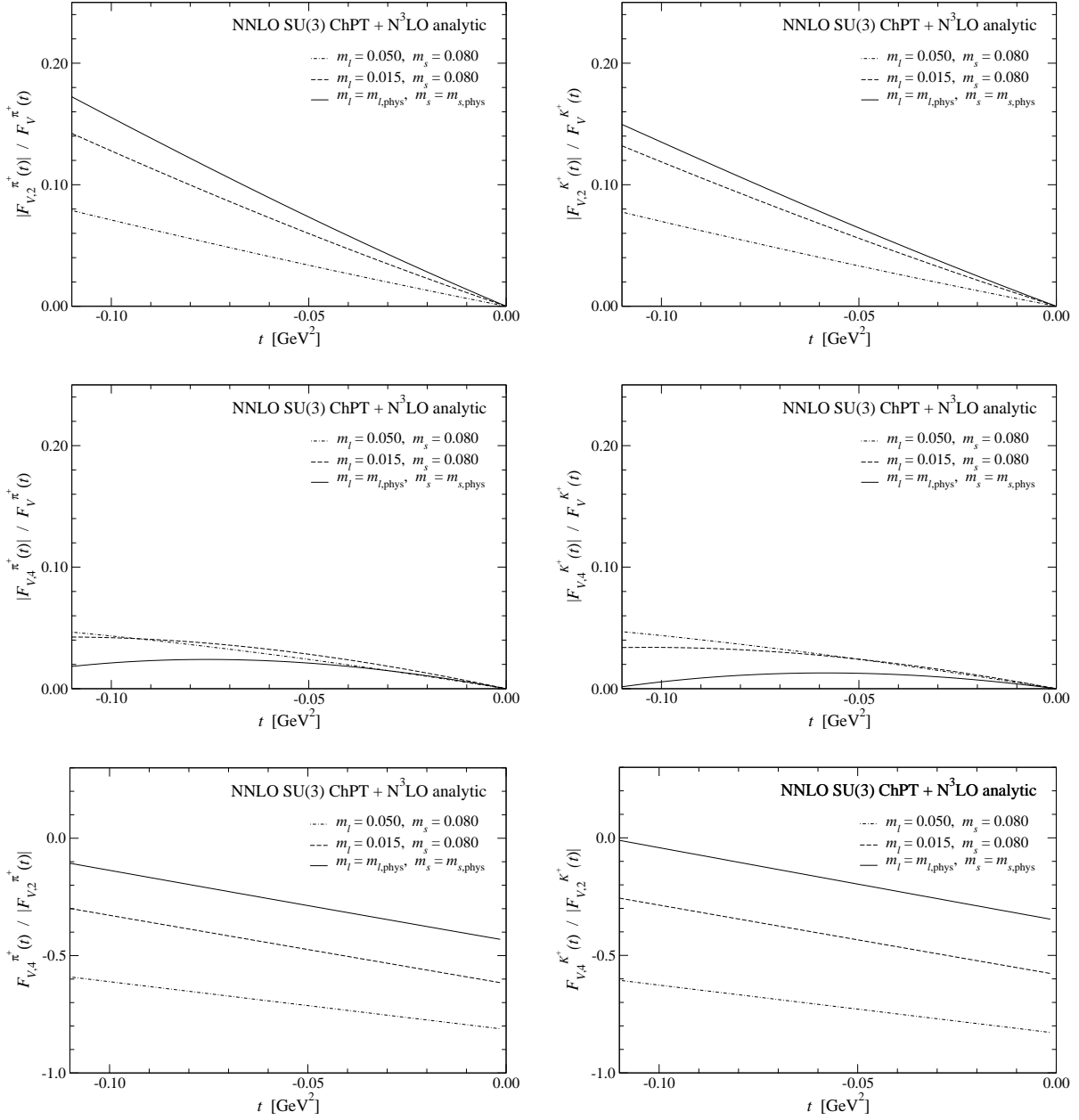

\begin{center}
   \includegraphics[angle=0,width=0.48\linewidth,clip]%
                   {pff_em_vs_q2_ms0080.phys.su3.w_Fpi.nnlo+n3lo-analy.be14.nlo_over_full.eps}
   \hspace{3mm}
   \includegraphics[angle=0,width=0.48\linewidth,clip]%
                   {k+ff_em_vs_q2_ms0080.phys.su3.w_Fpi.nnlo+n3lo-analy.be14.nlo_over_full.eps}
   \vspace{5mm}

   \includegraphics[angle=0,width=0.48\linewidth,clip]%
                   {pff_em_vs_q2_ms0080.phys.su3.w_Fpi.nnlo+n3lo-analy.be14.nnlo_over_full.eps}
   \hspace{3mm}
   \includegraphics[angle=0,width=0.48\linewidth,clip]%
                   {k+ff_em_vs_q2_ms0080.phys.su3.w_Fpi.nnlo+n3lo-analy.be14.nnlo_over_full.eps}
   \vspace{5mm}

   \includegraphics[angle=0,width=0.48\linewidth,clip]%
                   {pff_em_vs_q2_ms0080.phys.su3.w_Fpi.nnlo+n3lo-analy.be14.nnlo_over_nlo.eps}
   \hspace{3mm}
   \includegraphics[angle=0,width=0.48\linewidth,clip]%
                   {k+ff_em_vs_q2_ms0080.phys.su3.w_Fpi.nnlo+n3lo-analy.be14.nnlo_over_nlo.eps}
   \vspace{-3mm}
   \caption{
      Convergence of chiral expansion of $\pff$ (left panels)
      and $\kpff$ (right panels) near $m_{s, \rm phys}$.
      Top panels: ratio of the NLO contribution to the total 
                  $|\pkpffnlo|/\pkpff$.
                  The dot-dashed (dashed) line shows
                  data at $m_l\!=\!0.050$ (0.015) and $m_s\!=\!0.080$,
                  whereas the solid line is at $(m_{l,\rm phys},m_{s,\rm phys})$.
      Middle panels: ratio of the NNLO contribution to the total 
                     $|\pkpffnnlo|/\pkpff$. 
      Bottom panels: ratio of the NNLO and NLO contributions
                     $\pkpffnnlo/|\pkpffnlo|$.
   }
   \label{fig:chiral_fit:su3:convergence:pff+kpff}
\end{center}
\end{figure}

In Figures~\ref{fig:chiral_fit:su3:contribu:pff+kpff} 
and \ref{fig:chiral_fit:su3:convergence:pff+kpff}, 
we examine the convergence of the chiral expansion of $\pff$,
which is now explicitly depends on $m_s$ in {\it SU(3)} ChPT. 
Figure~\ref{fig:chiral_fit:su3:contribu:pff+kpff} shows 
a decomposition to the LEC-dependent and independent terms 
in Eqs.~(\ref{eqn:chiral_fit:su3:ff})\,--\,(\ref{eqn:chiral_fit:su3:ff:contribu}).
Similar to our SU(2) ChPT fit,
the NLO contribution $\pffnlo$ is largely dominated
by the analytic term $\pffnloL$ with $L_9^r$.
The loop term $\pffnloB$ is a small correction compared to $\pffnloL$, 
but increases towards the physical point
possibly due to the enhancement of the chiral logarithms 
$\propto\!\ln[M_\pi^2/\mu]$.

This can also be seen in Fig.~\ref{fig:chiral_fit:su3:convergence:pff+kpff},
where we plot ratios 
$|\pffnlo|/\pff$ (NLO/total),
$|\pffnnlo|/\pff$ (NNLO/total),
and 
$\pffnnlo/|\pffnlo|$ (NNLO/NLO).
We observe larger $|\pffnlo|/\pff$ at smaller $m_l$ 
not only due to the enhancement of $\pffnloB$
but also because $\pffnloL$ is enhanced by $F_\pi^{-2}$ in the $\xi$-expansion.
It turns out that, however, 
$\pffnlo$ is reasonably small correction
at most $\sim\!15$\,\%
at $m_l\!=\!m_{l,\rm phys}$ and $t\!\sim\!-(300~\mbox{MeV})^2$.
It decreases towards smaller $t$
because of the  vector current conservation $\pff(0)\!=\!\pfflo\!=1$.


We observe in Fig.~\ref{fig:chiral_fit:su3:convergence:pff+kpff} that
the NNLO contribution is even smaller in the whole region of 
$M_\pi^2$, $M_K^2$ and $t$. 
Figure~\ref{fig:chiral_fit:su3:contribu:pff+kpff} shows that
the analytic term $\pffnnloC$ is the largest NNLO contribution
at the largest $m_l$.
The first two terms in Eqs.~(\ref{eqn:chiral_fit:su3:pff:nnlo-c})\,--\,(\ref{eqn:chiral_fit:su3:kpff:nnlo-c}) largely contribute to $\pkpffnnloC$,
because we simulate $|t|\!\lesssim\!M_\pi^2, M_K^2$,
and the coefficients $\Cppt$, $\Cpkt$ and $\Ctt$ are of the same order.
Towards the chiral limit,
these terms are suppressed by the NG boson masses, $M_\pi^2$ and $M_K^2$,
and hence $\pffnnlo$ decreases,
whereas $\pffnlo$ increases in this limit.
This is why 
the magnitude of $\pffnnlo/|\pffnlo|$ rapidly decreases at smaller $m_l$
as shown in the bottom panels of 
Fig.~\ref{fig:chiral_fit:su3:convergence:pff+kpff}.
Namely,
the convergence between NNLO and NLO is largely improved 
towards the chiral limit.

While $\pffnnlo/|\pffnlo|$ $\gtrsim\!0.5$ at the largest $m_l$, 
we do not expect large N$^3$LO nor even higher order corrections.
We note that, around our largest $|t|\!\sim\!(300~\mbox{MeV})^2$,
the NNLO correction $\pffnnlo$ is statistically insignificant: 
namely, it has $\gtrsim\!50$\,\% statistical error.
Towards $t\!=\!0$, 
the error decreases
but its central value also decreases due to the vector current conservation:
at $|t|\!\lesssim\!(150~\mbox{MeV})^2$, for instance,
$\pffnnlo$ is sub-\% correction with the statistical accuracy of 
$\gtrsim\! 30$\,\%.
We therefore expect that 
even smaller N$^3$LO correction is insignificant within our accuracy,
and conclude that 
our data of $\pff$ are reasonably well described by NNLO SU(3) ChPT. 

A comparison with Figs.~\ref{fig:chiral_fit:su2:convergence:pff}
and \ref{fig:chiral_fit:su2:contribu:pff} suggests that 
the convergence of the chiral expansion of $\pff$ is not quite different
between SU(2) and SU(3) ChPT.
%


The right panels of Figs.~\ref{fig:chiral_fit:su3:contribu:pff+kpff} 
and \ref{fig:chiral_fit:su3:convergence:pff+kpff} suggest
similar convergence properties for $\kpff$,
which involves the strange quarks as the valence degree of freedom
in contrast to $\pff$.
This is mainly because 
the NLO contribution $\kpffnlo$ is dominated by the analytic term $\kpffnloL$,
which mildly depends on $m_l$ and $m_s$ only through the factor $F_\pi^{-2}$. 
At NNLO, in addition, 
a large part of $\kpffnnlo$ is composed of the analytic term $\kpffnnloC$, 
and the coefficients in Eqs.~(\ref{eqn:chiral_fit:su3:pff:nnlo-c})\,--\,(\ref{eqn:chiral_fit:su3:kpff:nnlo-c}) for $\pff$ and $\kpff$ 
are of the same magnitude: 
namely $\Cppt\!\approx\!\Ckpt$ and $\Cpkt\!\approx\!\Ckkt$.


Interestingly, 
we observe that 
the charged meson vector form factors, $\pff$ and $\kpff$,
are dominated by the NLO analytic term. 
A comparison between the analytic and loop terms in ChPT formulae 
leads to a naive order estimate 
$L_i^r\!=\!O((4\pi)^{-2})\!=\!O(6\!\times\!10^{-3})$
and $C_i^r\!=\!O((4\pi)^{-4})\!=\!O(4\!\times\!10^{-5})$~\cite{ChPT:LECs:SU2+SU3}.
Our fit results are roughly consistent with this order estimate
suggesting that 
the magnitude of the analytic terms $\pkpffnloL$ and $\pkpffnnloC$ 
is not unexpectedly large,
but loop terms are small. 
We in fact observe a large cancellation 
among the two-loop diagrams with the reducible, sunset and vertex integrals 
(see Appendix~\ref{sec:appndx:2-loop_int}, for their definition)
possibly to satisfy $\pkpffnnloB(0)\!=\!0$ 
required from the vector current conservation.


\begin{figure}[t]
\begin{center}
   \includegraphics[angle=0,width=0.48\linewidth,clip]%
                   {k0ff_em_vs_q2_mud0050_ms0080.phys.su3.w_Fpi.nnlo+n3lo-analy.be14.cntrb.eps}
   \hspace{3mm}
   \includegraphics[angle=0,width=0.48\linewidth,clip]%
                   {k0ff_em_vs_q2_mud0015_ms0080.phys.su3.w_Fpi.nnlo+n3lo-analy.be14.cntrb.eps}

   \vspace{5mm}
   \includegraphics[angle=0,width=0.48\linewidth,clip]%
                   {k0ff_em_vs_q2_mudphys_msphys.phys.su3.w_Fpi.nnlo+n3lo-analy.be14.cntrb.eps}

   \vspace{-3mm}
   \caption{ 
      Same as Fig.~\ref{fig:chiral_fit:su3:contribu:pff+kpff}, 
      but for $\knff$.
   }
   \label{fig:chiral_fit:su3:contribu:k0ff}
\end{center}
\end{figure}

\begin{figure}[t]
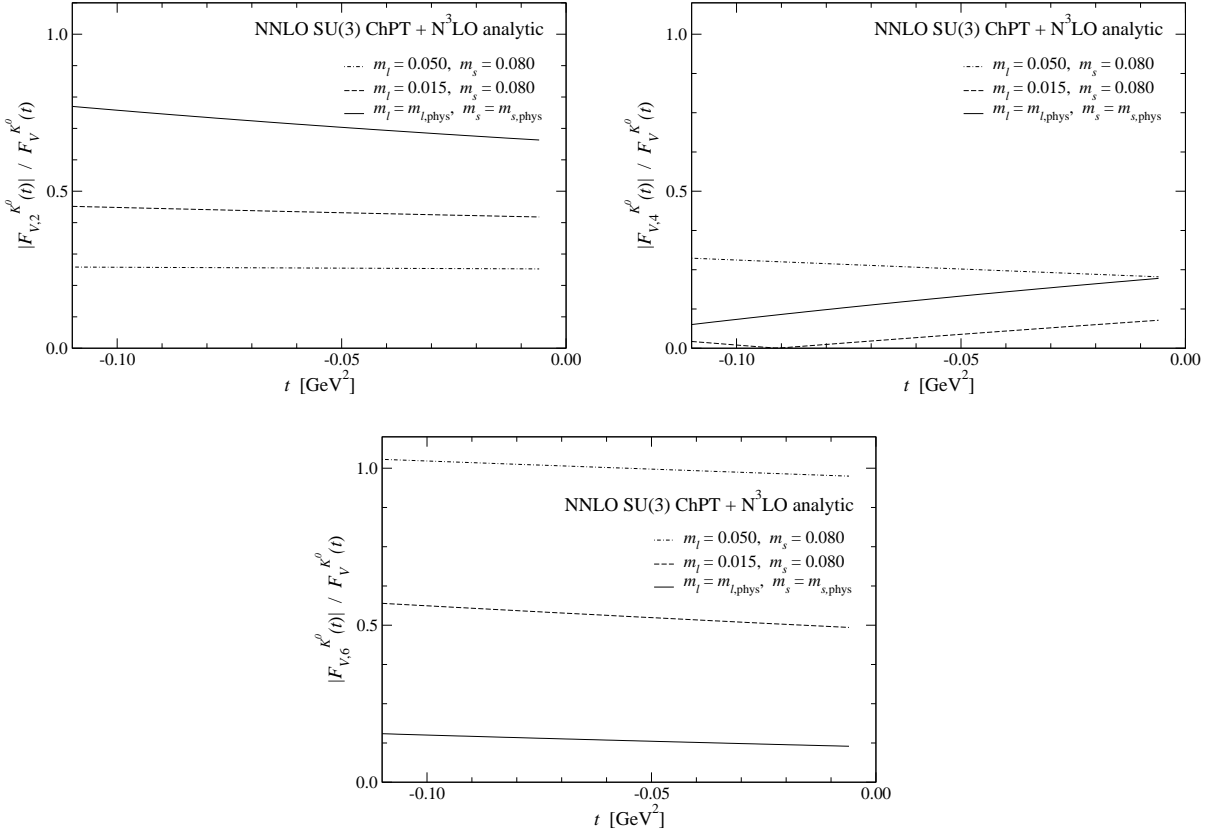

\begin{center}
   \includegraphics[angle=0,width=0.48\linewidth,clip]%
                   {k0ff_em_vs_q2_ms0080.phys.su3.w_Fpi.nnlo+n3lo-analy.be14.nlo_over_full.eps}
   \hspace{3mm}
   \includegraphics[angle=0,width=0.48\linewidth,clip]%
                   {k0ff_em_vs_q2_ms0080.phys.su3.w_Fpi.nnlo+n3lo-analy.be14.nnlo_over_full.eps}

   \vspace{5mm}
   \includegraphics[angle=0,width=0.48\linewidth,clip]%
                   {k0ff_em_vs_q2_ms0080.phys.su3.w_Fpi.nnlo+n3lo-analy.be14.n3lo_over_full.eps}

   \vspace{-3mm}
   \caption{ 
      Convergence of chiral expansion of $\knff$. 
      Top left, top right and bottom panels show 
      $|\knffnlo|/\knff$, $|\knffnnlo|/\knff$ and $|\knffnnnlo|/\knff$,
      respectively. 
      The dot-dashed (dashed) line shows
      data at $m_l\!=\!0.050$ (0.015) and $m_s\!=\!0.080$,
      whereas the solid line is at $(m_{l,\rm phys}, m_{s,\rm phys})$.
      Note that $\knfflo\!=\!0$ and the chiral expansion starts from $\knffnlo$.
   }
   \label{fig:chiral_fit:su3:convergence:k0ff}
\end{center}
\end{figure}

The neutral kaon form factor $\knff$ is the difference 
between the light and strange quark current contributions
as seen in Eq.~(\ref{eqn:ff:k0ff:current_contribu}).
While the LO and NLO analytic terms dominate $\pkpff$, 
these for $\knff$, namely $\knfflo$ and $\knffnloL$,
vanish even at $t\!\ne\!0$.
As a result,
$\knff$ shows much poorer convergence than $\pkpff$
as examined in Figs.~\ref{fig:chiral_fit:su3:contribu:k0ff}
and \ref{fig:chiral_fit:su3:convergence:k0ff}.
There is only the parameter-free 
term $\knffnloB$ within NLO. 
At the largest $m_l$, 
this term is rather small compared to our simulation results,
and hence the large part of $\knff$ is composed of
higher order corrections $\knffnnlo+\knffnnnlo$.
However, $\knffnloB$ increases as we approach to $m_{l,\rm phys}$ 
with $m_s$ held fixed.
This is in accordance with the VMD hypothesis~(\ref{eqn:ff:q2_interp:vmd:knff}):
larger $\knff$ with larger $M_\phi-M_\rho$.
As a result,
the convergence is rapidly improved towards the physical point,
where both NNLO and N$^3$LO corrections become small 
compared to the leading term $\knffnlo$. 

We also note that 
the large N$^3$LO contributions $\knffnnnlo$ 
may be partly attributed to the fact that 
the analytic NNLO and N$^3$LO contributions,
$\knffnnloC$ and $\knffnnnlo$, are difficult to distinguish
with our simulation set up,
and hence $\Ckn$ in Table~\ref{tbl:chiral_fit:su3:nnlo} is poorly determined. 
A better determination of $\Ckn$ and $\Dkn$ may need
simulations with a wider region and better resolution of $M_\pi$. 
We leave this for future work.


We also decompose the charge radii into the LEC-dependent and independent terms as 
\bea
   \crad
   & = &
   \cradnlo + \cradnnlo + \cradnnnlo,
   \label{eqn:chiral_fit:su3:radius}
   \\
   \cradnlo
   & = & 
   \cradnloL + \cradnloB,
   \hspace{5mm}
   \cradnnlo
   = 
   \cradnnloL + \cradnnloC + \cradnnloB
   \label{eqn:chiral_fit:su3:radius:contribu}, 
\eea
where $P\!=\!\pi^+$, $K^+$ or $K^0$.
The NLO terms are given by~\cite{PFF:ChPT:SU3:NLO} 
\bea
   \cradpffnloL
   & = & 
   \cradkpffnloL
   = 
   \frac{12}{F_\pi^2} L_9^r,
   \hspace{3mm}
   \cradknffnloL = 0,
   \label{eqn:chiral_fit:su3:radius:nlo_l}
   \\
   \cradpffnloB
   & = & 
  -\frac{1}{2NF_\pi^2} 
   \left( 
      2 \ln\left[ \frac{M_\pi^2}{\mu^2} \right]
    +   \ln\left[ \frac{M_K^2}{\mu^2} \right]
    + 3
   \right),
   \label{eqn:chiral_fit:su3:radius:pff:nlo_b}
   \\
   \cradkpffnloB
   & = & 
  -\frac{1}{2NF_\pi^2} 
   \left( 
        \ln\left[ \frac{M_\pi^2}{\mu^2} \right]
    + 2 \ln\left[ \frac{M_K^2}{\mu^2} \right]
    + 3
   \right),
   \label{eqn:chiral_fit:su3:radius:kpff:nlo_b}
\eea
\bea
   \cradknffnloB
   & = &
   \cradkpff - \cradpff
   = 
   \frac{1}{2NF_\pi^2} \ln \left[ \frac{M_\pi^2}{M_K^2} \right]. 
   \label{eqn:chiral_fit:su3:radius:knff:nlo_b}
\eea
The higher order analytic terms are obtained straightforwardly 
from Eqs.~(\ref{eqn:chiral_fit:su3:pff:nnlo-c})\,--\,(\ref{eqn:chiral_fit:su3:knff:nnlo-c}) and (\ref{eqn:chiral_fit:su3:n3lo})
through the definition~(\ref{intro:radius})
\bea
   F_\pi^4\, \cradpffnnloC
   & = &
   -24 \left( \Cppt M_\pi^2 + 2 \Cpkt M_K^2 \right),
   \\
   F_\pi^4\, \cradkpffnnloC
   & = &
   -24 \left( \Ckpt M_\pi^2 + \Ckkt M_K^2 \right),
   \\
   F_\pi^4\, \cradknffnnloC
   & = &
   -16 \Ckn \left( M_K^2 - M_\pi^2 \right)
\eea
and 
\bea
   \cradpffnnnlo = \cradkpffnnnlo = 0,
   \hspace{3mm}
   F_\pi^6 \cradknffnnnlo = 6\, \Dkn M_\pi^2 \left( M_K^2 - M_\pi^2 \right).
\eea
The NNLO non-analytic terms $\ffnnloL+\ffnnloB$ have rather 
complicated expression, and are not large as discussed above.
We therefore do not derive an explicit formula for 
the corresponding terms for the radii $\cradnnloL+\cradnnloB$, 
but estimate them by taking numerical derivative 
of $\ffnnloL+\ffnnloB$ with respect to $t$.


\begin{figure}[t]
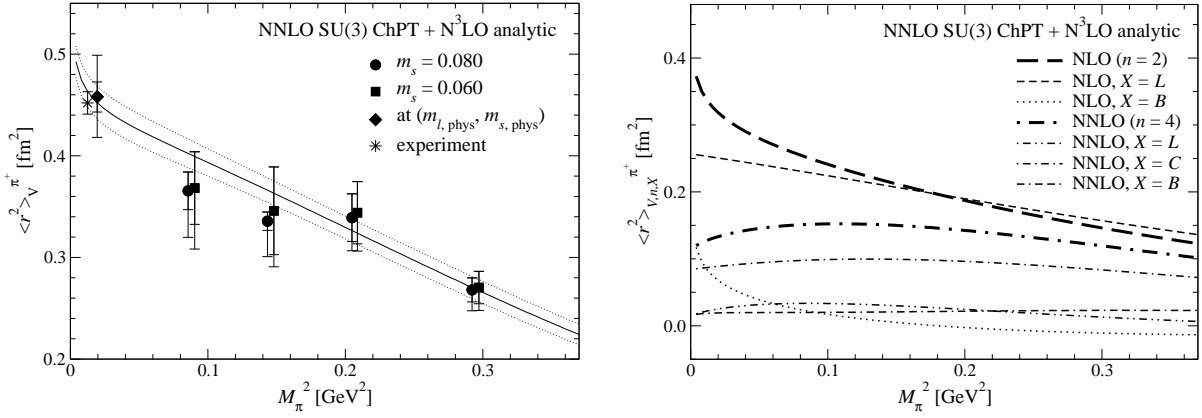

\begin{center}
   \includegraphics[angle=0,width=0.48\linewidth,clip]%
                   {r2_pff_em_vs_Mpi2.phys.su3.w_Fpi.nnlo+n3lo-analy.be14.eps}
   \hspace{3mm}
   \includegraphics[angle=0,width=0.48\linewidth,clip]%
                   {r2_pff_em_vs_Mpi2.ms0080.phys.su3.w_Fpi.nnlo+n3lo-analy.be14.cntrb.eps}

   \vspace{-3mm}
   \caption{
      Left panel:
      pion charge radius $\cradpff$ as a function of $M_\pi^2$. 
      The solid line represents $\cradpff$ at $m_s\!=\!0.080$
      reproduced from our chiral fit based on NNLO SU(3) ChPT.
      We plot the value extrapolated to the physical point by the diamond.
      The circles and squares are our estimate 
      at simulation points listed in Table~\ref{tbl:ff:q2_interp:radii}.
      The experimental value is plotted by the star.
      Right panel: 
      NLO and NNLO LEC-(in)dependent contributions to $\cradpff$.
   }
   \label{fig:chiral_fit:su3:r2_vs_Mpi2:pff}
\end{center}
\end{figure}

\begin{figure}[t]
\begin{center}
   \includegraphics[angle=0,width=0.48\linewidth,clip]%
                   {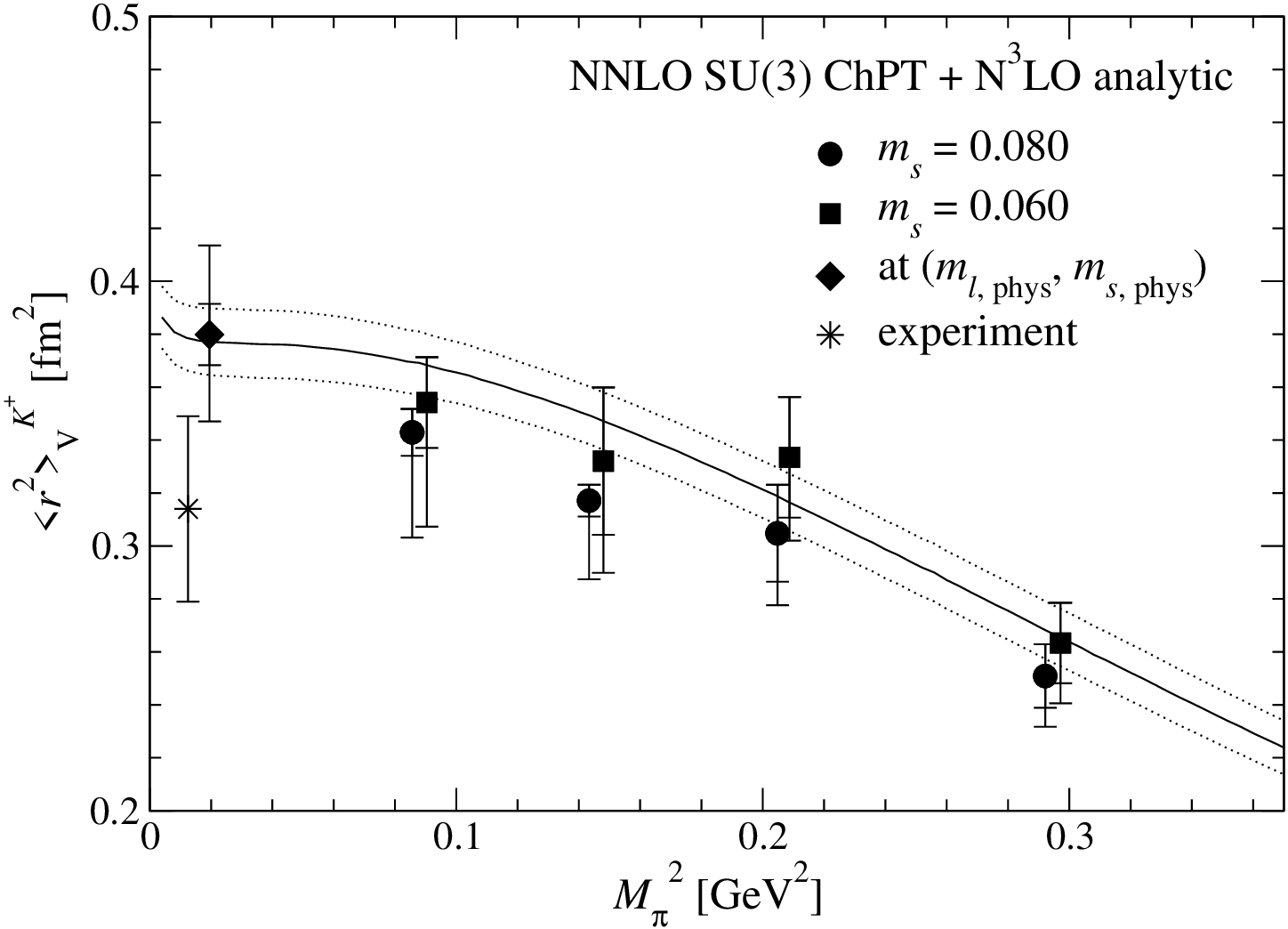}
   \hspace{3mm}
   \includegraphics[angle=0,width=0.48\linewidth,clip]%
                   {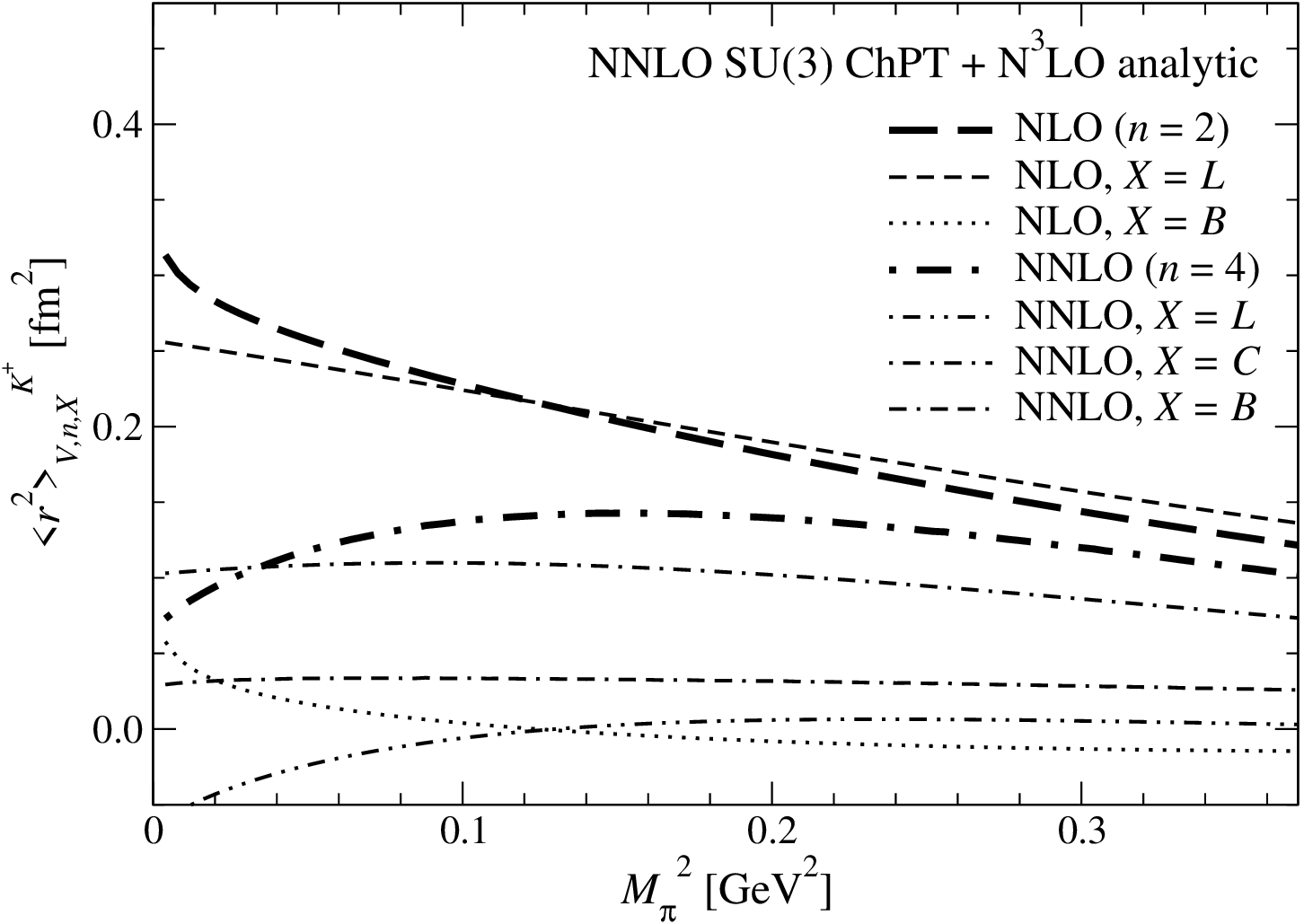}

   \vspace{-3mm}
   \caption{
      Same as Fig.~\ref{fig:chiral_fit:su3:r2_vs_Mpi2:pff}, but for $\kpff$.
   }
   \label{fig:chiral_fit:su3:r2_vs_Mpi2:k+ff}
\end{center}
\end{figure}

The chiral extrapolation of the pion charge radius $\cradpff$ is 
shown in the left panel of Fig.~\ref{fig:chiral_fit:su3:r2_vs_Mpi2:pff}.
In Subsection~\ref{subsec:r2}, 
we estimate $\cradpff$ at the simulation points
by assuming the phenomenological $t$ dependence 
Eq.~(\ref{eqn:ff:q2_interp:pff_vs_q2}).
These values are reproduced by our simultaneous chiral fit of 
$F_V^{\{\pi^+,K^+,K^0\}}$ reasonably well. 
This does not necessarily hold true: 
the non-analytic chiral behavior of $\pff$ may not be well described 
by our simple assumption~(\ref{eqn:ff:q2_interp:pff_vs_q2}),
which is essentially low-order polynomial in $t$
in our region $|t|\!\ll\!M_\rho^2$.
The reasonable consistency is partly because 
$\pff$ is largely dominated by the analytic terms $\pffnloL\!+\!\pffnnloC$.
In fact, the right panel of the same figure shows that 
$\cradpff$ is also dominated by the analytic terms
$\cradpffnloL\!+\!\cradpffnnloC$. 
This supports our strategy of the chiral fit:
namely, we determine $L_9^r$ and $O(p^6)$ couplings 
appearing in these large analytic terms from our simulations, 
whereas other $L_i^r$'s in the small loop corrections are fixed 
to the phenomenological estimate.

More importantly,
the value extrapolated to the physical point is 
in excellent agreement with the experimental value.  
The enhancement of the NLO chiral logarithm is important 
for this agreement.
It is however partly compensated by the decrease of the NNLO contribution,
similar to the analysis in SU(2) ChPT. 
The logarithmic singularity is therefore difficult to directly observe
at our simulation region of $M_\pi\!\gtrsim\!300$~MeV.


Also for the charged kaon radius,
we observe good agreement between simulation results and 
the experimental value $\cradkpff\!=\!0.314(35)~\mbox{fm}^2$~\cite{PDG:2014}
as plotted in the left panel of Fig.~\ref{fig:chiral_fit:su3:r2_vs_Mpi2:k+ff}.
A comparison of the right panels of 
Figs.~\ref{fig:chiral_fit:su3:r2_vs_Mpi2:pff} 
and \ref{fig:chiral_fit:su3:r2_vs_Mpi2:k+ff}
suggests that 
the difference between $\cradkpff$ and $\cradpff$ 
is mainly due to the suppression of the NLO chiral logarithms
in Eqs.~(\ref{eqn:chiral_fit:su3:radius:pff:nlo_b})\,--\,(\ref{eqn:chiral_fit:su3:radius:kpff:nlo_b}),
and because the NNLO term $\kpffnnloL$ becomes negative near the physical point
with our choice of the input $L_{\{1,\cdots,5\}}^r$.


\begin{figure}[t]
\begin{center}
   \includegraphics[angle=0,width=0.48\linewidth,clip]%
                   {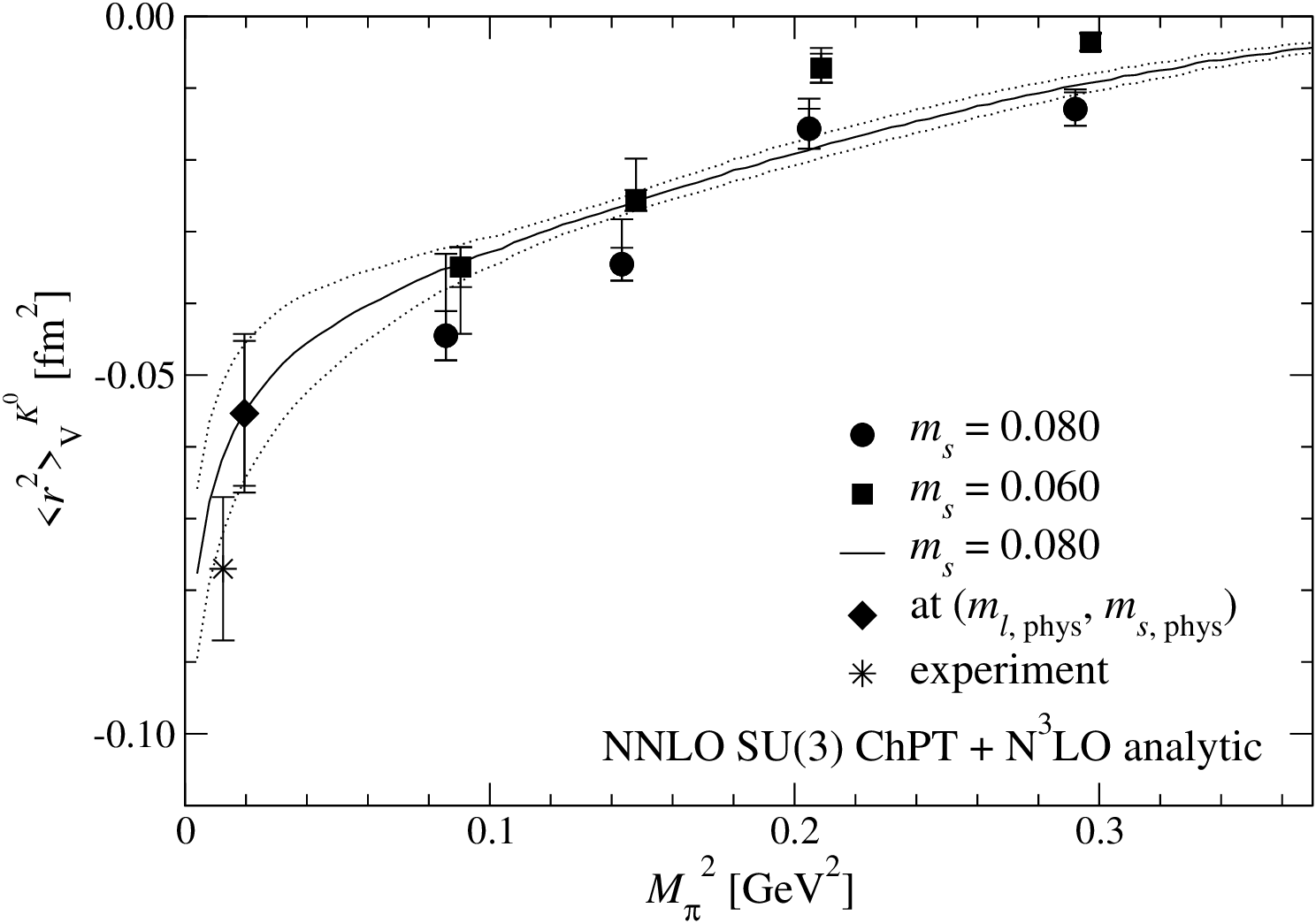}
   \hspace{3mm}
   \includegraphics[angle=0,width=0.48\linewidth,clip]%
                   {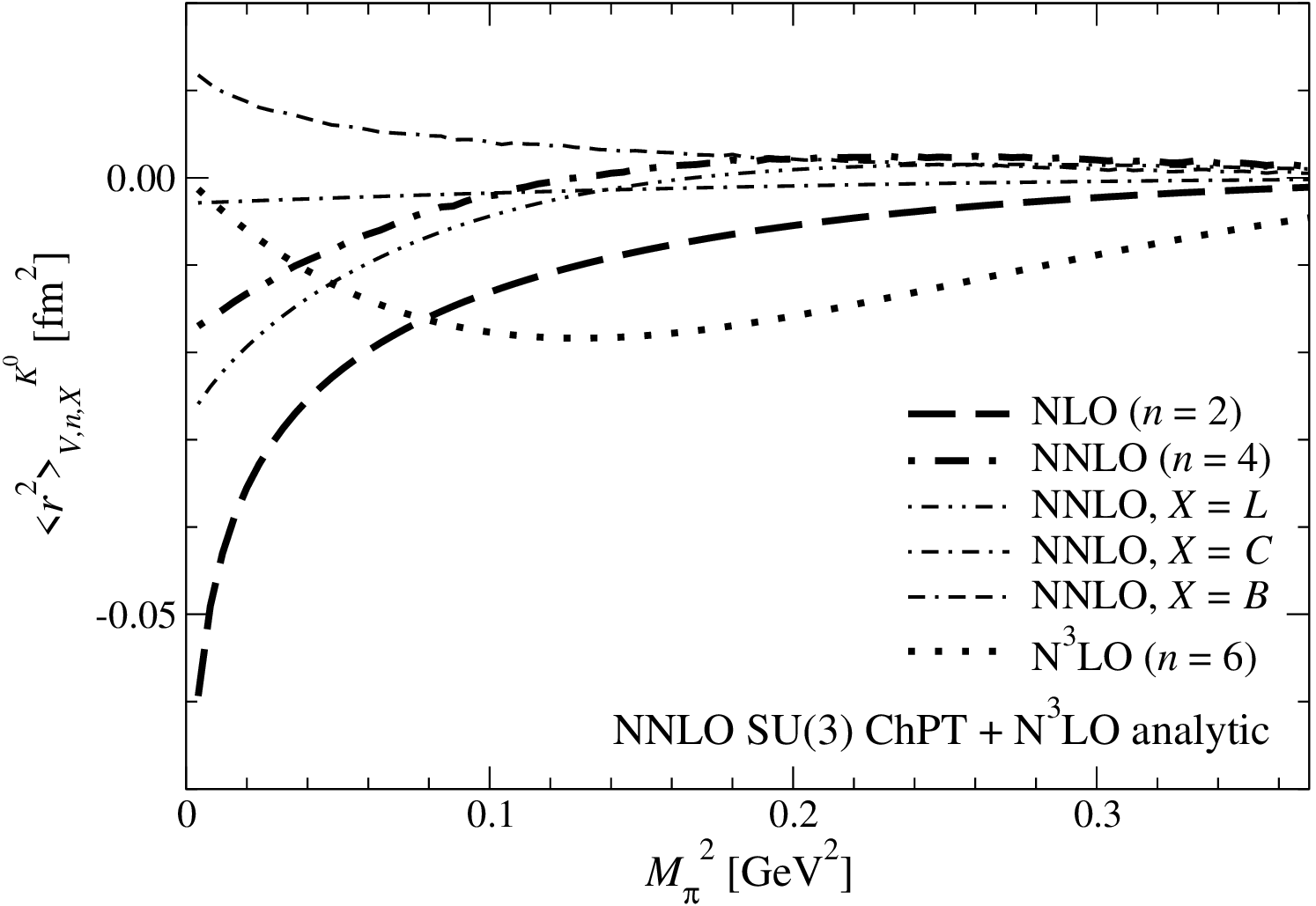}

   \vspace{-3mm}
   \caption{
      Same as Fig.~\ref{fig:chiral_fit:su3:r2_vs_Mpi2:pff}, but for $\knff$.
   }
   \label{fig:chiral_fit:su3:r2_vs_Mpi2:k0ff}
\end{center}
\end{figure}

Our chiral extrapolation also reproduces 
the experimental value of the neutral kaon radius 
$\cradknff\!=\!-0.077(10)~\mbox{fm}^2$
as shown in Fig.~\ref{fig:chiral_fit:su3:r2_vs_Mpi2:k0ff}.
Similar to $\knff$, 
the parameter-free leading term $\cradknffnlo$ becomes 
the largest contribution only at small pion masses $M_\pi\!\lesssim\!300$~MeV.
As already mentioned, 
the pion radius $\cradpff$ is considered as a good quantity 
to observe the one-loop chiral logarithm.
We note that 
$\cradknff$ does not have analytic term at this order ($\cradknffnloL\!=\!0$)
and could be another good candidate 
provided that
one simulates $M_\pi$ below 300~MeV 
with $m_s$ held fixed at a rather heavier value.


Since we simulate at a single lattice spacing,
we assign the discretization error to our numerical results 
by an order counting $O((a\Lambda_{\rm QCD})^2)\!\sim\!8$\,\%.
At the renormalization scale $\mu\!=\!M_\rho$, 
we obtain 
\bea
   L_9^r
   & = & 
   4.6(1.1)\left(^{+0.1}_{-0.5}\right)(0.4) \times 10^{-3},
   \\
   \Ctt
   & = & 
   -6.4(1.1)(0.1)(0.5) \times 10^{-5}.
\eea
These are in good agreement with 
$L_9^r\!=\!5.9(0.4)\!\times\!10^{-3}$ and 
$\Ctt\!=\!C_{88}^r\!-\!C_{90}^r\!=\!-5.5(0.5)\!\times\!10^5$
obtained from a phenomenological analysis of the experimental data of $\pff$
in NNLO SU(3) ChPT~\cite{PFF+KFF:ChPT:NNLO:Nf3}. 
Other $O(p^6)$ couplings 
\bea
   \Cppt
   & = &
   -1.95(84)\left(^{+38}_{-21}\right)(16) \times 10^{-5},
   \\
   \Cpkt
   & = & 
   -1.4(1.2)\left(^{+0.1}_{-0.7}\right)(0.1) \times 10^{-5},
   \\
   \Ckpt
   & = & 
   -1.3(1.2)\left(^{+0.1}_{-0.7}\right)(0.1) \times 10^{-5},
   \\
   \Ckkt
   & = &
   -3.4(1.9)\left(^{+0.1}_{-0.3}\right)(0.3) \times 10^{-5},
   \\
   \Ckn
   & = &
   0.15(62)\left(^{+12}_{-7}\right)(1) \times 10^{-5},
\eea
are poorly known phenomenologically,
and we obtain 
\bea
   \Dkn
   & = &
   -37(12)(2)(3) \times 10^{-7}
\eea
for the coefficient of the higher order correction to $\knff$.
Our numerical results for the light meson charge radii 
\bea
   \cradpff
   & = & 
   0.458(15)\left(^{+9}_{-1}\right)(37)~\mbox{fm}^2,
   \\
   \cradkpff
   & = & 
   0.380(12)\left(^{+7}_{-1}\right)(31)~\mbox{fm}^2,
   \\
   \cradknff
   & = &
   -0.055(10)(1)(4)~\mbox{fm}^2
\eea
are in reasonable agreement with experiment.

%% file: 6.conclusion.tex

\section{Conclusions} 
\label{sec:conclusion}


In this article,
we have presented our detailed study of 
the chiral behavior of the light meson EM form factors.
Chiral symmetry is exactly preserved in our simulations
for a direct comparison with continuum ChPT at NNLO.
Another salient feature is that
we precisely calculate the EM form factors 
by using the all-to-all quark propagator.


Our analyses in SU(2) and SU(3) ChPT suggest
reasonable convergence of the NNLO chiral expansion of 
the charged meson EM form factors $\pkpff$.
This is mainly because 
the non-trivial correction $\pkpff-1$ is largely dominated 
by the NLO analytic term, which mildly depends on the quark masses.
This term however vanishes in the neutral kaon form factor $\knff$.
Although the corresponding chiral expansion shows poorer convergence
at our simulated pion masses $M_\pi\!\gtrsim\!300$~MeV,
it is rapidly improved towards the physical pion mass. 

The NNLO tree diagrams with the $O(p^6)$ couplings also tend to 
compose of a large part of the NNLO contribution.
We observe small but non-negligible loop corrections,
which have non-analytic dependence on the quark masses and momentum transfer.
These confirm
the importance of the first-principle determination of the relevant LECs
based on the NNLO ChPT.


Our results for the LECs
$\bar{l}_6^r$, $L_9^r$ and $\Ctt\!=\!C_{88}^r\!-\!C_{90}^r$
are consistent with the phenomenological estimates.
We also observe a reasonable agreement of 
the charge radii with experiment.


Our results for the phenomenologically poorly-known $O(p^6)$ couplings
would be useful for studying different observables based on ChPT. 
An interesting application is the form factor of 
the $K\!\to\!\pi l \nu$ semileptonic decays,
since its vector form factor $f_+^{K\pi}(t)$
shares many $O(p^6)$ couplings 
with the EM form factors~\cite{KFF:weak:ChPT:SU3:NNLO:BT}.
These decays provide 
a precise determination of the CKM matrix element $|V_{us}|$
through a precision lattice calculation of the normalization $f_+^{K\pi}(0)$.
A comparison of the form factor shape with experiment
can demonstrate the reliability of such a precise calculation.
Our results of the LECs may enable us to study the normalization 
and shape simultaneously based on NNLO SU(3) ChPT.


Our analysis suggests that 
the charge radii show the one-loop chiral logarithm
below $M_\pi\!\approx\!300$~MeV.
Pushing simulations towards such small pion masses is interesting
for unambiguous observation of the logarithmic singularity in QCD.
Extension towards finer lattices is also important,
because the largest uncertainty in our numerical results 
is the discretization error.
Simulations in these directions are underway~\cite{JLQCD:Noaki:Lat14}
by using a computationally cheaper fermion formulation
with good chiral symmetry~\cite{JLQCD:TK:Lat15}.


\begin{acknowledgments}

We thank Johan Bijnens for making his code 
to calculate the EM form factors in NNLO SU(3) ChPT
available to us.
Numerical simulations are performed on Hitachi SR16000 and 
IBM System Blue Gene Solution 
at High Energy Accelerator Research Organization (KEK) 
under a support of its Large Scale Simulation Program (No.~15/16-09),
and on SR16000 at YITP in Kyoto University.
This work is supported in part by the Grant-in-Aid of the
Ministry of Education, Culture, Sports, Science and Technology (MEXT)
(No.~25287046, 26247043, 26400259 and 15K05065)
and by MEXT Strategic Programs for Innovative Research
and Joint Institute for Computational Fundamental Science 
as a priority issue 
(Elucidation of the fundamental laws and evolution of the universe) 
to be tackled by using Post ``K'' Computer.

\end{acknowledgments}

%% file: A.1-loop_integrals.tex
\section{One-loop integrals in SU(3) ChPT}
\label{sec:appndx:1-loop_int}

We summarize the expression of the one-loop integral functions
in SU(3) ChPT in this section,
as well as 
the expressions of the two-loop integrals and 
relevant two-loop contributions to $\pff$ 
in Appendix~\ref{sec:appndx:2-loop_int}.
We refer to the original papers~\cite{PFF+KFF:ChPT:NNLO:Nf3,Spec+DC:ChPT:SU3:NNLO:ABT}
for more detailed discussions.

\begin{figure}[t]
\begin{center}
   \vspace{10mm}
   \includegraphics[angle=0,width=0.15\linewidth,clip]%
                   {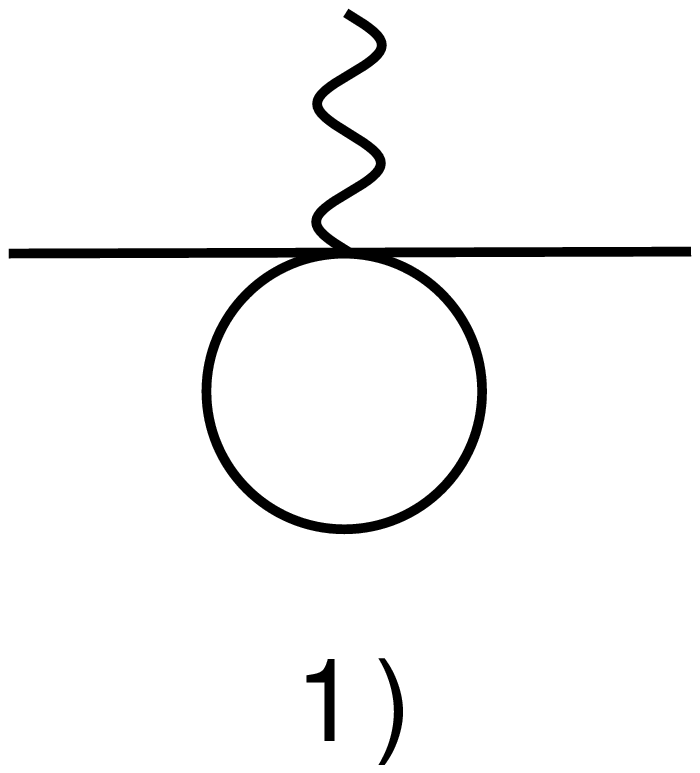}
   \hspace{5mm}
   \includegraphics[angle=0,width=0.15\linewidth,clip]%
                   {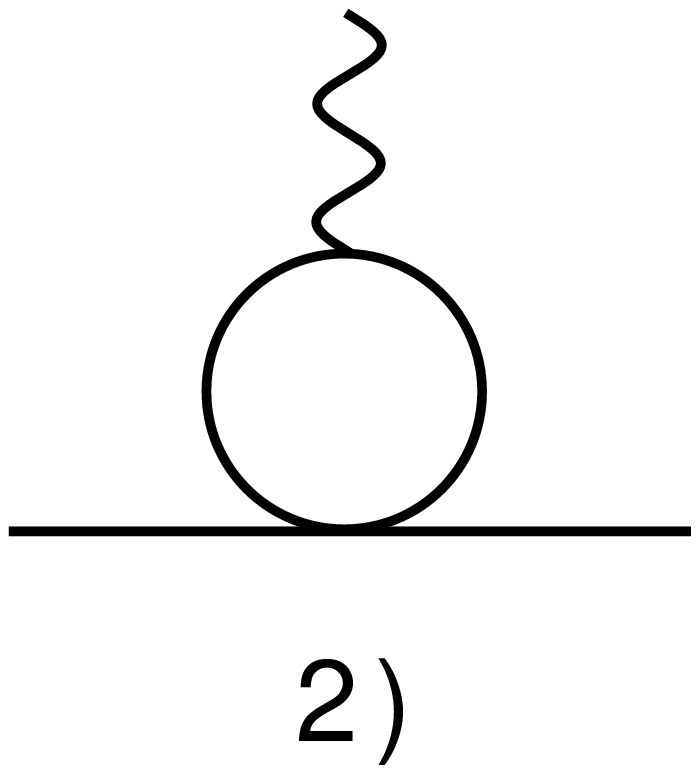}

   \vspace{0mm}
   \caption{
      Example of one-loop diagrams involving 
      momentum-transfer independent (1) and dependent loop integrals (2). 
   }
   \label{fig:appdxA:diagram:nlo} 
\end{center}
\vspace{0mm}
\end{figure}

The one-loop integral functions are defined as 
\bea
   A(M_1^2)
   & = &
   \frac{1}{i} 
   \int \frac{d^dk}{(2\pi)^d} 
      \frac{1}{k^2-M_1^2},
   \\
   B(M_1^2,M_2^2,t) 
   & = & 
   \frac{1}{i} 
   \int \frac{d^dk}{(2\pi)^d}
      \frac{1}{(k^2-M_1^2)\left\{(k-q)^2-M_2^2\right\}},
   \\
   B_\mu(M_1^2,M_2^2,t) 
   & = & 
   \frac{1}{i} 
   \int \frac{d^dk}{(2\pi)^d} 
      \frac{k_\mu}{(k^2-M_1^2)\left\{(k-q)^2-M_2^2\right\}},
   \\
   B_{\mu\nu}(M_1^2,M_2^2,t) 
   & = & 
   \frac{1}{i} 
   \int \frac{d^dk}{(2\pi)^d} 
      \frac{k_\mu k_\nu}{(k^2-M_1^2)\left\{(k-q)^2-M_2^2\right\}},
   \\
   B_{\mu\nu\alpha}(M_1^2,M_2^2,t)
   & = & 
   \frac{1}{i} 
   \int \frac{d^dk}{(2\pi)^d}
      \frac{k_\mu k_\nu k_\alpha}{(k^2-M_1^2)\left\{(k-q)^2-M_2^2\right\}},
\eea 
where $q^2\!=\!t$ and $d\!=\!4-2\epsilon$.
The scalar function $A$ is needed to evaluate diagrams 
such as shown in Fig.~\ref{fig:appdxA:diagram:nlo}\,--\,1,
and hence does not depend on $t$.
The $t$-dependent ``$B$'' functions are needed 
for Fig.~\ref{fig:appdxA:diagram:nlo}\,--\,2.

The Lorentz decomposition of the vector and tensor functions is given as 
\bea
   B_\mu(M_1^2,M_2^2,t)
   & = &
   q_\mu B_1(M_1^2,M_2^2,t),
   \\
   B_{\mu\nu}(M_1^2,M_2^2,t)
   & = &
   q_\mu q_\nu B_{21}(M_1^2,M_2^2,t) + g_{\mu\nu} B_{22}(M_1^2,M_2^2,t),
   \\
   B_{\mu\nu\alpha}(M_1^2,M_2^2,t)
   & = &
   q_\mu q_\nu q_\alpha B_{31}(M_1^2,M_2^2,t) 
   \nn \\
   & & 
   \hspace{5mm}
 + \left( q_\mu g_{\nu\alpha} + q_\nu g_{\alpha\mu} + q_\alpha g_{\mu\nu} \right)
   B_{32}(M_1^2,M_2^2,t).
\eea
The ``$B$'' functions in the right hand side are expressed 
in terms of the scalar functions $A$ and $B$
\bea
   B_1(M_1^2,M_2^2,t) 
   & = & 
   \frac{1}{2t}
   \left\{ 
      - A(M_1^2) + A(M_2^2) + (\Delta_{12}+t) B(M_1^2,M_2^2,t)
   \right\},
   \\
   B_{21}(M_1^2,M_2^2,t)   & = &
   \frac{1}{t}
   \left\{
      A(M_2^2) + M_1^2 B(M_1^2,M_2^2,t) - d B_{22}(M_1^2,M_2^2,t) 
   \right\},
   \\
   B_{22}(M_1^2,M_2^2,t)
   & = &
   \frac{1}{2(d-1)}
   \left\{
      A(M_2^2) + 2 M_1^2 B(M_1^2,M_2^2,t) 
   \right.
   \nn \\
   & &
   \hspace{60mm}
   \left.
    - (\Delta_{12} + t) B_1(M_1^2,M_2^2,t)
   \right\},
   \\
   B_{31}(M_1^2,M_2^2,t) 
   & = & 
   \frac{1}{2t}
   \left\{
      A(M_2^2) + (\Delta_{12} +t) B_{21}(M_1^2,M_2^2,t) - 4 B_{32}(M_1^2,M_2^2,t)
   \right\},
   \\
   B_{32}(M_1^2,M_2^2,t)
   & = &
   \frac{1}{2dt}
   \left\{
      -M_1^2 A(M_1^2) + M_2^2 A(M_2^2) + d (\Delta_{12}+t) B_{22}(M_1^2,M_2^2,t)
   \right\}
   \hspace{20mm}
\eea
with $\Delta_{12}=M_1^2-M_2^2$. 
The pole, finite and $O(\epsilon)$ parts of the one-loop integrals 
relevant to the EM form factors 
can be expressed in terms of those of $A$ and $B$ functions
\bea
   A(M_1)^2
   & = &
   A_{\rm pole}(M_1^2) + \bar{A}(M_1^2) 
 + \epsilon \bar{A}^{\epsilon}(M_1^2) + O(\epsilon^2),
   \\
   B(M_1^2,M_2^2,t)
   & = &
   B_{\rm pole}(M_1^2,M_2^2,t) + \bar{B}(M_1^2,M_2^2,t)
 + \epsilon \bar{B}^{\epsilon}(M_1^2,M_2^2,t) + O(\epsilon^2)
   \vspace{20mm}
\eea
with 
\bea
   A_{\rm pole}(M_1^2)
   & = & 
   \frac{M_1^2}{N} \lambda_0,
   \\
   \bar{A}(M_1^2) 
   & = &
  -\frac{M_1^2}{N} \ln\left[ \frac{M_1^2}{\mu^2} \right],
   \\
   \bar{A}^{\epsilon}(M_1^2)
   & = &
   \frac{M_1^2}{N}
   \left\{
      \frac{C^2}{2} + \frac{1}{2} + \frac{\pi^2}{12} 
    + \frac{1}{2} \ln^2\left[ \frac{M_1^2}{\mu^2} \right] 
    - C \ln\left[ \frac{M_1^2}{\mu^2} \right]
   \right\},
   \\
   B_{\rm pole}(M_1^2,M_2^2,t)
   & = &
   \frac{1}{N} \lambda_0,
\eea
\bea
   \bar{B}(M_1^2,M_2^2,t)
   & = &
  -\frac{1}{N}
   \frac{ M_1^2 \ln\left[ \frac{M_1^2}{\mu^2} \right]
         +M_2^2 \ln\left[ \frac{M_2^2}{\mu^2} \right]}
        {\Delta_{12}}
   \nn \\
   & & 
  +\frac{1}{2N}
   \left\{
      2 
    + \left( 
         - \frac{\Delta_{12}}{t} + \frac{\Sigma_{12}}{\Delta_{12}}
      \right)
      \ln\left[ \frac{M_1^2}{M_2^2} \right]
   \right.
   \nn \\[2mm]
   & & 
   \hspace{40mm}
   \left.
    - \frac{\nu_{12}(t)}{t} 
      \ln\left[ 
            \frac{(t+\nu_{12}(t))^2 - \Delta_{12}^2}
                 {(t-\nu_{12}(t))^2 - \Delta_{12}^2}
         \right]
   \right\},
   \\
   \bar{B}^{\epsilon}(M_1^2,M_2^2,t)
   & = &
   \frac{1}{N}
    \left\{
      \frac{C^2}{2} - \frac{1}{2} + \frac{\pi^2}{12}
    + (C-1)\bar{B}(M_1^2,M_2^2,t) 
   \right.
   \nn \\
   & &
   \hspace{20mm}
   \left. 
    + \frac{1}{2}
      \int_0^1 dx \ln^2\left[ 
                          \frac{ x M_1^2 + (1-x)M_2^2 - x(1-x)t}{\mu^2}
                      \right]
   \right\},
   \hspace{10mm}
\eea
where
\bea
   \Sigma_{12}
   & = & 
   M_1^2 + M_2^2, 
   \\
   \nu_{12}^2
   & = & 
   t^2 - 2 \Sigma_{12} t + \Delta_{12}^2,
   \\
   \lambda_0 
   & = & 
   \frac{1}{\epsilon} + \ln\left[ 4\pi \right] + 1 - \gamma
  =\frac{1}{\epsilon} + C.
\eea
The one-loop contributions 
in Eqs.~(\ref{eqn:chiral_fit:su3:pff:nlo_b})\,--\,(\ref{eqn:chiral_fit:su3:k0ff:nlo_b})
are expressed in terms of the finite parts $\bar{A}$ and $\bar{B}_{22}$.

%% file: B.2-loop_integrals.tex
\section{Two-loop integrals in SU(3) ChPT}
\label{sec:appndx:2-loop_int}

\begin{figure}[t]
\begin{center}
   \vspace{10mm}
   \includegraphics[angle=0,width=0.15\linewidth,clip]%
                   {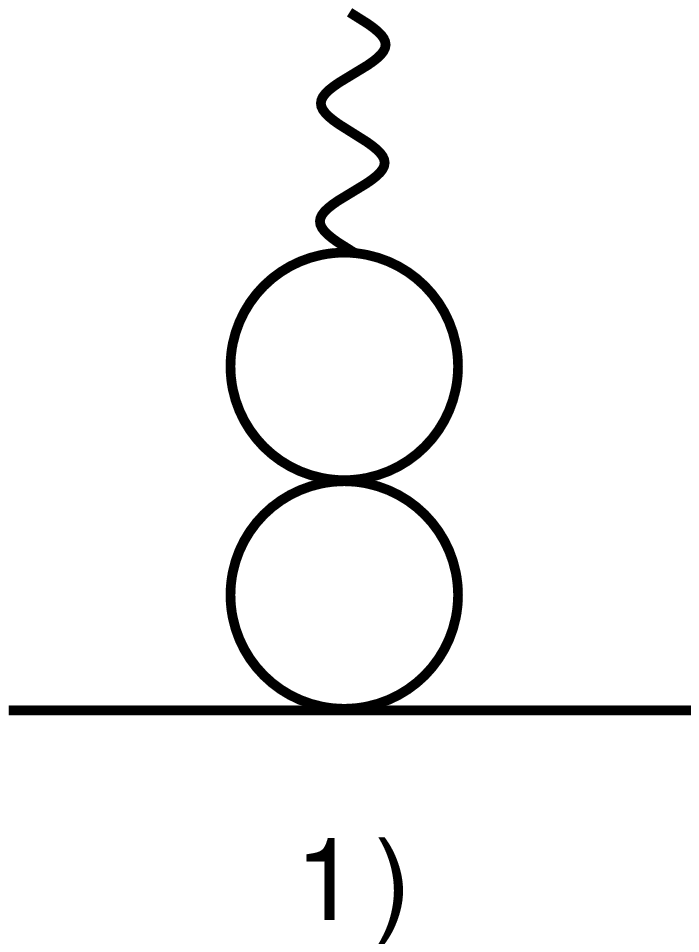}
   \hspace{5mm}
   \includegraphics[angle=0,width=0.15\linewidth,clip]%
                   {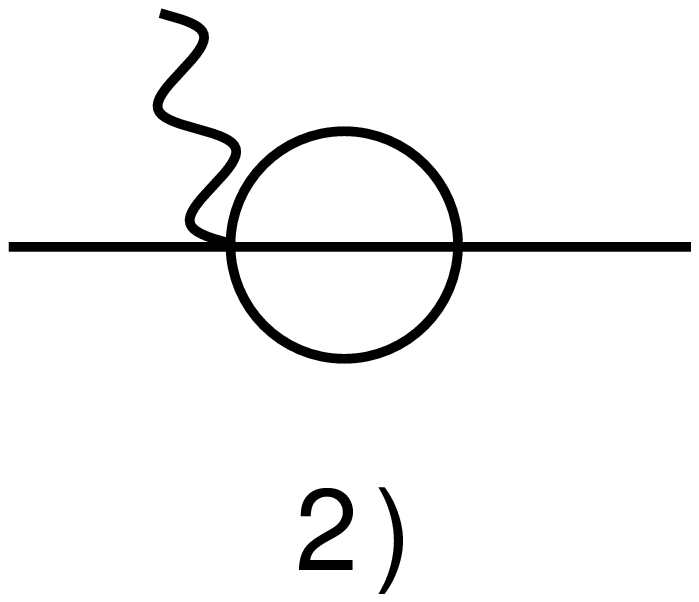}
   \hspace{5mm}
   \includegraphics[angle=0,width=0.15\linewidth,clip]%
                   {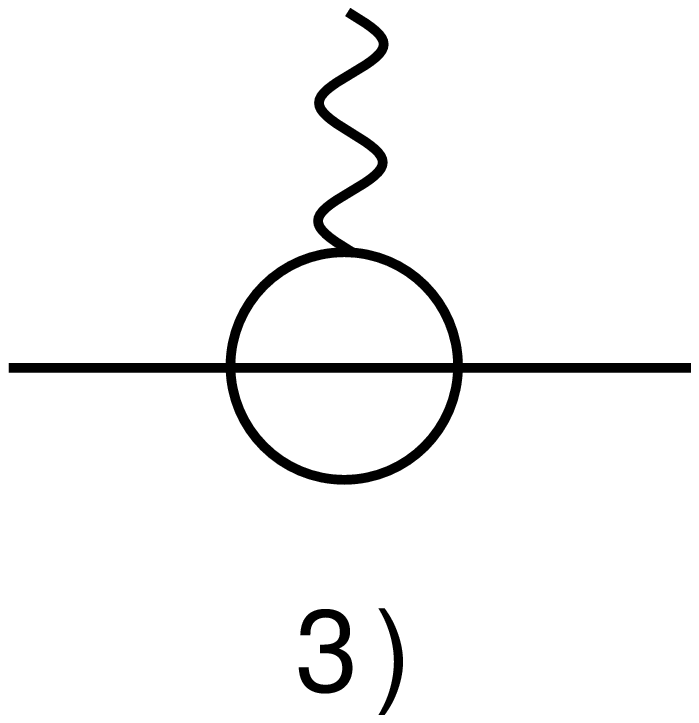}

   \vspace{0mm}
   \caption{
      Example of two-loop diagrams
      with reducible (1), sunset (2), and vertex integrals (3).
   }
   \label{fig:appdxB:diagram:nnlo} 
\end{center}
\vspace{0mm}
\end{figure}

We categorize the two-loop diagrams into three types:
those with the reducible, sunset and vertex integrals.
An example is shown in Fig.~\ref{fig:appdxB:diagram:nnlo}.

\subsection{Diagrams with reducible integral}
\label{sec:appndx:2-loop_int:reducible}

The diagram of Fig.~\ref{fig:appdxB:diagram:nnlo}\,--\,1 involves
two independent one-loop integrals. 
The contribution of this type of diagram can be written 
in terms of the one-loop integral functions
discussed in Appendix~\ref{sec:appndx:1-loop_int}.
The expression for the pion form factor is given by~\cite{PFF+KFF:ChPT:NNLO:Nf3}
\bea
   & &
   F_\pi^4\,
   F_{V,4,B,{\rm reducible}}^{\pi^+}(t)
   \nn \\[1mm]
   & = &
   \frac{1}{N}
   \left\{ -\frac{1}{2} M_\pi^2 \Ap - \frac{1}{4} M_\pi^2 \Ak \right\}
   \nn \\ 
   & &
  +\frac{1}{N^2}
   \left\{
     -\frac{\pi^2}{48} M_\pi^2 (10 M_\pi^2 + 3 M_K^2) 
     +\frac{35}{96} M_\pi^2 (M_\pi^2 - 2 M_K^2) 
   \right.
   \nn \\
   & & 
   \hspace{70mm}
   \left.
     -\frac{89}{48} M_\pi^4
     -\frac{1}{16}\left( 1 + \frac{\pi^2}{6} \right)\, (2M_\pi^2 + M_K^2 )\, t
   \right\}
   \nn \\
   & &
  +\frac{1}{N} \left( 5 M_\pi^2 - \frac{t}{2} \right)
   \bar{B}_{22}^\epsilon(M_\pi^2,M_\pi^2,t)
  +\frac{1}{N} \left( \frac{3}{2} M_\pi^2 - \frac{t}{4} \right)
   \bar{B}_{22}^\epsilon(M_K^2,M_K^2,t)
   \nn \\
   & &
  +4 \left\{ \Bttppt \right\}^2 + 4 \Bttppt \Bttkkt 
  +\left\{ \Bttkkt \right\}^2
   \nn \\
   & &
  -\left\{ 4 \Ap + 2 \Ak \right\} \Bttppt 
  -\left\{ 2 \Ap + \Ak \right\} \Bttkkt 
   \nn \\
   & &
  -\frac{1}{4} \left\{ \Ap \right\}^2 + \Ap \Ak
  +\left( \frac{1}{4} - \frac{3}{8}\frac{M_\pi^2}{M_K^2} \right)
   \left\{ \Ak \right\}^2
   \nn \\
   & &
  -\frac{1}{8}\frac{t}{M_\pi^2} \left\{ \Ap \right\}^2
  -\frac{1}{16} \frac{t}{M_K^2} \left\{ \Ak \right\}^2.
\eea

\subsection{Diagrams with sunset integral}
\label{sec:appndx:2-loop_int:sunset}

The diagram of Fig.~\ref{fig:appdxB:diagram:nnlo}\,--\,2 
involves the so-called sunset integral,
which is genuine two-loop integral.
This type of integral is $t$-independent, and hence 
also appears in the two-loop chiral expansion of 
the meson masses and decay constants~\cite{Spec+DC:ChPT:SU3:NNLO:ABT}.

A typical form of the sunset integral is 
\bea
   \langle \langle X(r,s) \rangle \rangle_{\rm s}
   & = &
   \frac{1}{i^2}
   \int \frac{d^2r}{(2\pi)^d} \frac{d^2s}{(2\pi)^d}
   \frac{X(r,s)}{(r^2-M_1^2)(s^2-M_2^2)\left\{(p-r-s)^2-M_3^2\right\}},
\eea
where $X(r,s)$ specifies the tensor structure 
in terms of the loop momenta $r$ and $s$.
We consider the following three integrals with 
$X(r,s)\!=\!1, r_\mu, r_\mu r_\nu$
\bea
   H(M_1^2,M_2^2,M_3^2;p^2)
   & = &    
   \langle \langle 1 \rangle \rangle_{\rm s},
   \\
   H_\mu(M_1^2,M_2^2,M_3^2;p^2)
   & = &    
   \langle \langle r_\mu \rangle \rangle_{\rm s},
   \\
   H_{\mu\nu}(M_1^2,M_2^2,M_3^2;p^2)
   & = &    
   \langle \langle r_\mu r_\nu \rangle \rangle_{\rm s}.
\eea
By redefining the momenta, 
other sunset integrals with $X(r,s)\!=\!s_\mu, s_\mu s_\nu, r_\mu s_\nu$ can 
be related to the above three functions~\cite{Spec+DC:ChPT:SU3:NNLO:ABT}.

The Lorentz decomposition of these ``$H$'' functions is given as
\bea
   H_\mu(M_1^2,M_2^2,M_3^2;p^2)
   & = &
   p_\mu H_1(M_1^2,M_2^2,M_3^2;p^2),
   \\
   H_{\mu\nu}(M_1^2,M_2^2,M_3^2;p^2)
   & = &
   p_\mu p_\nu H_{21}(M_1^2,M_2^2,M_3^2;p^2)
  +g_{\mu\nu} H_{22}(M_1^2,M_2^2,M_3^2;p^2).
\eea
It is possible to express $H_{22}$ as~\cite{PFF+KFF:ChPT:NNLO:Nf3}
\bea
   &&
   d H_{22}(M_1^2,M_2^2,M_3^2;p^2)
   \nn \\
   & = & 
 - p^2 H_{21}(M_1^2,M_2^2,M_3^2;p^2)
 + M_1^2 H(M_1^2,M_2^2,M_3^2;p^2) + A(M_2^2) A(M_3^2).
\eea
Therefore,
the contribution of the sunset diagrams to $\pff$ 
can be calculated with $H_X(M_1^2,M_2^2,M_3^2,p^2)$ with $X\!=\!"",1,21$
\bea
   &&
   F_\pi^4\,
   F_{V,4,{\rm sunset}}^{\pi^+}(t)
   \nn \\[1mm]
   & = &
   \frac{3}{2}  M_\pi^4  H^{F\prime}(M_\pi^2,M_\pi^2,M_\pi^2;M_\pi^2)
 - \frac{5}{8}  M_\pi^4  H^{F\prime}(M_\pi^2,M_K^2,M_K^2;M_\pi^2)
   \nn \\
   & & 
  +\frac{1}{18} M_\pi^4  H^{F\prime}(M_\pi^2,M_\eta^2,M_\eta^2;M_\pi^2)
  +M_\pi^2 M_K^2         H^{F\prime}(M_K^2,M_\pi^2,M_K^2;M_\pi^2)
   \nn \\
   & &
  -\frac{5}{6}  M_\pi^4  H^{F\prime}(M_K^2,M_K^2,M_\eta^2;M_\pi^2)   
  -\frac{1}{4}  M_\pi^2  \left( \frac{1}{2} M_\pi^2 - 2M_K^2 \right)
   H^{F\prime}(M_\eta^2,M_K^2,M_K^2;M_\pi^2)
   \nn \\
   & &
  -2 M_\pi^4 H_1^{F\prime}(M_\pi^2,M_\pi^2,M_\pi^2;M_\pi^2)   
  +M_\pi^4 H_1^{F\prime}(M_\pi^2,M_K^2,M_K^2;M_\pi^2)   
   \nn \\
   & &
  +2 M_\pi^4 H_1^{F\prime}(M_K^2,M_K^2,M_\eta^2;M_\pi^2)   
   \nn \\
   & &
  +3 M_\pi^4 H_{21}^{F\prime}(M_\pi^2,M_\pi^2,M_\pi^2;M_\pi^2)   
  -\frac{3}{8} M_\pi^4 H_{21}^{F\prime}(M_\pi^2,M_K^2,M_K^2;M_\pi^2)   
   \nn \\
   & &
  +3 M_\pi^4 H_{21}^{F\prime}(M_K^2,M_\pi^2,M_K^2;M_\pi^2)   
  +\frac{9}{8} M_\pi^4 H_{21}^{F\prime}(M_\eta^2,M_K^2,M_K^2;M_\pi^2)   
   \nn 
\eea
\bea
   & &
  +\left( \frac{5}{3} M_\pi^2 + \frac{1}{18}t \right)
   H^{F}(M_\pi^2,M_\pi^2,M_\pi^2;M_\pi^2)   
  +\frac{1}{12} ( M_\pi^2 + t ) H^{F}(M_\pi^2,M_K^2,M_K^2;M_\pi^2)
   \nn \\
   & &
  +\left\{
      \frac{15}{32} M_\pi^2 - \frac{5}{96} (M_\pi^2-2M_K^2) - \frac{5}{48}t 
   \right\}
   H^{F}(M_K^2,M_\pi^2,M_K^2;M_\pi^2)
   \nn \\
   & &
  -\frac{5}{48} M_K^2 H^{F}(M_K^2,M_K^2,M_\pi^2;M_\pi^2)
   \nn \\
   & &
  -\left( \frac{1}{12} M_\pi^2 + \frac{1}{16} t \right) 
   H^{F}(M_K^2,M_K^2,M_\eta^2;M_\pi^2)
   \nn \\
   & &
  -\left( 3 M_\pi^2 + \frac{1}{3} t \right) 
   H_1^{F}(M_\pi^2,M_\pi^2,M_\pi^2;M_\pi^2)
  +\frac{1}{24} t H_1^{F}(M_\pi^2,M_K^2,M_K^2;M_\pi^2)
   \nn \\
   & &
  -\left( M_\pi^2 + \frac{1}{8} t \right) 
   H_1^{F}(M_K^2,M_\pi^2,M_K^2;M_\pi^2)
  +\left( M_\pi^2 + \frac{1}{8} t \right) 
   H_1^{F}(M_K^2,M_K^2,M_\eta^2;M_\pi^2)
   \nn \\
   & &
  +\left( 3 M_\pi^2 + \frac{1}{6} t \right) 
   H_{21}^{F}(M_\pi^2,M_\pi^2,M_\pi^2;M_\pi^2)
  -\left( \frac{3}{8} M_\pi^2 + \frac{1}{48} t \right) 
   H_{21}^{F}(M_\pi^2,M_K^2,M_K^2;M_\pi^2)
   \nn \\
   & &
  +\left( \frac{53}{16} M_\pi^2 + \frac{1}{16} t \right) 
   H_{21}^{F}(M_K^2,M_\pi^2,M_K^2;M_\pi^2)
   \nn \\
   & &
  -\left( \frac{5}{16} M_\pi^2 - \frac{5}{48} t \right) 
   H_{21}^{F}(M_K^2,M_K^2,M_\pi^2;M_\pi^2)
   \nn \\
   & &
  +\left( \frac{9}{8} M_\pi^2 + \frac{1}{16} t \right) 
   H_{21}^{F}(M_\eta^2,M_K^2,M_K^2;M_\pi^2).
\eea
Here $H_X^F(M_1^2,M_2^2,M_3^2;p^2)$ and $H_X^{F \prime}(M_1^2,M_2^2,M_3^2;p^2)$
($X\!=\!"",1,21$) represent
the finite part of $H_X(M_1^2,M_2^2,M_3^2;p^2)$
and its derivative with respective to $p^2$. 
We refer to Ref.~\cite{Spec+DC:ChPT:SU3:NNLO:ABT}
for the explicit expression of these ``$H$'' functions.

\subsection{Diagrams with vertex integral}
\label{sec:appndx:2-loop_int:vertex}

The vertex integral 
is $t$-dependent genuine two-loop integral
involved in diagrams such as Fig.~\ref{fig:appdxB:diagram:nnlo}\,--\,3. 
It is defined as~\cite{PFF+KFF:ChPT:NNLO:Nf3}
\bea
   &&
   \langle \langle X(r,s) \rangle \rangle_{\rm v}
   \nn \\[2mm]
   & = &
   \frac{1}{i^2}
   \int \frac{d^2r}{(2\pi)^d} \frac{d^2s}{(2\pi)^d}
   \frac{X(r,s)}{(r^2-M_1^2) \left\{ (r-q)^2 - M_2^2 \right\}
                 (s^2-M_3^2) \left\{ (p-r-s)^2 - M_4^2 \right\} }.
   \hspace{10mm}
\eea
The Lorentz decomposition of the vertex integrals can be expressed as 
\bea
   \langle \langle r_\mu \rangle \rangle_{\rm v}
   & = & 
   p_\mu V_{1,1} + q_\mu V_{1,2},
   \\
   \langle \langle r_\mu r_\nu \rangle \rangle_{\rm v}
   & = & 
   g_{\mu\nu} V_{2,1} + p_\mu p_\nu V_{2,2} + q_\mu q_\nu V_{2,3}   
  +(p_\mu q_\nu + q_\mu p_\nu) V_{2,4},
\eea
\bea
   \langle \langle r_\mu r_\nu r_\alpha \rangle \rangle_{\rm v}
   & = & 
   (g_{\mu\nu} p_\alpha + g_{\nu\alpha} p_\mu + g_{\alpha\mu} p_\nu ) V_{3,1}
  +(g_{\mu\nu} q_\alpha + g_{\nu\alpha} q_\mu + g_{\alpha\mu} q_\nu ) V_{3,2}
   \nn \\
   & &
  + p_\mu p_\nu p_\alpha V_{3,3} 
  + q_\mu q_\nu q_\alpha V_{3,4} 
  +(p_\mu p_\nu q_\alpha + p_\mu q_\nu p_\alpha + q_\mu p_\nu p_\alpha ) V_{3,5}
   \hspace{5mm}
   \nn \\
   & &
  +(p_\mu q_\nu q_\alpha + q_\mu p_\nu q_\alpha + q_\mu q_\nu p_\alpha ) V_{3,6}
\eea
using the integral functions 
$V_{i,j}(M_1^2,M_2^2,M_3^2,M_4^2;p^2,q^2,(p-q)^2)$, 
where $i$ represents the number of the momentum appearing in 
$X(r,s)$ and $j$ is the index of the integral function for a given $i$.
For $i\!\leq\!3$,
there exist 44 scalar functions with
\bea
   (i,j) 
   & = &
   (0,0), \hspace{2mm}
   (1,1), \ldots, (1,4),  \hspace{2mm}
   (2,1), \ldots, (2,13), \hspace{2mm}
   (3,1), \ldots, (3,26).
\eea
The explicit expression of these "$V$" functions is 
given in Ref.~\cite{PFF+KFF:ChPT:NNLO:Nf3}.
The two-loop contribution to $\pff$ with the vertex integral
can be written with a subset of these functions 
in a rather involved form: 
\bea
   &&
   F_\pi^4\,
   F_{V,4,{\rm vertex}}^{\pi^+}(t)
   \nn \\[2mm]
   & = &
   %
   %
   \left( \frac{5}{2} M_\pi^4 - \frac{7}{3} M_\pi^2 t \right)
   V_{1,1}(M_\pi^2,M_\pi^2,M_\pi^2,M_\pi^2;M_\pi^2,t,M_\pi^2)
   \nn \\
   & &
  +\left( 
      M_\pi^4 - \frac{2}{3} M_\pi^2 t + \frac{1}{12} t^2 
   \right)
   V_{1,1}(M_\pi^2,M_\pi^2,M_K^2,M_K^2;M_\pi^2,t,M_\pi^2)
   \nn \\  
   & &
  +\frac{1}{18} M_\pi^4 
   V_{1,1}(M_\pi^2,M_\pi^2,M_\eta^2,M_\eta^2;M_\pi^2,t,M_\pi^2)
   \nn \\ 
   & &
  +\left(
      \frac{3}{2} M_\pi^4 - \frac{17}{12} M_\pi^2 t + \frac{1}{6} t^2
   \right)
   V_{1,1}(M_K^2,M_K^2,M_\pi^2,M_K^2;M_\pi^2,t,M_\pi^2)
   \nn \\ 
   & &
  +\left( 
      \frac{2}{3} M_\pi^4 - \frac{2}{3} M_\pi^2 t + \frac{1}{8} t^2
   \right)
   V_{1,1}(M_K^2,M_K^2,M_K^2,M_\eta^2;M_\pi^2,t,M_\pi^2)
   \nn \\
   %
   %
   & &
  +\left( -6 M_\pi^2 + t \right)
   V_{2,1}(M_\pi^2,M_\pi^2,M_\pi^2,M_\pi^2;M_\pi^2,t,M_\pi^2)
   \nn \\
   & &
  +\left( -2 M_\pi^2 + \frac{2}{3} t \right)
   V_{2,1}(M_\pi^2,M_\pi^2,M_K^2,M_K^2;M_\pi^2,t,M_\pi^2)
   \nn \\  
   & &
  +\left( -4 M_\pi^2 + \frac{4}{3} t \right) 
   V_{2,1}(M_K^2,M_K^2,M_\pi^2,M_K^2;M_\pi^2,t,M_\pi^2)
   \nn \\ 
   & &
  +\left( -2 M_\pi^2 + t \right)
   V_{2,1}(M_K^2,M_K^2,M_K^2,M_\eta^2;M_\pi^2,t,M_\pi^2)
   \nn \\
   %
   %
   & &
  +\left( -6 M_\pi^4 + \frac{10}{3} M_\pi^2 t \right)
   V_{2,2}(M_\pi^2,M_\pi^2,M_\pi^2,M_\pi^2;M_\pi^2,t,M_\pi^2)
   \nn \\
   & &
  +\left( 
      -2 M_\pi^4 + \frac{4}{3} M_\pi^2 t - \frac{1}{6} t^2
   \right)
   V_{2,2}(M_\pi^2,M_\pi^2,M_K^2,M_K^2;M_\pi^2,t,M_\pi^2)
   \nn \\  
   & &
  +\left( 
      -4 M_\pi^4 + \frac{11}{4} M_\pi^2 t - \frac{1}{3} t^2
   \right) 
   V_{2,2}(M_K^2,M_K^2,M_\pi^2,M_K^2;M_\pi^2,t,M_\pi^2)
   \nn \\ 
   & &
  +\left( 
      -2 M_\pi^4 + \frac{5}{3} M_\pi^2 t  - \frac{1}{4} t^2 
   \right)
   V_{2,2}(M_K^2,M_K^2,M_K^2,M_\eta^2;M_\pi^2,t,M_\pi^2)
   \nn 
\eea
\bea
   %
   %
   & &
  +\left( \frac{5}{3} M_\pi^2 t + \frac{1}{2} t^2 \right)
   V_{2,4}(M_\pi^2,M_\pi^2,M_\pi^2,M_\pi^2;M_\pi^2,t,M_\pi^2)
   \nn \\
   & &
  +\frac{1}{3} M_\pi^2 t 
   V_{2,4}(M_\pi^2,M_\pi^2,M_K^2,M_K^2;M_\pi^2,t,M_\pi^2)
  +\frac{5}{6} M_\pi^2 t 
   V_{2,4}(M_K^2,M_K^2,M_\pi^2,M_K^2;M_\pi^2,t,M_\pi^2)
   \nn \\  
   & &
  +\frac{1}{3} M_\pi^2 t 
   V_{2,4}(M_K^2,M_K^2,M_K^2,M_\eta^2;M_\pi^2,t,M_\pi^2)
   \nn \\
   %
   %
   & &
  +\left( -4 M_\pi^2 + \frac{4}{3} t \right)
   V_{2,5}(M_\pi^2,M_\pi^2,M_\pi^2,M_\pi^2;M_\pi^2,t,M_\pi^2)
   \nn \\
   & &
  +\left( -2 M_\pi^2 + \frac{1}{2} t \right)
   V_{2,5}(M_\pi^2,M_\pi^2,M_K^2,M_K^2;M_\pi^2,t,M_\pi^2)
   \nn \\  
   & &
  +\left( -3 M_\pi^2 + \frac{17}{12} t \right) 
   V_{2,5}(M_K^2,M_K^2,M_\pi^2,M_K^2;M_\pi^2,t,M_\pi^2)
   \nn \\ 
   & &
  +\left( -2 M_\pi^2 + t \right)
   V_{2,5}(M_K^2,M_K^2,M_K^2,M_\eta^2;M_\pi^2,t,M_\pi^2)
   \nn \\
   %
   %
   & &
  +\left( 
      -4 M_\pi^4 + \frac{7}{3} M_\pi^2 t + \frac{1}{3} t^2
   \right)
   V_{2,6}(M_\pi^2,M_\pi^2,M_\pi^2,M_\pi^2;M_\pi^2,t,M_\pi^2)
   \nn \\
   & &
  +\left( -2 M_\pi^4 + M_\pi^2 t \right)
   V_{2,6}(M_\pi^2,M_\pi^2,M_K^2,M_K^2;M_\pi^2,t,M_\pi^2)
   \nn \\  
   & &
  +\left( 
      -3 M_\pi^4 + \frac{13}{6} M_\pi^2 t - \frac{5}{24} t^2
   \right) 
   V_{2,6}(M_K^2,M_K^2,M_\pi^2,M_K^2;M_\pi^2,t,M_\pi^2)
   \nn \\ 
   & &
  +\left( 
      -2 M_\pi^4 + \frac{3}{2} M_\pi^2 t - \frac{1}{8} t^2 
   \right)
   V_{2,6}(M_K^2,M_K^2,M_K^2,M_\eta^2;M_\pi^2,t,M_\pi^2)
   \nn \\
   %
   %
   & &
  +\frac{4}{3} t^2 
   V_{2,9}(M_\pi^2,M_\pi^2,M_\pi^2,M_\pi^2;M_\pi^2,t,M_\pi^2)
  +\frac{1}{4} t^2 
   V_{2,9}(M_\pi^2,M_\pi^2,M_K^2,M_K^2;M_\pi^2,t,M_\pi^2)
   \nn \\  
   & &
  +\frac{7}{24} t^2
   V_{2,9}(M_K^2,M_K^2,M_\pi^2,M_K^2;M_\pi^2,t,M_\pi^2)
  +\frac{1}{4} t^2
   V_{2,9}(M_K^2,M_K^2,M_K^2,M_\eta^2;M_\pi^2,t,M_\pi^2)
   \nn \\
   %
   %
   & &
  +\left( 6 M_\pi^2 - 2 t \right)
   V_{3,1}(M_\pi^2,M_\pi^2,M_\pi^2,M_\pi^2;M_\pi^2,t,M_\pi^2)
   \nn \\
   & &
  +\left( 3 M_\pi^2 - t \right)
   V_{3,1}(M_\pi^2,M_\pi^2,M_K^2,M_K^2;M_\pi^2,t,M_\pi^2)
   \nn \\  
   & &
  +\left( 6 M_\pi^2 - 2 t \right) 
   V_{3,1}(M_K^2,M_K^2,M_\pi^2,M_K^2;M_\pi^2,t,M_\pi^2)
   \nn \\ 
   & &
  +\left( \frac{9}{2} M_\pi^2 - \frac{3}{2} t \right)
   V_{3,1}(M_K^2,M_K^2,M_K^2,M_\eta^2;M_\pi^2,t,M_\pi^2)
   \nn \\
   %
   %
   & &
  -\frac{1}{3} t
   V_{3,2}(M_\pi^2,M_\pi^2,M_K^2,M_K^2;M_\pi^2,t,M_\pi^2)
  -\frac{2}{3} t
   V_{3,2}(M_K^2,M_K^2,M_\pi^2,M_K^2;M_\pi^2,t,M_\pi^2)
   \nn \\  
   & &
  -\frac{1}{2} t
   V_{3,2}(M_K^2,M_K^2,M_K^2,M_\eta^2;M_\pi^2,t,M_\pi^2)
   \nn \\
   %
   %
   & &
  +\left( 2 M_\pi^4 - M_\pi^2 t \right)
   V_{3,3}(M_\pi^2,M_\pi^2,M_\pi^2,M_\pi^2;M_\pi^2,t,M_\pi^2)
   \nn \\
   & &
  +\left( 
      M_\pi^4 - \frac{2}{3} M_\pi^2 t + \frac{1}{12} t^2 
   \right)
   V_{3,3}(M_\pi^2,M_\pi^2,M_K^2,M_K^2;M_\pi^2,t,M_\pi^2)
   \nn \\  
   & &
  +\left( 
      2 M_\pi^4 - \frac{4}{3} M_\pi^2 t + \frac{1}{6} t^2
   \right) 
   V_{3,3}(M_K^2,M_K^2,M_\pi^2,M_K^2;M_\pi^2,t,M_\pi^2)
   \nn \\ 
   & &
  +\left( 
      \frac{3}{2} M_\pi^4 - M_\pi^2 t + \frac{1}{8} t^2 
   \right)
   V_{3,3}(M_K^2,M_K^2,M_K^2,M_\eta^2;M_\pi^2,t,M_\pi^2)
   \nn 
\eea
\bea
   %
   %
   & &
  -\frac{1}{2} t^2
   V_{3,5}(M_\pi^2,M_\pi^2,M_\pi^2,M_\pi^2;M_\pi^2,t,M_\pi^2)
  -\frac{1}{3} M_\pi^2 t
   V_{3,5}(M_\pi^2,M_\pi^2,M_K^2,M_K^2;M_\pi^2,t,M_\pi^2)
   \nn \\  
   & &
  -\frac{2}{3} M_\pi^2 t
   V_{3,5}(M_K^2,M_K^2,M_\pi^2,M_K^2;M_\pi^2,t,M_\pi^2)
  -\frac{1}{2} M_\pi^2 t 
   V_{3,5}(M_K^2,M_K^2,M_K^2,M_\eta^2;M_\pi^2,t,M_\pi^2)
   \nn \\
   %
   %
   & &
  -\frac{1}{2} t^2
   V_{3,6}(M_\pi^2,M_\pi^2,M_\pi^2,M_\pi^2;M_\pi^2,t,M_\pi^2)
  -\frac{1}{12} t^2
   V_{3,6}(M_\pi^2,M_\pi^2,M_K^2,M_K^2;M_\pi^2,t,M_\pi^2)
   \nn \\  
   & &
  -\frac{1}{6} t^2
   V_{3,6}(M_K^2,M_K^2,M_\pi^2,M_K^2;M_\pi^2,t,M_\pi^2)
  -\frac{1}{8} t^2
   V_{3,6}(M_K^2,M_K^2,M_K^2,M_\eta^2;M_\pi^2,t,M_\pi^2)
   \nn \\
   %
   %
   & &
  +4 M_\pi^2 
   V_{3,7}(M_\pi^2,M_\pi^2,M_\pi^2,M_\pi^2;M_\pi^2,t,M_\pi^2)
   \nn \\
   & &
  +\left( 2 M_\pi^2 - \frac{1}{2} t \right)
   V_{3,7}(M_\pi^2,M_\pi^2,M_K^2,M_K^2;M_\pi^2,t,M_\pi^2)
   \nn \\  
   & &
  +\left( 4 M_\pi^2 - t \right) 
   V_{3,7}(M_K^2,M_K^2,M_\pi^2,M_K^2;M_\pi^2,t,M_\pi^2)
   \nn \\ 
   & &
  +\left( 3 M_\pi^2 - \frac{3}{4} t \right)
   V_{3,7}(M_K^2,M_K^2,M_K^2,M_\eta^2;M_\pi^2,t,M_\pi^2)
   \nn \\
   %
   %
   & &
  +2 t
   V_{3,8}(M_\pi^2,M_\pi^2,M_\pi^2,M_\pi^2;M_\pi^2,t,M_\pi^2)
   \nn \\
   %
   %
   & &
  +\left( 8 M_\pi^2 - 4 t \right)
   V_{3,9}(M_\pi^2,M_\pi^2,M_\pi^2,M_\pi^2;M_\pi^2,t,M_\pi^2)
   \nn \\
   & &
  +\left( 4 M_\pi^2 - \frac{3}{2} t \right)
   V_{3,9}(M_\pi^2,M_\pi^2,M_K^2,M_K^2;M_\pi^2,t,M_\pi^2)
   \nn \\  
   & &
  +\left( 8 M_\pi^2 - 3 t \right) 
   V_{3,9}(M_K^2,M_K^2,M_\pi^2,M_K^2;M_\pi^2,t,M_\pi^2)
   \nn \\ 
   & &
  +\left( 6 M_\pi^2 - \frac{9}{4} t \right)
   V_{3,9}(M_K^2,M_K^2,M_K^2,M_\eta^2;M_\pi^2,t,M_\pi^2)
   \nn \\ 
   %
   %
   & &
  -\frac{2}{3} t
   V_{3,10}(M_\pi^2,M_\pi^2,M_\pi^2,M_\pi^2;M_\pi^2,t,M_\pi^2)
  -\frac{5}{6} t
   V_{3,10}(M_K^2,M_K^2,M_\pi^2,M_K^2;M_\pi^2,t,M_\pi^2)
   \nn \\
   & & 
  -\frac{1}{2} t 
   V_{3,10}(M_K^2,M_K^2,M_K^2,M_\eta^2;M_\pi^2,t,M_\pi^2)
   \nn \\
   %
   %
   & &
  +\left( 
      4 M_\pi^4 - \frac{4}{3} M_\pi^2 t - \frac{1}{3} t^2 
   \right)
   V_{3,11}(M_\pi^2,M_\pi^2,M_\pi^2,M_\pi^2;M_\pi^2,t,M_\pi^2)
   \nn \\
   & &
  +\left( 2 M_\pi^4 - M_\pi^2 t \right)
   V_{3,11}(M_\pi^2,M_\pi^2,M_K^2,M_K^2;M_\pi^2,t,M_\pi^2)
   \nn \\  
   & &
  +\left( 
      4 M_\pi^4 - \frac{29}{12} M_\pi^2 t + \frac{5}{24} t^2
   \right) 
   V_{3,11}(M_K^2,M_K^2,M_\pi^2,M_K^2;M_\pi^2,t,M_\pi^2)
   \nn \\ 
   & &
  +\left( 
      3 M_\pi^4 - \frac{7}{4} M_\pi^2 t + \frac{1}{8} t^2 
   \right)
   V_{3,11}(M_K^2,M_K^2,M_K^2,M_\eta^2;M_\pi^2,t,M_\pi^2)
   \nn \\
   %
   %
   & &
  +\left( 2 M_\pi^2 t - \frac{4}{3} t^2 \right)
   V_{3,13}(M_\pi^2,M_\pi^2,M_\pi^2,M_\pi^2;M_\pi^2,t,M_\pi^2)
  -\frac{1}{4} t^2 
   V_{3,13}(M_\pi^2,M_\pi^2,M_K^2,M_K^2;M_\pi^2,t,M_\pi^2)
   \nn \\  
   & &
  -\frac{7}{24} t^2 
   V_{3,13}(M_K^2,M_K^2,M_\pi^2,M_K^2;M_\pi^2,t,M_\pi^2)
  -\frac{1}{4} t^2 
   V_{3,13}(M_K^2,M_K^2,M_K^2,M_\eta^2;M_\pi^2,t,M_\pi^2)
   \nn \\ 
   %
   %
   & &
  +\left( 
      - \frac{2}{3} M_\pi^2 t - \frac{2}{3} t^2 
   \right)
   V_{3,15}(M_\pi^2,M_\pi^2,M_\pi^2,M_\pi^2;M_\pi^2,t,M_\pi^2)
   \nn \\
   & &
  -\frac{1}{4} t^2
   V_{3,15}(M_\pi^2,M_\pi^2,M_K^2,M_K^2;M_\pi^2,t,M_\pi^2)
   \nn \\  
   & &
  +\left( 
      - \frac{5}{6} M_\pi^2 t - \frac{1}{12} t^2
   \right) 
   V_{3,15}(M_K^2,M_K^2,M_\pi^2,M_K^2;M_\pi^2,t,M_\pi^2)
   \nn \\ 
   & &
  +\left( 
      -\frac{1}{2} M_\pi^2 t - \frac{1}{8} t^2
   \right)
   V_{3,15}(M_K^2,M_K^2,M_K^2,M_\eta^2;M_\pi^2,t,M_\pi^2)
   \nn 
\eea
\bea
   %
   %
   & &
  -\frac{5}{3} t^2 
   V_{3,16}(M_\pi^2,M_\pi^2,M_\pi^2,M_\pi^2;M_\pi^2,t,M_\pi^2)
  -\frac{1}{2} t^2 
   V_{3,16}(M_\pi^2,M_\pi^2,M_K^2,M_K^2;M_\pi^2,t,M_\pi^2)
   \nn \\  
   & &
  -\frac{7}{12} t^2 
   V_{3,16}(M_K^2,M_K^2,M_\pi^2,M_K^2;M_\pi^2,t,M_\pi^2)
  -\frac{1}{2} t^2 
   V_{3,16}(M_K^2,M_K^2,M_K^2,M_\eta^2;M_\pi^2,t,M_\pi^2)
   \nn \\ 
   %
   %
   & &
  +\left( 4 M_\pi^2 - 2 t \right)
   V_{3,17}(M_\pi^2,M_\pi^2,M_\pi^2,M_\pi^2;M_\pi^2,t,M_\pi^2)
   \nn \\
   & &
  +\left( 2 M_\pi^2 - t \right)
   V_{3,17}(M_\pi^2,M_\pi^2,M_K^2,M_K^2;M_\pi^2,t,M_\pi^2)
   \nn \\  
   & &
  +\left( 
      \frac{3}{2} M_\pi^2 - \frac{3}{4} t
   \right) 
   V_{3,17}(M_K^2,M_K^2,M_\pi^2,M_K^2;M_\pi^2,t,M_\pi^2)
   \nn \\ 
   & &
  +\left( 
      \frac{3}{2} M_\pi^2 - \frac{3}{4} t
   \right)
   V_{3,17}(M_K^2,M_K^2,M_K^2,M_\eta^2;M_\pi^2,t,M_\pi^2)
   \nn \\
   %
   %
   & &
  +\left( 8 M_\pi^2 - 2 t \right)
   V_{3,19}(M_\pi^2,M_\pi^2,M_\pi^2,M_\pi^2;M_\pi^2,t,M_\pi^2)
   \nn \\
   & &
  +\left( 4 M_\pi^2 - t \right)
   V_{3,19}(M_\pi^2,M_\pi^2,M_K^2,M_K^2;M_\pi^2,t,M_\pi^2)
   \nn \\  
   & &
  +\left( 
      3 M_\pi^2 - \frac{3}{4} t
   \right) 
   V_{3,19}(M_K^2,M_K^2,M_\pi^2,M_K^2;M_\pi^2,t,M_\pi^2)
   \nn \\ 
   & &
  +\left( 
      3 M_\pi^2 - \frac{3}{4} t
   \right)
   V_{3,19}(M_K^2,M_K^2,M_K^2,M_\eta^2;M_\pi^2,t,M_\pi^2)
   \nn \\ 
   %
   %
   & &
  +\left( 4 M_\pi^4 - 2 M_\pi^2 t \right)
   V_{3,21}(M_\pi^2,M_\pi^2,M_\pi^2,M_\pi^2;M_\pi^2,t,M_\pi^2)
   \nn \\
   & &
  +\left( 2 M_\pi^4 - M_\pi^2 t \right)
   V_{3,21}(M_\pi^2,M_\pi^2,M_K^2,M_K^2;M_\pi^2,t,M_\pi^2)
   \nn \\  
   & &
  +\left( 
      \frac{3}{2} M_\pi^4 - \frac{3}{4} M_\pi^2 t
   \right) 
   V_{3,21}(M_K^2,M_K^2,M_\pi^2,M_K^2;M_\pi^2,t,M_\pi^2)
   \nn \\ 
   & &
  +\left( 
      \frac{3}{2} M_\pi^4 - \frac{3}{4} M_\pi^2 t
   \right)
   V_{3,21}(M_K^2,M_K^2,M_K^2,M_\eta^2;M_\pi^2,t,M_\pi^2)
   \nn \\ 
   %
   %
   & &
  -t^2 
   V_{3,23}(M_\pi^2,M_\pi^2,M_\pi^2,M_\pi^2;M_\pi^2,t,M_\pi^2)
  -\frac{1}{2} t^2 
   V_{3,23}(M_\pi^2,M_\pi^2,M_K^2,M_K^2;M_\pi^2,t,M_\pi^2)
   \nn \\  
   & &
  -\frac{3}{8} t^2 
   V_{3,23}(M_K^2,M_K^2,M_\pi^2,M_K^2;M_\pi^2,t,M_\pi^2)
  -\frac{3}{8} t^2 
   V_{3,23}(M_K^2,M_K^2,M_K^2,M_\eta^2;M_\pi^2,t,M_\pi^2)
   \nn \\ 
   %
   %
   & &
  -t^2 
   V_{3,25}(M_\pi^2,M_\pi^2,M_\pi^2,M_\pi^2;M_\pi^2,t,M_\pi^2)
  -\frac{1}{2} t^2 
   V_{3,25}(M_\pi^2,M_\pi^2,M_K^2,M_K^2;M_\pi^2,t,M_\pi^2)
   \nn \\  
   & &
  -\frac{3}{8} t^2 
   V_{3,25}(M_K^2,M_K^2,M_\pi^2,M_K^2;M_\pi^2,t,M_\pi^2)
  -\frac{3}{8} t^2 
   V_{3,25}(M_K^2,M_K^2,M_K^2,M_\eta^2;M_\pi^2,t,M_\pi^2).
   \hspace{15mm}
\eea